%% file: id_tutorial_v7.tex
\documentclass[journal]{IEEEtran}
\usepackage{url}
\usepackage[utf8]{inputenc}
\usepackage{xcolor}
\usepackage{amsmath}
\usepackage{amssymb}

\usepackage[acronyms,nonumberlist,nopostdot,nomain,nogroupskip]{glossaries}
\usepackage{tablefootnote}
\usepackage{booktabs}
\usepackage{tabularx}
\usepackage{tikz}
\usepackage{pgfplots}
\pgfplotsset{compat=newest} 
\pgfplotsset{plot coordinates/math parser=false} 
\newlength\fheight
\newlength\fwidth
\usetikzlibrary{plotmarks,patterns,decorations.pathreplacing,backgrounds,calc,arrows,arrows.meta,spy,matrix}
\usepgfplotslibrary{patchplots,groupplots}
\usepackage{tikzscale}
\usepackage{hyperref}

\usepackage{multirow}
\usepackage{tkz-kiviat}

\usepackage[font=scriptsize]{subcaption}
\usepackage[font=footnotesize]{caption}

\usepackage{mathtools}

\usepackage{dblfloatfix}    
\usepackage{colortbl}

\input{acronymsSorted.tex}
\input{myTikz.tex}

\def\si{\tikz\fill[scale=0.4](0,.35) -- (.25,0) -- (1,.7) -- (.25,.15) -- cycle;}

\makeglossaries

\begin{document}

\title{A Tutorial on Beam Management for \\ 3GPP NR at mmWave Frequencies}

\author{\IEEEauthorblockN{Marco Giordani, \IEEEmembership{Student Member, IEEE}, Michele Polese, \IEEEmembership{Student Member, IEEE},  Arnab Roy, \IEEEmembership{Member, IEEE}, Douglas Castor, \IEEEmembership{Member, IEEE}, Michele Zorzi, \IEEEmembership{Fellow, IEEE}}
\thanks{Marco Giordani, Michele Polese and Michele Zorzi are with the Department of Information Engineering (DEI), University of Padova, Italy, and Consorzio Futuro in Ricerca (CFR), Italy.
Email:\{giordani,polesemi,zorzi\}@dei.unipd.it.

Arnab Roy and Douglas Castor are with InterDigital Communications, Inc., USA. 
Email:\{arnab.roy,douglas.castor\}@interdigital.com.

\noindent Part of this work has been submitted for publication at Med-Hoc-Net 2018~\cite{giordani2018initial}.
}
}

\maketitle

\glsunset{nr}

\begin{abstract}
The millimeter wave (mmWave) frequencies offer
the availability of huge bandwidths to provide unprecedented
data rates to next-generation cellular mobile terminals. However,
 mmWave links are highly susceptible to rapid channel
variations and suffer from severe free-space pathloss and atmospheric absorption. 
To address these challenges, the base stations and the mobile
terminals will use highly directional antennas to achieve sufficient link budget in wide area networks.
The consequence is the need for precise alignment of the transmitter and the receiver beams, an operation which may increase the latency of establishing a link, and has important implications for
control layer procedures, such as initial access, handover and beam tracking.
This tutorial provides an overview of recently
proposed measurement techniques for beam and mobility management in mmWave cellular networks, and gives insights into the design of accurate, reactive and robust control schemes suitable for a 3GPP \gls{nr} cellular network.
We will illustrate that the best strategy depends on the specific environment in which the nodes are deployed, and give guidelines to inform the optimal
choice as a function of the system parameters.
\end{abstract}

\begin{IEEEkeywords}
5G, NR, mmWave, 3GPP, beam management.
\end{IEEEkeywords}

\section{Introduction}

From analog through \gls{lte}, each generation of
mobile technology has been motivated by the need to address the challenges not overcome by its predecessor.
 The \gls{5g} of mobile technology is positioned to address the demands and
business contexts of 2020 and beyond. It is expected to enable a fully mobile and connected society, related to the tremendous growth in connectivity and density/volume of traffic that will be required in the near future \cite{cisco2017}, to provide and guarantee: (i) very high throughput (1 Gbps or more), to support ultra-high definition video and virtual reality applications; (ii) very low latency (even less than 1 ms in some cases), to support real-time mobile control and Device-to-Device (D2D) applications and communications; (iii) ultra high reliability; (iv) low energy consumption; and (v) ultra high connectivity resilience and robustness \cite{boccardi2014five} to support advanced safety applications and services.
In order to meet these complex and sometimes contradictory
requirements, 5G will
encompass both an evolution of traditional 4G-\gls{lte} networks and the addition of a new radio access technology, globally
standardized by the \gls{3gpp} as \gls{nr} \cite{ericsson2017review, 38802}.

In this context, the \gls{mmwave} spectrum -- roughly above 10 GHz\footnote{Although strictly speaking mmWave bands include frequencies between
30 and 300 GHz, industry has loosely defined it to include any frequency
above 10 GHz.} -- has
been considered as an enabler of the 5G performance requirements
in micro and picocellular networks \cite{rangan2017potentials, rappaport2013millimeter}.
These frequencies offer much more bandwidth than current cellular
systems in the congested bands below 6 GHz, and initial capacity estimates have suggested that
networks operating at mmWaves can offer orders of magnitude higher bit-rates than 4G systems~\cite{akoum2012coverage}.
Nonetheless, the higher carrier frequency makes the propagation conditions harsher than at the lower frequencies traditionally used for wireless services,
especially in terms of robustness~\cite{pi2011introduction}.
Signals propagating in the mmWave band suffer from increased
pathloss and severe channel intermittency, and are blocked by many common
materials such as brick or mortar~\cite{lu2012modeling}, and even the changing position of
the body relative to the mobile device can lead to rapid drops in signal strength.

To deal with these impairments, next-generation cellular
networks must provide a set of mechanisms by which \glspl{ue} and mmWave \gls{gnb} stations\footnote{Notice that \gls{gnb} is the \gls{nr} term for a base station.} establish highly directional transmission links, typically using high-dimensional phased
arrays, to benefit from the resulting beamforming gain
and sustain an acceptable communication quality.
Directional links, however, require fine alignment of the transmitter and receiver beams, achieved through a set of operations known as \emph{beam management}. They are fundamental to 
perform a variety of
control tasks including 
(i) \gls{ia} \cite{giordani2016comparative, giordani2016initial} for idle users, which
allows a mobile \gls{ue} to establish a physical
link connection with a \gls{gnb}, and (ii)
beam tracking, for connected users, which enable beam adaptation schemes, or handover, path selection and radio link failure recovery procedures \cite{giordani2017tracking,polese2017jsac}.
In current \gls{lte} systems, these control procedures are performed
using omnidirectional signals, and beamforming or
other directional transmissions can only be performed after
a physical link is established, for data plane transmissions. 
On the other hand, in
the mmWave bands, it may be essential to exploit the
antenna gains even during initial access and, in general, for control operations. Omnidirectional control signaling at such high frequencies, indeed, may generate a mismatch between the relatively short range at which a cell can be detected or the control signals can be received (control-plane range),
and the much longer range at which a user could
send and receive data when using beamforming (data-plane range).
However, directionality can significantly delay the access procedures and make the performance more sensitive to the beam alignment. These are particularly important issues in \gls{5g} networks,
and motivate the need to extend current \gls{lte} control procedures with innovative mmWave-aware beam management algorithms and methods.

\subsection{Contributions}


This paper is a tutorial on the design and dimensioning of beam management frameworks for mmWave cellular networks. In particular, we consider the parameters of interest for 3GPP \gls{nr} networks, which will support carrier frequencies up to 52.6 GHz~\cite{38802}. We also report an evaluation of beam management techniques, including initial access and tracking strategies, for cellular networks operating at \glspl{mmwave} under realistic \gls{nr} settings and channel configurations, and describe how to optimally design fast, accurate and robust control-plane management schemes through measurement reports in different scenarios.
More specifically, in this tutorial we:
\begin{itemize}
\item Provide an overview of the most effective measurement collection frameworks for \gls{5g} systems operating at mmWaves. We focus on \gls{dl} and \gls{ul} frameworks, according to whether the  reference signals are sent from the \glspl{gnb} to the \glspl{ue} or vice versa, respectively, and  on \gls{nsa} and \gls{sa} architectures, according to whether the control plane is managed with the support of an \gls{lte} overlay or not, respectively.
A \gls{dl} configuration is in line with the 3GPP
specifications for NR and reduces the energy consumption at the
\gls{ue} side, but it may be lead to a worse beam management performance than in the \gls{ul}.
Moreover, when considering stable and dense scenarios which
are marginally affected by the variability of the mmWave channel,
an \gls{sa} architecture is preferable for the design of fast \gls{ia} procedures, while an \gls{nsa} scheme may be preferable for reducing the impact of the overhead
on the system performance and enable more robust and stable communication capabilities. 

\item Simulate the performance of the presented measurement frameworks in terms of \emph{signal detection accuracy}, using a realistic mmWave channel model based on real-world measurements conducted in a dense, urban scenario in which environmental obstructions (i.e.,
urban buildings) can occlude the path between the transmitter and the receiver. 
The tutorial shows that accurate beam management operations can be guaranteed when configuring narrow beams for the transmissions, small subcarrier spacings, denser network deployments and by adopting \emph{frequency diversity} schemes.

\item Analyze the \emph{reactiveness} (i.e., how quickly a mobile user gets access to the network and how quickly the framework is able to detect an updated channel condition), and the \emph{overhead} (i.e., how many time and frequency resources should be allocated for the measurement operations). 
In general, fast initial access and tracking schemes are ensured by allocating a large number of time/frequency resources to the users in the system, at the expense of an increased overhead, and by using advanced beamforming capabilities (e.g., digital or hybrid beamforming), which allow the transceiver to sweep multiple directions at any given time.

\item Illustrate some of the complex and interesting trade-offs to be considered when designing solutions for next-generation  cellular networks by examining a wide set of parameters based on \gls{3gpp} \gls{nr} considerations and agreements (e.g., the frame structure and other relevant physical-layer aspects).
\end{itemize}

In general, the results prove that the optimal design choices for implementing efficient and fast initial access and reactive tracking of the mobile user strictly depend on the specific environment in which the users are deployed, and must
account for several specific features such as the base stations density, the antenna geometry, the beamforming configuration and the level of integration and harmonization of different technologies.

\subsection{Organization}

The sections of this tutorial are organized as follows. Sec.~\ref{sec:rel_work} reports the related work on beam management at mmWave frequencies.
Sec.~\ref{sec:3gpp_specifications} provide basic information on the \gls{3gpp} Release 15 frame structure for \gls{nr}, and presents the candidate \gls{dl} and \gls{ul} measurement signals that can be collected by the \gls{nr} nodes for the beam management operations.
Sec.~\ref{sec:meas_frameworks} describes the beam management frameworks whose performance will be analyzed, simulated and compared in the remainder of the work.
Sec.~\ref{sec:sim_params} defines the parameters that affect the performance of beam management in \gls{nr}.
Sec.~\ref{sec:results} reports a performance evaluation and some considerations on the trade-offs and on which are the best configurations for beam management frameworks. Additional considerations and final remarks, aiming at providing guidelines for selecting the optimal IA and tracking configuration settings as a function of the system parameters, are stated in Sec.~\ref{sec:considerations}. 
Finally, Sec.~\ref{sec:conclusions} concludes the paper.

\begin{table*}
\small
  \centering
  \def\tabularxcolumn#1{m{#1}}
  \setlength\belowcaptionskip{-0.4cm}
    \renewcommand{\arraystretch}{1.2}
  \begin{tabularx}{0.95\textwidth}{@{}l|X@{}}
  \toprule
  Topic & Relevant References \\ \hline
  IEEE 802.11ad \cite{nitsche201460ghz} & \cite{wang2009beam, santosa2006adhoc,chandra2014adaptive}. Not suitable for long-range, dynamic and outdoor scenarios.  \\ 	\hline
Initial Access \cite{giordani2016comparative, giordani2016initial, liu2016user} &  \cite{jeong2015random, barati2015directional, barati2015directionalasilomar} Exhaustive search. 
\newline  \cite{desai2014initial,wei2017exhaustive,choi2015beam} More advanced searching schemes.
\newline  \cite{capone2015context, capone2015obstacle, li2013anchor,abbas2016context} Context-aware initial access. 
\newline \cite{alkhateeb2017initial,li2017beamformed} Performance comparison. \\\hline
Beam Management \cite{polese2017jsac} & \cite{cacciapuoti2017mobility,palacios2017tracking,jayaprakasam2017robust} Mobility-aware strategies.
\newline \cite{giordani2016efficient, giordani2016multi, polese2016performance,tesema2015mobility,semiari2017caching,prelcic2017millimeter} Multi-connectivity solutions.\\
  \bottomrule    
  \end{tabularx}
  \caption{Relevant literature on measurement reporting, initial access and beam management strategies for mmWave networks.}
  \label{table:related_work}
\end{table*}

\section{Related Work }
\label{sec:rel_work}

Measurement reporting is quite  straightforward in \gls{lte} \cite{schwarz2010calculation}: the \gls{dl} channel
quality is estimated from an omnidirectional signal  called the \gls{crs}, which is regularly monitored by each \gls{ue} in connected state to create a wideband channel estimate that can
be used both for demodulating downlink transmissions and for
estimating the channel quality \cite{giordani2016channel}.
 However, when considering mmWave networks, in addition to the rapid
variations of the channel,  \gls{crs}-based estimation is challenging  due to the directional nature of the communication, thus requiring 
the network and the \gls{ue} to constantly monitor the direction of transmission of each potential
link. 
Tracking changing directions can decrease the rate at which the network can adapt, and can
be a major obstacle in providing robust and ubiquitous service in the face of variable link quality. 
In addition,
the \gls{ue} and the \gls{gnb} may only be able to listen to one direction at a time, thus making it hard
to receive the control signaling necessary to switch paths.

To overcome these limitations, several approaches in the literature, as summarized in Table~\ref{table:related_work}, have proposed directional-based schemes to enable efficient control procedures for both the idle and the connected mobile terminals, as surveyed in the following paragraphs.

 Papers on \gls{ia}\footnote{We refer to works \cite{giordani2016comparative, giordani2016initial, liu2016user} for a  detailed  taxonomy of recent IA strategies.} and tracking in \gls{5g} mmWave cellular systems are very
recent.
Most  literature refers to challenges that have been analyzed in the past at lower frequencies in ad hoc wireless network scenarios or, more recently, referred to the 60 GHz IEEE 802.11ad
WLAN and WPAN scenarios (e.g., \cite{nitsche201460ghz,wang2009beam, santosa2006adhoc}).
However, most of the proposed
solutions are unsuitable for  next-generation cellular network
requirements and present many limitations (e.g., they
are appropriate for short-range, static and indoor scenarios,
which do not match well the requirements of \gls{5g} systems).
Therefore, new specifically designed solutions for cellular
networks need to be found.

In \cite{jeong2015random, barati2015directional}, the authors propose an exhaustive method that performs directional communication
over mmWave frequencies by periodically transmitting
synchronization signals to scan the angular
space.
The result of this
approach is that the growth of the number of antenna elements at either the transmitter or
the receiver provides a large performance gain compared to the case of an omnidirectional
antenna. 
However, this solution leads to a long duration of the \gls{ia} with respect
to LTE, and poorly reactive tracking.
Similarly, in \cite{barati2015directionalasilomar}, measurement reporting  design options are compared, considering different scanning
and signaling procedures, to evaluate access delay and system overhead. The channel structure and multiple access issues are also considered. The analysis demonstrates significant
benefits of low-resolution fully digital architectures in comparison to single stream analog
beamforming.
Additionally, more sophisticated discovery techniques (e.g., \cite{desai2014initial,wei2017exhaustive}) alleviate the exhaustive search delay through the implementation of a multi-phase hierarchical
procedure based on the access signals being initially sent in few directions over wide beams, which are iteratively refined until the communication is sufficiently directional.
In \cite{choi2015beam} a low-complexity beam selection
method by low-cost analog beamforming is derived by exploiting a certain sparsity of mmWave channels. It is shown that beam
selection can be carried out without explicit channel estimation, using the notion of compressive sensing.

The issue of designing efficient beam management solutions for mmWave networks is addressed in~\cite{cacciapuoti2017mobility}, in which the author designs a mobility-aware user association strategy to overcome the limitations of the conventional power-based association schemes in a mobile 5G scenario.
Other relevant papers on this topic include~\cite{palacios2017tracking}, in which the authors propose smart beam tracking strategies for fast mmWave link establishment and maintenance under node mobility.
In \cite{jayaprakasam2017robust},  the authors proposed the use of an extended Kalman filter to enable a static
base station, equipped with a digital beamformer, to effectively track
a mobile node equipped with an analog beamformer after 
initial channel acquisition, with the goal of reducing the alignment error and guarantee a more durable connectivity.
Recently, robust \gls{ia} and tracking schemes have been designed by leveraging out-of-band information to estimate the mmWave channel. 
In \cite{polese2017jsac,giordani2016efficient, giordani2016multi, polese2016performance}
an approach where \gls{5g} cells operating
at mmWaves (offering much higher rates) and traditional 4G
cells below 6 GHz (providing much more robust operation)
are  employed in parallel have been proved to enable fast and resilient tracking operations.
In~\cite{tesema2015mobility}, a framework which integrates both LTE and 5G interfaces is proposed as a solution for mobility-related link failures
and throughput degradation of cell-edge users, relying on coordinated transmissions from
cooperating cells are coordinated for both data and control signals.
In~\cite{semiari2017caching}, a novel approach for analyzing and managing mobility in joint sub-6GHz--mmWave networks is proposed by leveraging on device caching along with the capabilities of dual-mode base stations to minimize handover failures, reduce inter-frequency measurement, reduce energy consumption, and provide seamless mobility in emerging dense heterogeneous networks.
Moreover, the authors in \cite{prelcic2017millimeter} illustrate how to exploit spatial congruence between signals in different frequency bands and extract mmWave channel parameters from side information obtained in another band.
Despite some advantages, the use of out-of-band information for the 5G control plane management poses new challenges that remain unsolved and which deserve further investigation.

Context information  can also be exploited
to improve the cell discovery procedure and minimize the delay \cite{capone2015context, capone2015obstacle}, while capturing the effects of position inaccuracy
in the presence of obstacles.
In the
scheme proposed in \cite{li2013anchor}, booster cells (operating at mmWave) are deployed under the coverage of an anchor cell
(operating at \gls{lte} frequencies). The anchor base station gets control
over \gls{ia} informing the booster cell about user locations, in order
to enable mmWave \gls{gnb} to directly steer towards the user
position.
Finally, in \cite{abbas2016context}, the authors studied how the
performance of analog beamforming degrades in the presence
of angular errors in the available Context Information during
the initial access or tracking procedures, according to the status of the \gls{ue} (connected or non-connected, respectively).

\begin{table*}[b!]
\small
  \centering
  \def\tabularxcolumn#1{m{#1}}
    \renewcommand{\arraystretch}{1}
  \begin{tabularx}{0.95\textwidth}{@{}lXX@{}}
  \toprule
  & Initial Access (Idle \gls{ue}) & Tracking (Connected \gls{ue}) \\ \midrule
  Downlink & \gls{ss} blocks (carrying the 
  \gls{pss}, the
  \gls{sss}, and the
  \gls{pbch}). \newline
  See references~\cite{38802,38211,samsung2017ss,moto2017ss,nokia2017ss,huawei2017beam}.
  & \glspl{csirs} and \gls{ss} blocks. \newline
  See references~\cite{38802,38211,ericsson2017CSI,samsung2017periodicity,huawei2017summary,huawei2017measurement,ericsson2017cellquality,ZTE2017beammanagement}. \\ \midrule
  Uplink & 3GPP does not use  uplink signals for initial access, but the usage of \glspl{srs} has been proposed in~\cite{giordani2016multi,giordani2016efficient,polese2017jsac} & \glspl{srs}.
  See references~\cite{38802,38211,interdigital2017srs,catt2017srs}. \\
  \bottomrule
    
  \end{tabularx}
  \caption{Reference signals for beam management operations, for users in idle and connected states, in downlink or uplink.}
  \label{table:signals}
\end{table*}

The performance of the  association techniques also depends on the beamforming architecture implemented in the transceivers.
Preliminary works aiming at finding the optimal beamforming strategy refer to WLAN scenarios.
 For example, the algorithm proposed in \cite{chandra2014adaptive} takes into account the spatial distribution of nodes to
allocate the beamwidth of each antenna pattern in an adaptive fashion and satisfy the required link budget criterion.
 Since the proposed algorithm
minimizes the collisions, it also minimizes the average time required to transmit a data packet from the source to the destination through a specific direction.
In \gls{5g} scenarios, papers \cite{jeong2015random, barati2015directional,desai2014initial} give some insights on trade-offs among different beamforming architectures  in terms of users' communication quality.
More recently, articles \cite{alkhateeb2017initial,li2017beamformed} evaluate the mmWave cellular network performance while accounting for the beam training, association overhead and beamforming architecture.
The  results show that, although employing wide beams, initial beam training with full pilot reuse is nearly as good as perfect beam alignment. 
However, they lack considerations on the latest  3GPP specifications for NR. Finally, paper~\cite{liu2018initial} provides an overview of the main features of \gls{nr} with respect to initial access and multi-beam operations, and article~\cite{onggosanusi2018modular} reports the details on the collection of channel state information in \gls{nr}. However, both these papers only present a high level overview, and do not include a comprehensive performance evaluation of \gls{nr} beam management frameworks at mmWave frequencies.

The above discussion makes it apparent how next-generation mmWave cellular networks
should support a mechanism by which the users and the infrastructure can quickly determine the best directions to establish the mmWave links, an operation which may increase
the latency and the overhead of the communication and have a substantial impact on the overall network performance.
In the remainder of this paper we will provide guidelines to characterize the
optimal beam management strategies  as a function of a variety of realistic system parameters.

\section{Frame Structure and Signals for\\3GPP NR at mmWave Frequencies}
\label{sec:3gpp_specifications}
Given that \gls{nr} will support communication at mmWave frequencies, it is necessary to account for beamforming and directionality in the design of its \gls{phy} and \gls{mac} layers. 
The \gls{nr} specifications will thus include a set of parameters for the frame structure dedicated to high carrier frequencies, as well as synchronization and reference signals that enable beam management procedures~\cite{38802}.
In this regard, in Sec.~\ref{sec:frame} and Sec.~\ref{sec:meas} 
we introduce the 3GPP frame structure and measurement signals proposed for \gls{nr}, respectively, which
will provide the necessary background for the remainder of this tutorial. 

\subsection{NR Frame Structure}
\label{sec:frame}
The \gls{3gpp} technical specification in~\cite{38211} and the report in~\cite{38802} provide the specifications for the \gls{phy} layer. Both \gls{fdd} and \gls{tdd} will be supported. 

The \textit{waveform} is \gls{ofdm} with a cyclic prefix. Different numerologies\footnote{The term numerology refers to a set of parameters for the waveform, such as subcarrier spacing and cyclic prefix duration for \gls{ofdm}~\cite{zaidi2016waveform}.} will be used, in order to address the different use cases of 5G~\cite{osseiran2014scenarios}. 
The frame structure  follows a time and frequency grid similar to that of \gls{lte}, with a higher number of configurable parameters. 
The subcarrier spacing is $15\times2^n$ kHz, $n \in \mathbb{Z}, n \le 4$. In Release 15, there will be at most 3300 subcarriers, for a maximum bandwidth of 400 MHz. 
A \emph{frame} lasts 10 ms, with 10 subframes of 1 ms. 
It will be possible to multiplex different numerologies for a given carrier frequency, and the whole communication must be aligned on a subframe basis. 
A \textit{slot} is composed of 14 \gls{ofdm} symbols. There are multiple slots in a subframe, and their number is given by the numerology used, since the symbol duration is inversely proportional to the subcarrier spacing~\cite{ericsson2017review}. \emph{Mini-slots} are also supported: they can be as small as 2 \gls{ofdm} symbol and have variable length, and can be positioned asynchronously with respect to the beginning of the slot (so that low-latency data can be sent without waiting for the whole slot duration). 

\begin{figure}[t!]
\setlength\belowcaptionskip{-0.4cm}
	\centering
	\includegraphics[width=0.25\textwidth]{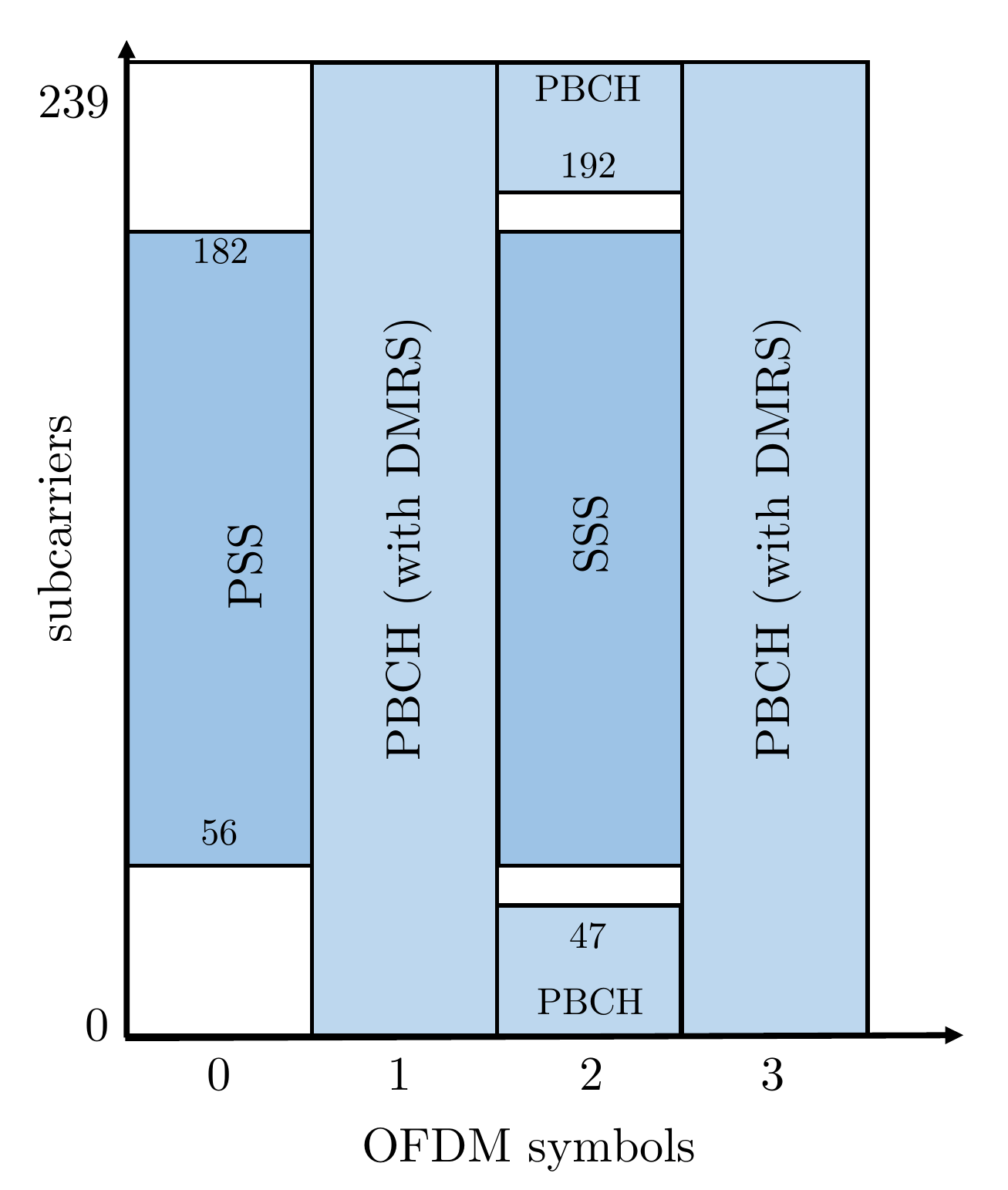}
	\caption{\gls{ss} block structure~\cite{38300}.}
	\label{fig:ssblock_struct}
\end{figure}

\begin{figure*}[t]
	\centering
	\begin{subfigure}[t]{0.99\columnwidth}
		\centering
		\includegraphics[width=0.9\textwidth]{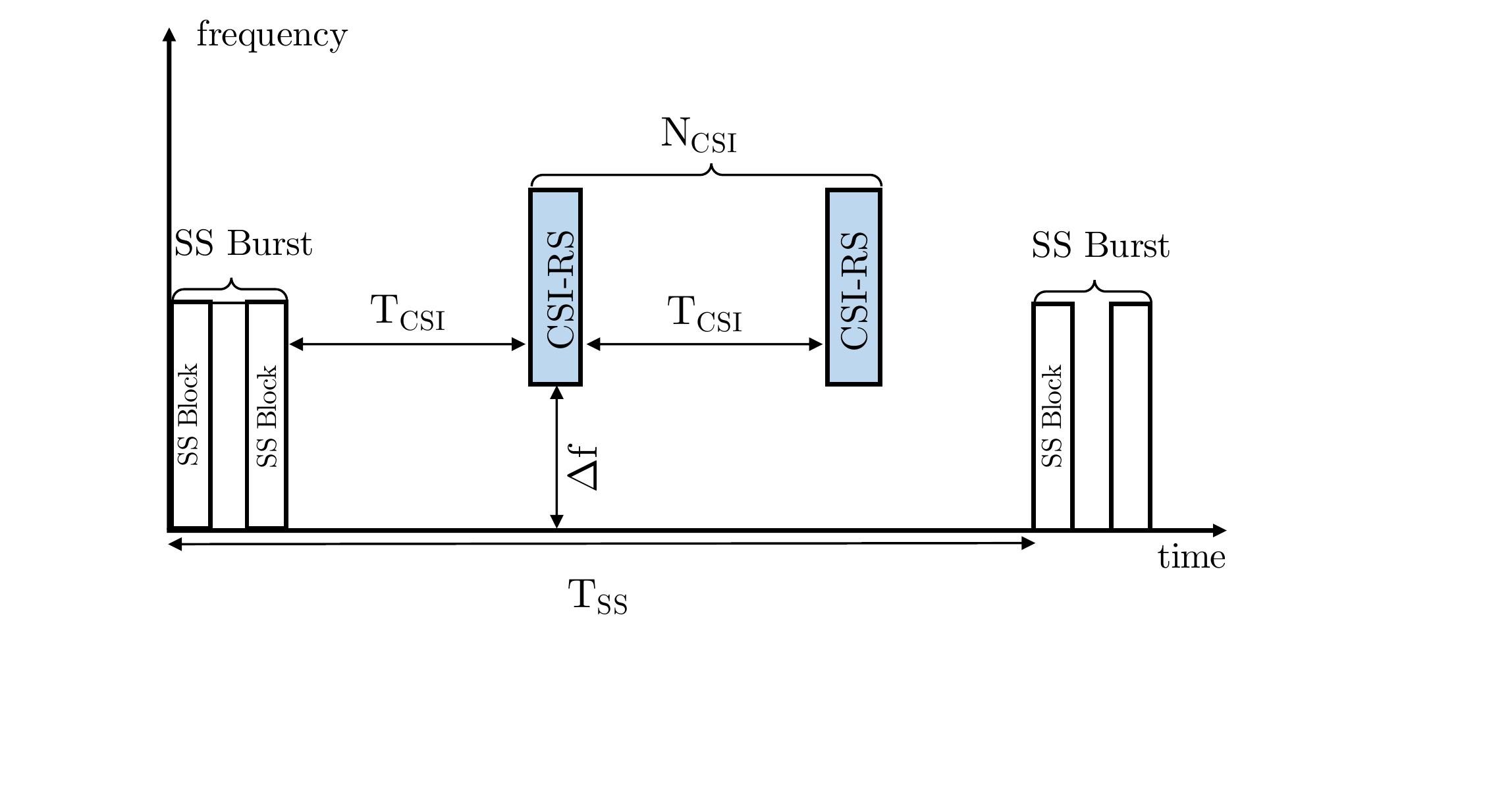}
		\caption{Option 1: the first CSI-RS is sent $T_{\rm CSI}$ ms after an \gls{ss} burst.}
		\label{fig:csi1}
	\end{subfigure}
	\begin{subfigure}[t]{0.99\columnwidth}
		\centering
		\includegraphics[width=0.9\textwidth]{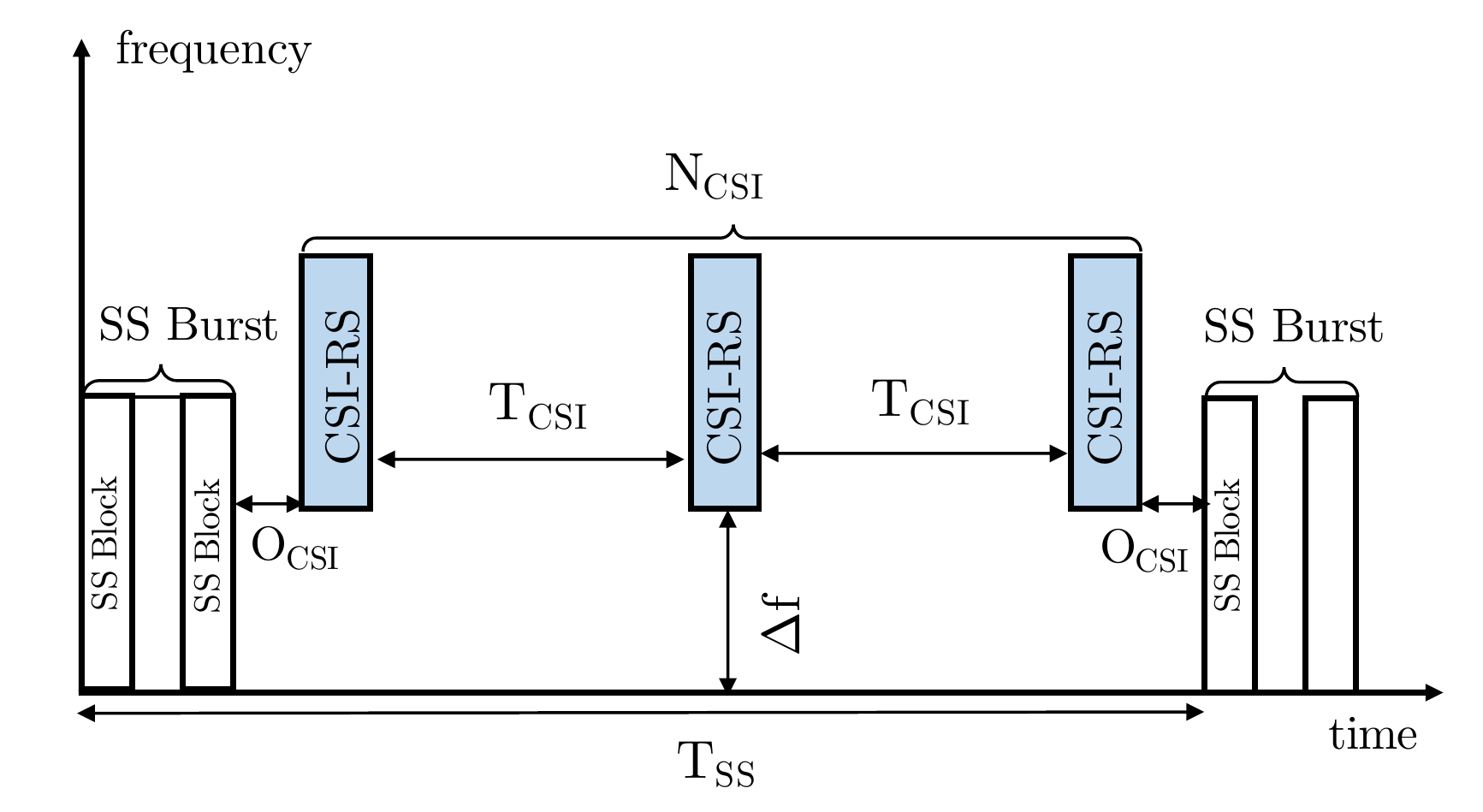}
		\caption{Option 2: the first CSI-RC is sent $O_{\rm CSI}$ ms after an \gls{ss} burst.}
		\label{fig:csi2}
	\end{subfigure}
	\caption{Examples of CSI-RS measurement window and periodicity configurations. \gls{ss} blocks are sent every $T_{\rm SS}$ ms, and they embed time and frequency offsets indicating the time and frequency allocation of CSI-RS signals within the frame structure.}
	\label{fig:offset_ss_csi}
\end{figure*}

\subsection{NR Measurements for Beam Management}
\label{sec:meas}
Regular beam management operations are based on the control messages which are periodically exchanged between the transmitter and the receiver nodes. 
In the following paragraphs we will review the most relevant \gls{dl} and \gls{ul} measurement signals supported by 3GPP \gls{nr} for beam management purposes, as summarized in Table~\ref{table:signals}.

\textbf{Downlink Measurements: SS Blocks.}
In the most recent versions of the 3GPP specifications~\cite{38211}, the concept of \gls{ss} block and burst emerged for periodic synchronization signal transmission from the \glspl{gnb}. 
An \gls{ss} block is a group of 4 \gls{ofdm} symbols~\cite[Sec. 7.4.3]{38211} in time and 240 subcarriers in frequency (i.e., 20 resource blocks)~\cite{samsung2017ss}, as shown in Fig.~\ref{fig:ssblock_struct}. 
It carries the \gls{pss}, the \gls{sss} and the \gls{pbch}. The \gls{dmrs} associated with the \gls{pbch} can be used to estimate the \gls{rsrp} of the \gls{ss} block. In a slot of 14 symbols, there are two possible locations for \gls{ss} blocks: symbols 2-5 and symbols 8-11.

The \gls{ss} blocks are grouped into the first 5 ms of an \gls{ss} burst~\cite{moto2017ss}, which can have different periodicities $T_{\rm SS}$. At the time of writing, the value of $T_{\rm SS}$ is still under discussion in \gls{3gpp}, and the candidates are $T_{\rm SS} \in \{5, 10, 20, 40, 80, 160\}$ ms~\cite{38331}. When accessing the network for the first time, the \gls{ue} should assume a periodicity $T_{\rm SS} = 20$ ms~\cite{38213}.

The maximum number $L$ of \gls{ss} blocks in a burst is frequency-dependent~\cite{moto2017ss}, and above 6 GHz there could be up to 64
blocks per burst.
When considering frequencies for which beam operations are required~\cite{samsung2016framework}, each \gls{ss} block can be mapped to a certain angular direction. 
To reduce the impact of SS
transmissions, SS can be sent through wide beams, while
data transmission for the active UE is usually performed
through narrow beams, to increase the gain produced by
beamforming~\cite{huawei2017beam}.\\

\textbf{Downlink Measurements: CSI-RS.}
It has been agreed that \glspl{csirs} can be used for \gls{rrm} measurements for mobility management purposes in connected mode \cite{38802}. 
As in LTE, it shall be possible to configure multiple CSI-RS to the same \gls{ss} burst, in such a way that the UE can first obtain synchronization with a given cell using the \gls{ss} bursts, and then use that as a reference to search for CSI-RS resources~\cite{ericsson2017CSI}. Therefore, the CSI-RS measurement window configuration should contain at least the periodicity and time/frequency offsets relative to the associated \gls{ss} burst. 
Fig.~\ref{fig:offset_ss_csi} shows the two options we consider for the time offset of the \gls{csirs} transmissions. The first option, shown in Fig.~\ref{fig:csi1}, allows the transmission of the first \gls{csirs} $T_{\rm CSI}$ ms after the end of an \gls{ss} burst. The second one, shown in Fig.~\ref{fig:csi2}, has an additional parameter, i.e., an offset in time $O_{\rm CSI}$, which represents the time interval between the end of the \gls{ss} burst and the first \gls{csirs}.  
The \glspl{csirs}, which may not necessarily be broadcast through all the available frequency resources~\cite{samsung2017periodicity},  
may span $N = $1, 2 or 4 \gls{ofdm} symbols \cite{qualcomm2017csi}. 
For periodic \gls{csirs} transmissions, the supported periodicities are $T_{\rm CSI, slot} \in \{5, 10, 20, 40, 80, 160, 320, 640\}$~slots~\cite{38211}, thus the actual periodicity in time depends on the slot duration.

As we assessed in the previous sections of this work, when considering  directional communications, the best directions for the beams of the transceiver need to be periodically identified (e.g., through  beam search operations), in order to maintain the alignment between the communicating nodes. 
For this purpose, SS- and CSI-based measurement results can be jointly used to reflect the different coverage which can be achieved through different beamforming architectures~\cite{huawei2017measurement}.
As far as CSI signals are concerned,  the communication quality can be derived by averaging the signal quality from the $N_{\rm CSI, RX}$ best beams among all the available ones, where the value of $N_{\rm CSI, RX}$ can be configured to 1 or more than 1  \cite{ericsson2017CSI}\footnote{The maximum value for $N_{\rm CSI, RX}$ has not been standardized yet. In~\cite{38331} it is specified that, for the derivation of the quality of a cell, the \glspl{ue} should consider an absolute threshold, and average the beams with quality above the threshold, up to $N_{\rm CSI, RX}$ beams. If there are no beams above threshold, then the best one (regardless of its absolute quality) should be selected for the cell quality derivation.}.
Nevertheless, to avoid the high overhead associated with wide spatial domain coverage with a huge number of very narrow beams, on which \glspl{csirs} are transmitted, it is reasonable to consider transmitting only subsets of those beams, based on the locations of the active UEs. This is also important for UE power consumption considerations~\cite{ZTE2017beammanagement}.
 For example, the measurement results based on SS blocks (and referred to a subset of transmitting directions) can be used to narrow down the CSI-RS resource sets based on which a UE performs measurements for beam management, thereby increasing the energy efficiency.\\



\textbf{Uplink Measurements: SRS}
The \glspl{srs} are used to monitor the uplink channel quality, and are transmitted by the \gls{ue} and received by the \glspl{gnb}. According to~\cite{interdigital2017srs}, their transmission is scheduled by the \gls{gnb} to which the \gls{ue} is attached, which also signals to the \gls{ue} the resource and direction to use for the transmission of the \gls{srs}. 
The \gls{ue} may be configured with multiple \glspl{srs} for beam management. Each resource may be periodic (i.e., configured at the slot level), semi-persistent (also at the slot level, but it can be activated or deactivated with messages from the \gls{gnb}) and a-periodic (the \gls{srs} transmission is triggered by the \gls{gnb})~\cite{catt2017srs}.
The \glspl{srs} can span 1 to 4 \gls{ofdm} symbols, and a portion of the entire bandwidth available at the \gls{ue}~\cite{interdigital2017srs}.



\begin{table*}
\small
\def\tabularxcolumn#1{m{#1}}
\begin{center}
\renewcommand{\arraystretch}{1.3}

  \begin{tabularx}{0.9\textwidth}{|>{\hsize=0.88\hsize}l|>{\hsize=1.04\hsize}X|>{\hsize=1.04\hsize}X|>{\hsize=1.04\hsize}X|}
  \hline
   \rowcolor{gray!25} & \gls{sa}-\gls{dl} & \gls{nsa}-\gls{dl} & \gls{nsa}-\gls{ul} \\  \hline

  \cellcolor{gray!10} Multi-\gls{rat} connectivity & Not available & \multicolumn{2}{>{\hsize=2.08\hsize}X|}{\gls{lte} overlay available for robust control operations and quick data fallback~\cite{tesema2015mobility,giordani2016multi,polese2016performance}.} \\\hline
  \cellcolor{gray!10} Reference signal transmission & Downlink & Downlink & Uplink \\\hline

  \cellcolor{gray!10} Network coordination & Not available & \multicolumn{2}{>{\hsize=2.08\hsize}X|}{Possibility of using a centralized controller~\cite{polese2017jsac}.} \\ 

  \hline
  \multicolumn{4}{c}{\vspace{0.3cm}}\\
  \hline

  \rowcolor{gray!25} Beam management phase   & \gls{sa}-\gls{dl} & \gls{nsa}-\gls{dl} & \gls{nsa}-\gls{ul} \\  \hline

  \cellcolor{gray!10} Beam sweep & \multicolumn{2}{>{\hsize=2.08\hsize}X|}{Exhaustive search based on  \gls{ss} blocks~\cite{jeong2015random}.} &  Based on \gls{srs}~\cite{giordani2016efficient}. \\
  \hline

  \cellcolor{gray!10} Beam measurement & \gls{ue}-side & \gls{ue}-side & \gls{gnb}-side \\  \hline

  \cellcolor{gray!10} Beam determination & \multicolumn{2}{>{\hsize=2.08\hsize}X|}{The \gls{ue} selects the optimal communication direction.} & Each \gls{gnb} sends information on the received beams to a central controller, which selects the best beam pair~\cite{giordani2016multi}.\\  \hline

  \cellcolor{gray!10} Beam reporting & Exhaustive search at the gNB side ~\cite{ericsson2016prach}. 
  & The \gls{ue} signals the best beam pair using \gls{lte}, a \gls{rach} opportunity in that direction is then scheduled. & The \gls{gnb} signals the best beam pair using \gls{lte}, a \gls{rach} opportunity in that direction is then scheduled.  \\

  \hline
  \end{tabularx}
\end{center}

\caption{Comparison of the beam management frameworks.}
\label{table:frameworks}
\end{table*}

\section{Beam Management Frameworks for\\5G Cellular Systems}
\label{sec:meas_frameworks}

In this section, we present three measurement frameworks for both initial access and tracking purposes, whose performance will be investigated and compared in Sec. \ref{sec:results}. 

As we introduced in the above sections of this tutorial, the \gls{nr} specifications include a set of basic beam-related procedures~\cite{38802} for the control of multiple beams at frequencies above 6 GHz and the related terminologies, which are based on the reference signals described in Sec.~\ref{sec:3gpp_specifications}. The different operations are categorized under the term \textit{beam management}, which is composed of four different operations:
\begin{itemize}
  \item \textit{Beam sweeping}, i.e., covering a spatial area with a set of beams transmitted and received according to pre-specified intervals and directions.
  \item \textit{Beam measurement}, i.e., the evaluation of the quality of the received signal at the \gls{gnb} or at the \gls{ue}. Different metrics could be used~\cite{38215}. In this paper, we consider
the Signal to Noise Ratio (SNR), which is the  average of
the received power on synchronization
signals divided by the noise power.
  \item \textit{Beam determination}, i.e., the selection of the suitable beam or beams either at the \gls{gnb} or at the \gls{ue}, according to the measurements obtained with the beam measurement procedure.
  \item \textit{Beam reporting}, i.e., the procedure used by the  \gls{ue} to send beam quality and beam decision information to the \gls{ran}. 
 \end{itemize}
These procedures are periodically repeated  to update the optimal transmitter and receiver beam pair over time.

We consider a \textit{\gls{nsa}} or a \textit{	standalone (SA)} architecture. 
Non-standalone is a deployment configuration in which a NR \gls{gnb} uses an LTE cell as support for the control plane management \cite{38801} and mobile terminals exploit \emph{multi-connectivity} to maintain multiple possible connections (e.g., 4G and 5G overlays) to different cells
so that drops in one link can be overcome by switching data paths \cite{giordani2016multi, giordani2016efficient,polese2017jsac,tesema2015mobility,polese2016performance,37340}. 
Mobiles in a \gls{nsa} deployment can benefit from both the
high bit-rates that can be provided by the mmWave links and the more robust, but lower-
rate, legacy channels, thereby opening up new ways of solving capacity issues, as well as new
ways of providing good mobile network performance and robustness.
Conversely,  with the standalone option, there is no \gls{lte} control plane, therefore the integration between \gls{lte} and NR is not~supported.

The measurement frameworks can be also based on a \textit{downlink} or an \textit{uplink} beam management architecture. 
In the first case, the \glspl{gnb} transmit synchronization and reference signals (i.e., \gls{ss} blocks and \glspl{csirs}) which are collected by the surrounding \glspl{ue}, while in the second case the measurements are based on \glspl{srs} forwarded by the mobile terminal instead.
Notice that the increasing heterogeneity in cellular networks is dramatically changing our traditional notion of
a communication cell \cite{boccardi2014five}, making the role of the uplink important~\cite{oueis2016uplink} and calling for the design of innovative \gls{ul}-driven solutions for both the data and the control~planes.	

In the following, we will describe in detail the three considered measurement schemes\footnote{Notice that we do not consider the \gls{sa}-\gls{ul} configuration for both \gls{ia} and tracking applications. In fact, we believe that uplink-based architectures will likely necessitate the support of the \gls{lte} overlay for the management of the control plane and the implementation of efficient measurement operations.}. Table~\ref{table:frameworks} provides a summary of the main features of each framework.

\subsection{Standalone-Downlink (SA-DL) Scheme}
The SA-DL configuration scheme is shown in Fig.~\ref{fig:sadl_signals}. 
No support from the LTE overlay is provided in this configuration.
The beam management procedure is composed of the following phases:


\begin{figure}[t!]
	\centering
	\includegraphics[width=0.45\textwidth]{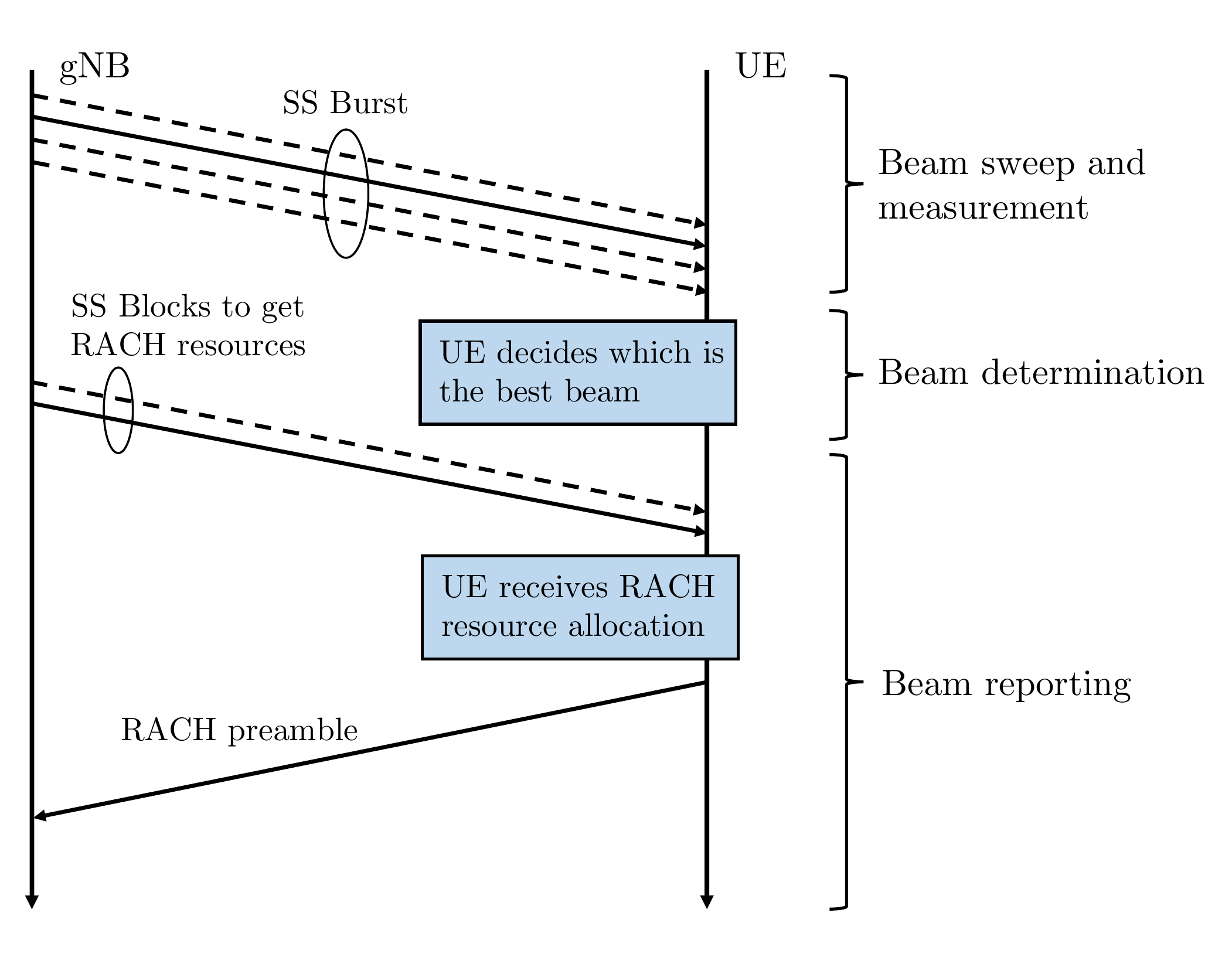}
	\caption{Signals and messages exchanged during the SA-DL beam management procedure (with the beam reporting step of the \gls{ia}). Notice that the duration of the three phases is not in scale, since it depends on the actual configuration of the network parameters.}
	\label{fig:sadl_signals}
\end{figure}

\begin{itemize}
\item[(i)] \textit{Beam sweeping. }The measurement process  is carried out with an exhaustive search, i.e., both users and base stations have a predefined codebook of directions (each identified by a beamforming
vector) that cover the whole angular space and are used
sequentially to transmit/receive synchronization and reference signals~\cite{jeong2015random}.

\item[(ii)] \textit{Beam measurements.} The mmWave-based measurements for  \gls{ia} are based on the \gls{ss} blocks. The tracking is done using both the measurements collected with the \gls{ss} bursts and the \glspl{csirs}. These last elements cover a set of directions which may or may not cover the entire set of available directions according to the users' needs, as explained in Sec.~\ref{sec:3gpp_specifications}. No support from the \gls{lte} overlay is provided in this configuration.

\item[(iii)]\textit{Beam determination.} The mobile terminal selects the beam through which it experienced the maximum SNR, if above a predefined threshold. The corresponding sector will be chosen for the subsequent transmissions and receptions and benefit from the resulting antenna gain.

\item[(iv)]\textit{Beam reporting.}
For  \gls{ia}, as proposed by 3GPP, after beam determination the mobile terminal has to wait for the \gls{gnb} to schedule the \gls{rach} opportunity towards the best direction that the \gls{ue} just determined, for performing random access and implicitly informing the selected serving infrastructure of the optimal direction (or set of directions) through which it has to steer its beam, in order to be properly aligned. It has been agreed that for each \gls{ss} block the \gls{gnb} will specify one or more \gls{rach} opportunities with a certain time and frequency offset and direction, so that the \gls{ue} knows when to transmit the \gls{rach} preamble~\cite{ericsson2016prach}.
This may require an additional complete directional scan of the \gls{gnb}, thus further increasing the time it takes to access the network. 
For the tracking in connected mode, the \gls{ue} can provide  feedback using the mmWave control channel it has already established, unless there is a link failure and no directions can be recovered using \gls{csirs}. In this case the \gls{ue} must repeat the \gls{ia} procedure or try to recover the link using the \gls{ss} bursts while the user experiences a service unavailability.
\end{itemize}

\subsection{Non-Standalone-Downlink (NSA-DL) Scheme}

\begin{figure}[t!]
	\centering
	\includegraphics[width=0.45\textwidth]{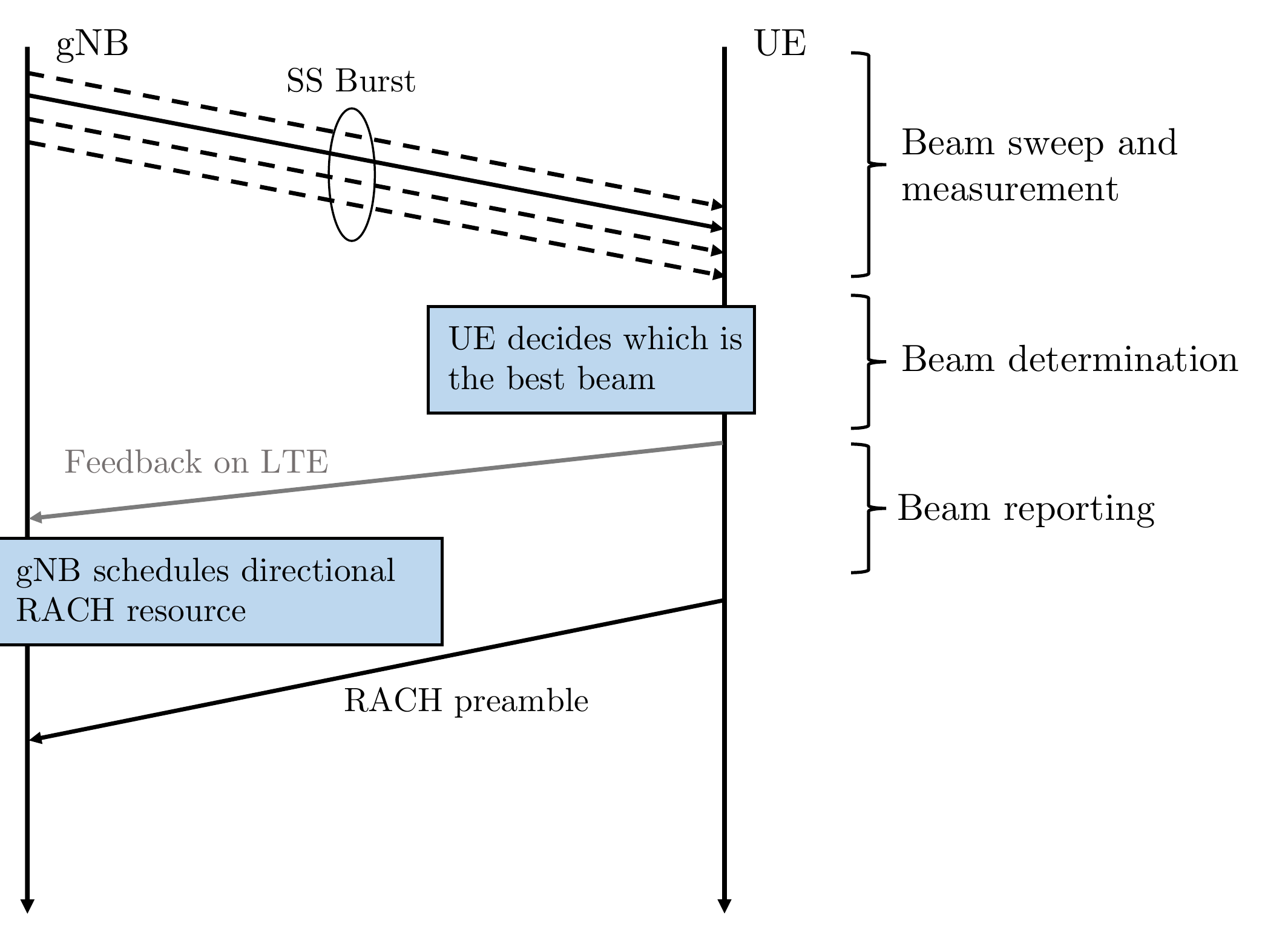}
	\caption{Signals and messages exchanged during the \gls{nsa}-DL beam management procedure (with the beam reporting step of the \gls{ia}). Notice that the duration of the three phases is not in scale, since it depends on the actual configuration of the network parameters.}
	\label{fig:mcdl_signals}
\end{figure}

The sub-6-GHz overlay can be used with different levels of integration. 
As shown in Fig.~\ref{fig:mcdl_signals}, the first three procedures are as in the SA-DL scheme. However, non-standalone enables an improvement in the beam reporting phase.
Thanks to the control-plane integration with the overlay, the \gls{lte} connection can be used to report the optimal set of directions to the \glspl{gnb}, so that the \gls{ue} does not need to wait for an additional beam sweep from the \gls{gnb} to perform the beam reporting or the \gls{ia} procedures. 
Thanks to this signaling, a random access opportunity can therefore be immediately scheduled for that direction with the full beamforming~gain. Moreover, the \gls{lte} link can be also used to immediately report a link failure, and allow a quick data-plane fallback to the sub-6-GHz connection, while the \gls{ue} recovers the mmWave link. 


\subsection{Non-Standalone-Uplink (NSA-UL) Scheme}
Unlike in traditional \gls{lte} schemes, this framework (first proposed in \cite{giordani2016multi} and then used in~\cite{polese2017jsac}) is based on the channel quality of the \gls{ul} rather than that of the \gls{dl} signals and, with the joint support of a central coordinator (i.e., an \gls{lte} \gls{enb} operating at sub-6 GHz frequencies), it enables efficient measurement operations. In this framework, a user searches for synchronization signals from conventional 4G cells. This detection is fast since it can be performed omnidirectionally and there is no need for directional scanning. Under the assumption that the 5G mmWave eNBs are roughly time synchronized to the 4G cell, and the round trip propagation times are not large, an uplink transmission from the UE will be roughly time aligned at any closeby mmWave cell\footnote{For example, if the cell radius is 150 m (a typical mmWave cell), the round trip delay is only 1 $\mu$s.}~\cite{giordani2016efficient}. 
The NSA-UL procedure\footnote{Unlike the conventional DL-based measurement configuration, the uplink scheme has not been considered by 3GPP. Nevertheless, we will freely adapt the same NR frame structure proposed for the downlink case to the \gls{nsa}-UL scheme, using for the uplink \glspl{srs} the resources that would be allocated to \gls{ss} blocks in a downlink framework.} is shown Fig.~\ref{fig:mcul_signals} with a detailed breakout of the messages exchanged by the different parties. In detail, it is composed of:


\begin{figure}[t!]
	\centering
	\includegraphics[width=0.45\textwidth]{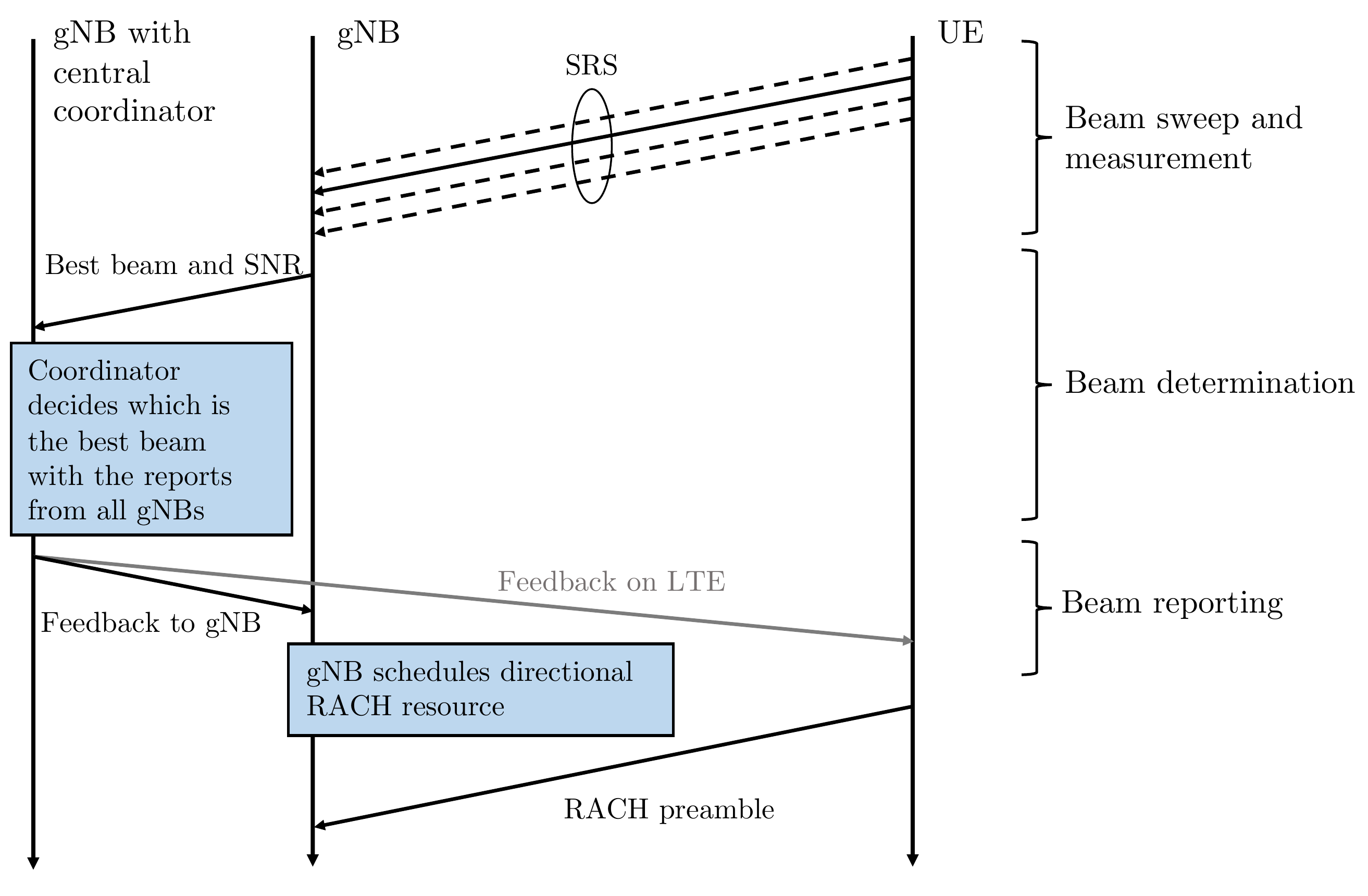}
	\caption{Signals and messages exchanged during the \gls{nsa}-UL beam management procedure (with the beam reporting step of the \gls{ia}). Notice that the duration of the three phases is not in scale, since it depends on the actual configuration of the network parameters.}
	\label{fig:mcul_signals}
\end{figure}

\begin{itemize}
\item[(i-ii)] \textit{Beam sweeping and beam measurements.} 
Each UE directionally broadcasts \glspl{srs} in the mmWave bands in time-varying directions that continuously sweep the angular space. Each potential serving \gls{gnb} scans all its angular directions as well, monitoring the strength of the received \glspl{srs} and building a \textit{report table} based on the channel quality of each receiving direction, to capture the dynamics of the channel.

\item[(iii)]\textit{Beam determination.} Once the report table of each mmWave \gls{gnb} has been filled for each \gls{ue}, each mmWave cell sends this information to  the \gls{lte} \gls{enb} which, due to the knowledge gathered on the signal quality in each angular direction for each \gls{gnb}-\gls{ue} pair, obtains complete directional knowledge over the cell it controls. Hence, it is able to match the beams of the transmitters and the receivers to provide maximum performance.

\item[(iv)]\textit{Beam reporting.}
The coordinator reports to the UE, on a legacy \gls{lte} connection, which  \gls{gnb} yields the best performance, together with the optimal direction in which the \gls{ue} should steer its beam, to reach the candidate serving cell in the optimal way. 
The choice of using the LTE control link  during the tracking is motivated by the fact that the UE may not be able to receive from the optimal mmWave link if not properly aligned, thereby removing a possible point of failure in the control signaling path. 
Moreover, since path switches and cell additions in the mmWave regime are common due to link failures, the control link to the serving mmWave cell may not be available either.
Finally, the coordinator notifies the designated \gls{gnb}, through a backhaul high-capacity link, about the optimal direction in which to steer the beam for serving each \gls{ue}.
\end{itemize}

\section{Performance Metrics and 3GPP Frameworks Parameters }
\label{sec:sim_params}

In this section we define the metrics that will be used to compare and characterize the performance of the different beam management frameworks. Moreover, we will list the relevant parameters that affect the performance of the frameworks in 3GPP \gls{nr}.

\subsection{Performance Metrics}
The performance of the different architectures and beam management procedures for \gls{ia} and tracking will be assessed using three different metrics. The \textit{detection accuracy} is measured in terms of probability of misdetection $P_{\rm MD}$, defined as the probability that the \gls{ue} is not detected by the base station (i.e., the  \gls{snr} is below a threshold $\Gamma$) in an uplink scenario, or, vice versa, the base station is not detected by the \gls{ue} in a downlink scenario. 
The \textit{reactiveness} differs according to the purpose of the measurement framework. For non-connected users, i.e., for \gls{ia}, it is represented by the average time to find the best beam pair. For connected users, i.e., for tracking, it is the time required to receive the first \gls{csirs} after an \gls{ss} burst, and thus react to channel variations or mobility in order to eventually switch beams, or declare a \gls{rlf}. Moreover, we also consider the time it takes to react to the \gls{rlf}. 
Finally, the \textit{overhead} is the amount of time and frequency resources allocated to the framework with respect to the total amount of available resources, taking into account both the \gls{ia} (i.e., \gls{ss} blocks or \glspl{srs} and the \gls{rach}) and the tracking (i.e., \glspl{csirs}).

\subsection{3GPP Framework Parameters}
\label{sec:3gpp_params}
In this section, we list the parameters that affect the performance of the measurement architectures, as summarized in Table \ref{table:pTh}. 
Moreover, we provide insights on the impact of each parameter on the different metrics.

\begin{table*}[t!]
\small
  \centering
    \renewcommand{\arraystretch}{1}
  \begin{tabular}{@{}lccccccccccc@{}}
  \toprule
    Parameter & $\Delta_f$ & $D$ & $N_{\rm SS}$ & $T_{\rm SS}$ & CSI & $N_{\rm CSI, RX}$ & $K_{\rm BF}$ & $M$, $N_{\theta}$ and $N_{\phi}$ & $N_{\rm user}$ & $\lambda_b$ \\ \midrule
    Accuracy & \si & \si & x & x & \si$^*$ & x & x & \si & x & \si \\
    Reactiveness & \si & x & \si & \si & \si$^*$ & \si$^*$ & \si & \si & \si & x \\
    Overhead & \si & \si & \si & \si & \si & x & \si & \si & \si & x \\
    \bottomrule
    \end{tabular}
  \caption{Relation among performance metrics and parameters. \\\scriptsize $^*$This depends on the tracking strategy.}
  \label{table:pTh}
\end{table*}

\begin{figure}[t!]
  \centering
  \includegraphics[width=0.4\textwidth]{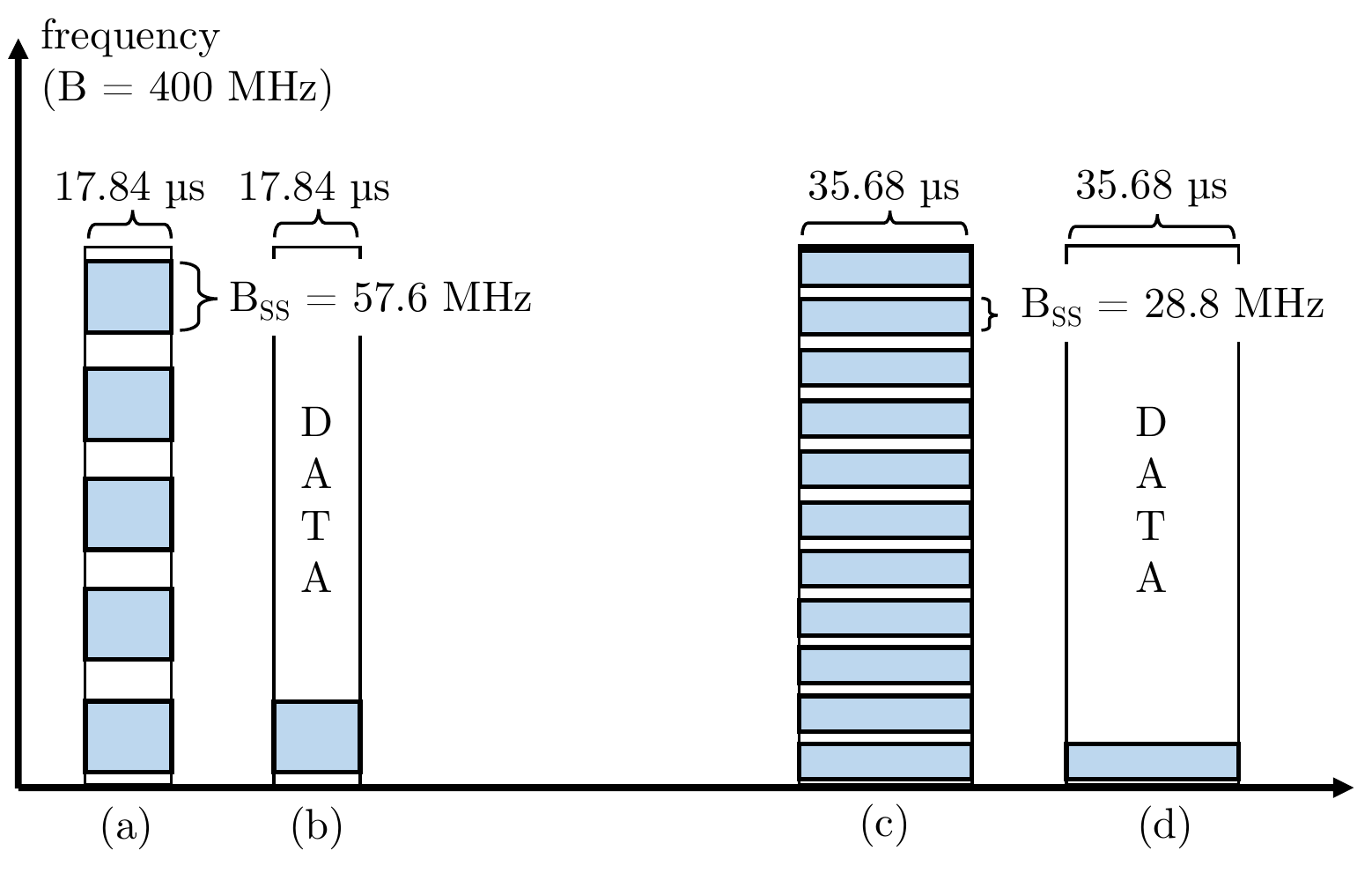}
  \caption{\gls{ss} block structure. For configurations (a) and (b), each blue rectangle is an \gls{ss} block (with 4 OFDM symbols) of duration 17.84 $\mu$s (i.e., $\Delta_f = 240$ kHz) and bandwidth $B_{\rm SS}=57.6$ MHz. For configurations (c) and (d) (for which $\Delta_f=$ 120 kHz), instead, the blocks last 35.68 $\mu$s and have bandwidth $B_{\rm SS}=28.8$ MHz.
  Cases (a) and (c) implement a \emph{frequency repetition} scheme (with $N_{rep}=5$ and $11$, respectively) while, for cases (b) and (d), a \emph{data} solution (i.e., $N_{rep}=1$) is preferred.}
  \label{fig:ssblock}
\end{figure}

\textbf{Frame Structure -- } As depicted in Fig. \ref{fig:ssblock}, we consider the {frame structure} of 3GPP \gls{nr}, with different subcarrier spacings $\Delta_f$. Given that in~\cite{moto2017ss} the only subcarrier spacings considered for \gls{ia} at frequencies above 6 GHz are $\Delta_f = 120$ and  $240$ kHz, i.e., $15 \times 2^n$ kHz, with $n \in [3,4]$, we will only consider these cases. The slot duration in ms is given by~\cite{ericsson2017review}
\begin{equation}\label{eq:slot}
  T_{\rm slot} = \frac{1}{2^n},
\end{equation} 
while the duration of a symbol in $\mu$s is~\cite{ericsson2017review}
\begin{equation}\label{eq:symbol}
  T_{\rm symb} = \frac{71.35}{2^n}.
\end{equation}
Therefore, for $n = 3$ and $4$ the slot duration is $125~\mu$s or $62.5~\mu$s, respectively. 
Moreover, according to the \gls{3gpp} specifications~\cite{38211}, the maximum number of subcarriers allocated to the \gls{ss} blocks is 240, thus the bandwidth reserved for the \gls{ss} blocks would be respectively 28.8 and 57.6 MHz. As mentioned in Sec.~\ref{sec:3gpp_specifications}, we consider a maximum channel bandwidth $B=400$ MHz per carrier~\cite{38802}.

\textbf{Frequency Diversity -- } It is possible to configure the system to exploit {frequency diversity}, $D$. Given that 240 subcarriers are allocated in frequency to an \gls{ss}, the remaining bandwidth in the symbols which contain an \gls{ss} block is $B - 240 \Delta_f$. Therefore, it is possible to adopt two different strategies: (i) \textit{data} (as represented in Figs. \ref{fig:ssblock}(b) and (d)), i.e., the remaining bandwidth $B - 240 \Delta_f$ is used for data transmission towards users which are in the same direction in which the \gls{ss} block is transmitted, or (ii) \textit{repetition} (as displayed in Figs. \ref{fig:ssblock}(a) and (c)), i.e., the information in the first 240 subcarriers is repeated in the remaining subcarriers to increase the robustness against noise and enhance the detection capabilities. The number of repetitions is therefore $N_{rep} = 1$ if frequency diversity is not used (i.e., $D=0$, and a single chunk of the available bandwidth is used for the \gls{ss} block), and $N_{rep} = 11$ or $N_{rep} = 5$ when repetition is used (i.e., $D=1$) with $\Delta_f = 120$~kHz or $\Delta_f = 240$~kHz, respectively. There is a guard interval
in frequency among the different repetitions of the \gls{ss} blocks, to provide a good trade-off between frequency diversity and coherent combining~\cite{barati2015directional}.

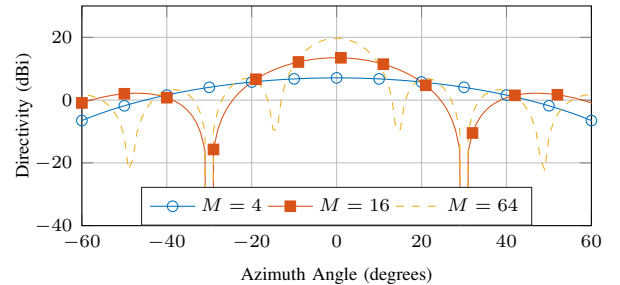
\begin{figure}[b!]
  \centering
  \setlength\abovecaptionskip{-0cm}
  \setlength\fwidth{0.8\columnwidth}
  \setlength\fheight{0.33\columnwidth}
  \input{figures/beamwidth.tex}
  \caption{Relationship between beamwidth and antenna array size.}
  \label{fig:BF}
\end{figure}

\begin{table}[b!]
  \footnotesize
  \renewcommand{\arraystretch}{1}
  \centering
  \begin{tabular}{@{}llll@{}}
  \toprule
  $M$ & $\theta$ [deg]& $N_{\theta}$ gNB & $N_{\theta}$ UE  \\ \midrule
  4       & 60        & 2                          & 6                            \\
  16      & 26        & 5                          & 14                           \\
  64      & 13        & 10                         & 28          \\                \bottomrule
  \end{tabular}
  \caption{Relationship between $M$, $\theta$ and $N_{\theta}$, for the azimuth case. Each gNB sector sweeps through $\Delta_{\theta,\rm gNB}=120^\circ$, while the UE scans over $\Delta_{\theta,\rm UE}=360^\circ$. In our evaluation, we consider a single antenna array at the \gls{ue} modeled as a uniform rectangular array with isotropic antenna elements, following the approach of the literature~\cite{bai2015coverage}. Real handheld devices~will be equipped with multiple patch antennas able to cover the whole angular~space.}
  \label{tab:BF}
\end{table}

\textbf{SS Block Configuration -- } We  consider different configurations of the \gls{ss} blocks and bursts. The maximum {number $N_{\rm SS}$ of \gls{ss} blocks in a burst} for our frame structure and carrier frequencies is $L = 64$. We assume that, if $N_{\rm SS} < L$, the \gls{ss} blocks will be transmitted in the first $N_{\rm SS}$ opportunities. The actual maximum duration of an \gls{ss} burst is $D_{\max, \rm SS} = 2.5$~ms for $\Delta_f = 240$~kHz and $D_{\max, \rm SS} = 5$~ms for $\Delta_f = 120$~kHz. We will also investigate all the possible values for the {\gls{ss} burst periodicity $T_{\rm ss}$}, as defined in~\cite{nokia2017ss}, i.e., $T_{\rm SS} \in \{5, 10, 20, 40, 80, 160\}$~ms.

\textbf{CSI-RS Configuration -- } As for the tracking, there are different options for the configuration of the {\gls{csirs} structure.} These options include (i) the number $N_{\rm CSI}$ of \gls{csirs} per \gls{ss} burst period, (ii) the  \gls{csirs} periodicity $T_{\rm CSI, slot} \in \{5,10,20,40,80,160,320,640\}$~slots, and (iii) the offset $O_{\rm CSI}$ with respect to the end of an \gls{ss} burst. In the analysis in Sec.~\ref{sec:results} we will also refer to $T_{\rm CSI} = T_{\rm CSI, slot}  T_{\rm slot}$, which represents the absolute \gls{csirs} periodicity in ms.
These settings will be specified by the system information carried by the \gls{ss} blocks of each burst. Other CSI-related parameters are the number of symbols of each \gls{csirs} transmission, i.e., $N_{\rm symb, CSI} \in \{1, 2, 4\}$, and the portion of bandwidth $\rho B$ allocated to the \glspl{csirs}. Moreover, the user will listen to $N_{\rm CSI, RX}$ \glspl{csirs} through an equivalent number of directions, when in connected state. We will consider $N_{\rm CSI, RX} \in \{1,4\}$.

\textbf{Array Geometry -- }  As  shown in Fig.~\ref{fig:BF}~and~Table~\ref{tab:BF}, another fundamental parameter is the {array geometry}, i.e., the number of antenna elements $M$ at the \gls{gnb} and \gls{ue} and the number of directions that need to be covered, both in azimuth $N_{\theta}$ and in elevation $N_{\phi}$. At the \gls{gnb} we consider a single sector in a three sector site, i.e., the azimuth $\theta$ varies from $-60$ to $60$ degrees, for a total of $\Delta_{\theta} = 120$ degrees. 
The elevation $\phi$ varies between $-30$ and $30$ degrees, for a total of $\Delta_{\phi} = 60$ degrees, and also includes a fixed mechanical tilt of the array pointing towards the ground. 
There exists a strong correlation among beamwidth, number of antenna elements and BF gain. The more antenna elements in the system, the narrower the beams, the higher the  gain that can be achieved by beamforming, and the more precise and directional the transmission. 
Thus, given the array geometry, we compute the beamwidth $\Delta_{\rm beam}$ at 3 dB of the main lobe of the beamforming vector, and then $N_{\theta} = \Delta_{\theta} / \Delta_{\rm beam}$ and $N_{\phi} = \Delta_{\phi} / \Delta_{\rm beam}$.

\textbf{Beamforming Architecture -- } Different {beamforming architectures}, i.e., analog, hybrid or digital, can be used both at the \gls{ue} and at the \gls{gnb}.
\textit{Analog beamforming} shapes the beam through a single \gls{rf} chain for all the antenna elements, therefore the processing is performed in the analog domain and it is possible to transmit/receive in only one direction at any given time. This model saves power by using only a single pair of \glspl{adc}, but has a little flexibility since the transceiver can only beamform in one direction.
\textit{Hybrid beamforming} uses $K_{\rm BF}$ \gls{rf} chains (with $K_{\rm BF}\leq M$), thus is equivalent to $K_{\rm BF}$ parallel analog beams and enables the transceiver to transmit/receive in $K_{\rm BF}$ directions simultaneously. 
Nevertheless, when hybrid beamforming is used for transmission, the power available at each transmitting beam is the total node power constraint divided by $K_{\rm BF}$, thus potentially reducing the received power.
\textit{Digital beamforming} requires a separate \gls{rf} chain and data converters for each
antenna element and therefore  allows the processing of the received signals in the digital domain,  potentially enabling the transceiver to direct beams at infinitely many directions. Indeed, the availability of a sample for each antenna allows the transceiver to apply arbitrary weights to the received signals, and perform a more powerful and flexible processing than that in the analog domain. 
As in the hybrid case, the use of digital beamforming to transmit multiple beams simultaneously leads to a reduced transmit power being available to each (i.e., the total power constraint applies to the sum of all beams, not to each of them individually). 
Moreover, the digital transceiver can process at most $M$ simultaneous and orthogonal beams without any inter-beam interference (i.e., through a zero-forcing beamforming structure \cite{yoo2005optimality}).  
For this reason, we limit the number of parallel  beams that can be generated to $M$. 
Furthermore, as previously mentioned, we implement a digital beamforming scheme only at the receiver side to avoid higher energy consumption in tranmsission.
For the sake of completeness, we also consider an omnidirectional strategy at the \gls{ue} i.e., without any beamforming gain but allowing the reception through the whole angular space at any given time.

\textbf{Network Deployment -- } Finally, the last parameters are the {number of users} $N_{\rm user} \in \{5, 10, 20\}$  per sector of the \glspl{gnb} and the {density of base stations }$\lambda_b$, expressed in \gls{gnb}/km$^2$.

\renewcommand{\arraystretch}{1}
\begin{table}[!t]
\setlength{\belowcaptionskip}{-0.2cm}
\small
\centering
\begin{tabular}{ @{}lll@{}}
\toprule
Parameter & Value & Description \\ \midrule
 $B$ & $400$ MHz & Total bandwidth of each mmWave gNB\\ 
$f_{c}$ & $28$ GHz & mmWave carrier frequency \\
$P_{\rm TX}$ & $30$ dBm & Transmission power  \\
$\Gamma$ & $ -5$ dB & SNR threshold \\
\bottomrule
\end{tabular}
\caption{Main simulation parameters.}
\label{tab:params}
\end{table}

\input{figures/params_table.tex}

\subsection{Channel Model}

The simulations for the detection accuracy performance evaluation are based on realistic system design configurations. 
Our results are derived through a Monte Carlo approach, where multiple independent simulations are repeated, to get different statistical quantities of interest. 
The channel model is based on recent real-world measurements
at 28 GHz in New York City, to provide a realistic
assessment of mmWave micro and picocellular networks in
a dense urban deployment.  A complete description of the channel parameters
can be found in \cite{akdeniz2014millimeter}, while the main simulation parameters for this paper are reported in Table~\ref{tab:params}.

\section{Results and Discussion}
\label{sec:results}
In this section, we present some simulation results aiming at 
(i) evaluating the performance of the presented initial access schemes in terms of detection accuracy (i.e., probability of misdetection), as reported in Sec.~\ref{sec:accuracy};
(ii) describing the analysis and the results related to the performance of the measurement frameworks for the reactiveness and the overhead, respectively in Sec.~\ref{sec:react_ia}-\ref{sec:react_tr} and Sec.~\ref{sec:overhead}.
Table~\ref{table:symbols} reports the notation used in this section.

\subsection{Detection Accuracy Results}
\label{sec:accuracy}

\begin{figure}[t!]
  \centering
    \setlength\abovecaptionskip{0.1cm}
    \setlength\belowcaptionskip{-0.3cm}
  \setlength\fwidth{0.7\columnwidth}
  \setlength\fheight{0.34\columnwidth}
  \input{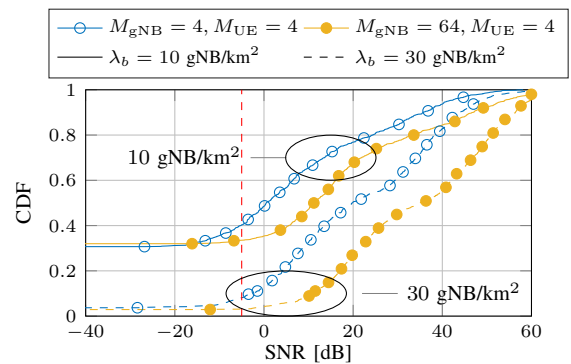}
  \caption{CDF of the SNR, for different antenna configurations. $\Delta_f=120$ kHz, $N_{rep}=0$. The red dashed line represents the SNR threshold $\Gamma=-5$ dB that has been considered throughout this work.}
    \label{fig:CDF_SNR}
\end{figure}

\textbf{Array size and \gls{gnb} density -- }Fig.~\ref{fig:CDF_SNR} shows the \gls{cdf} of the SNR between the mobile terminal and the gNB it is associated to, for different antenna configurations and considering two density values. Notice that the curves are not smooth because of the progressive transitions of the \gls{snr} among the different path loss regimes, i.e., \gls{los}, \gls{nlos} and outage.
We see that better detection accuracy performance can be achieved when densifying the network and when using larger arrays. 
In the first case, the endpoints are progressively closer, thus ensuring better signal quality and, in general, stronger received power. 
In the second case, narrower beams can be steered thus guaranteeing higher gains produced by beamforming.
We also notice that, for good SNR regimes, the $ M_{\rm gNB} = 4, M_{\rm UE} = 4$ and $ M_{\rm gNB} = 64, M_{\rm UE} = 4$  configurations present good enough SNR values: in these regions, the channel conditions are sufficiently good to ensure satisfactory signal quality (and, consequently, acceptable misdetection) even when considering small antenna factors. 
Finally, the red line represents the SNR threshold $\Gamma=-5$ dB that we will consider in this~work.

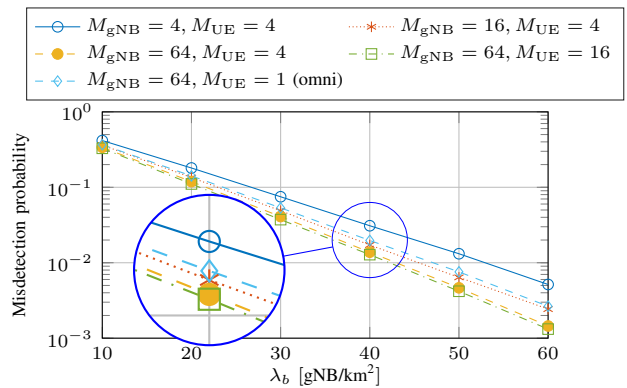
\begin{figure}[t]
  \centering
  \setlength\abovecaptionskip{0.1cm}
  \setlength\fwidth{0.7\columnwidth}
  \setlength\fheight{0.34\columnwidth}
  \input{figures/Perr_N.tex}
  \caption{$P_{\rm MD}$ as a function of $\lambda_{b}$, for different antenna configurations.}
  \label{fig:P_err_N}
\end{figure}

Similar considerations can be deduced from Fig. \ref{fig:P_err_N}, which illustrates how the misdetection probability monotonically decreases when the gNB density $\lambda_b$ progressively increases or when the transceiver is equipped with a larger number of antenna elements, since more focused beams can be generated in this case. 
Moreover, we notice that the beamforming strategy in which the UE transmits or receives omnidirectionally, although guaranteeing fast access operations, does not ensure accurate \gls{ia} performance  and leads to degraded detection capabilities. More specifically, the gap with a fully directional architecture (e.g., $ M_{\rm gNB} = 64, M_{\rm UE} = 16$) is quite remarkable for very dense scenarios, and increases as the \gls{gnb} density increases. For example, the configuration with 16 antennas (i.e., $M_{\rm UE} =16$) and that with a single omnidirectional antenna at the \gls{ue} reach the same $P_{\rm MD}$, but at different values of \gls{gnb} density $\lambda_b$, respectively 30 and 35 $\text{gNB/km}^\text{2}$: the omnidirectional configuration requires a higher density (i.e., 5 $\text{gNB/km}^\text{2}$ more) to compensate for the smaller beamforming gain.

\begin{figure}[t!]
  \centering    
    \setlength\abovecaptionskip{0.1cm}
  \setlength\belowcaptionskip{-.4cm}
  \setlength\fwidth{0.7\columnwidth}
  \setlength\fheight{0.34\columnwidth}
  \input{figures/D_4x4.tex}
  \caption{$P_{\rm MD}$ as a function of $\lambda_{b}$, for different subcarrier spacings $\Delta_f$ and repetition strategies and for different antenna configurations.  $ M_{\rm gNB} = 4, M_{\rm UE} = 4$, $\Gamma=-5$ dB.}
  \label{fig:D_4x4}
\end{figure}
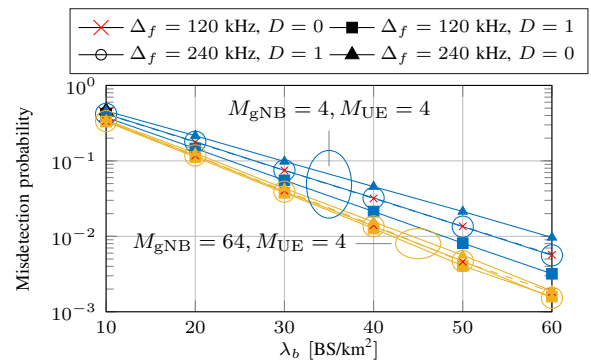

\textbf{Subcarrier spacing and frequency diversity -- }Fig. \ref{fig:D_4x4} reports the misdetection probability related to $\lambda_b$, for different subcarrier spacings $\Delta_f$ and repetition strategies $D$.
First, we see that, if no repetitions are used (i.e., $D=0$), lower detection accuracy performance is associated with the $\Delta_f=240$ kHz configuration, due to the resulting larger impact of the thermal noise and the consequent SNR degradation. 
Furthermore, the detection efficiency can be enhanced by repeating the SS block information embedded in the first 240 subcarriers in the remaining subcarriers (i.e., $D=1$), to increase the robustness of the communication and mitigate the effect of the noise in the detection process. 
In fact, if a frequency diversity approach is preferred, the UE (in the DL measurement technique) or the gNB (in the UL measurement technique) has $N_{rep}>1$ attempts to properly collect the synchronization signals exchanged during the beam sweeping phase, compared to the single opportunity the nodes would have had if they had not implemented any repetition strategy.
We also observe that the  $\Delta_f=120$ kHz with no frequency diversity configuration and the $\Delta_f=240$ kHz scheme with $N_{rep}=5$ produce the same detection accuracy results, thus showing how the effect of increasing the subcarrier spacing  and the number of repetitions of the SS block information in multiple frequency subbands is similar in terms of misdetection capabilities.
Finally, we observe that the impact of the frequency diversity $D$ and the subcarrier spacing $\Delta_f$ is less significant when increasing the array factor, as can be seen from the reduced gap between the curves plotted in Fig. \ref{fig:D_4x4} for the $M_{\rm gNB} = 4, M_{\rm UE} = 4$ and $M_{\rm gNB} = 64, M_{\rm UE} = 4$ configurations.
The reason is that, when considering larger arrays, even the configuration with $\Delta_f = 240$~kHz and no repetitions has an average \gls{snr} which is high enough to reach small misdetection probability values.

\subsection{Reactiveness Results for \gls{ia}}
\label{sec:react_ia}
\textbf{Analysis -- }
For initial access, reactiveness is defined as the delay required to perform a full iterative search in all the possible combinations of the directions. 
The \gls{gnb} and the \gls{ue} need to scan respectively $N_{\theta,\rm gNB}N_{\phi,\rm gNB}$ and $N_{\theta,\rm UE}N_{\phi,\rm UE}$ directions to cover the whole horizontal and vertical space. Moreover, they can transmit or receive respectively $K_{\rm BF, gNB}$ and $K_{\rm BF, UE}$ beams simultaneously. Notice that, as mentioned in Sec. \ref{sec:3gpp_params},  for digital and omnidirectional architectures $K_{\rm BF} = \min\{N_{\theta}N_{\phi}, M\}$,
for hybrid $K_{\rm BF} = \min\{N_{\theta}N_{\phi},M\}/\nu$, where $\nu$ is a factor that limits the number of directions in which it is possible to transmit or receive at the same time, and for analog $K_{\rm BF} = 1$ \cite{sun2014MIMO}.

Then the total number of \gls{ss} blocks needed is\footnote{We recall that hybrid or digital architectures consume more power than analog ones, if the same number of bits in the \glspl{adc} is used, and thus are more likely to be implemented only at the receiver side. Nevertheless, some \gls{adc} configurations enable energy efficient digital beamforming (e.g., 3 bits \gls{adc}~\cite{dutta2017fully}), with a power consumption comparable to that of an analog implementation.}
\begin{equation}
  S_D = \left\lceil\frac{N_{\theta,\rm gNB}N_{\phi,\rm gNB}}{K_{\rm BF, gNB}}\right\rceil\left\lceil\frac{N_{\theta,\rm UE}N_{\phi,\rm UE}}{K_{\rm BF, UE}}\right\rceil.
\label{eq:S_D}
\end{equation}

Given that there are $N_{\rm SS}$ blocks in a burst, the total delay from the beginning of an \gls{ss} burst transmission in a \gls{gnb} to the completion of the sweep in all the possible directions~is
\begin{equation}\label{eq:reactIa}
  T_{\rm IA} = T_{\rm SS}\left(\left\lceil\frac{S_D}{N_{SS}}\right\rceil - 1\right) + T_{last},
\end{equation}
where $T_{last}$ is the time required to transmit the remaining \gls{ss} blocks in the last burst (notice that there may be just one burst, thus the  first term in Eq.~\eqref{eq:reactIa} would be 0). This term depends on the subcarrier spacing and on the number of remaining \gls{ss} blocks which is given by 
\begin{equation}
  N_{\rm SS, left} = S_D - N_{\rm SS}\left(\left\lceil\frac{S_D}{N_{\rm SS}}\right\rceil - 1\right).
\end{equation}
Then, $T_{last}$ is
\begin{equation}
T_{last} = 
  \begin{cases}
    \frac{N_{\rm SS, left}}{2}T_{slot} - 2T_{symb} & \mbox{ if }  N_{\rm SS, left}\bmod 2   = 0\\
    \left\lfloor\frac{N_{\rm SS, left}}{2}\right\rfloor T_{slot} + 6T_{symb} & \mbox{ otherwise,}
  \end{cases}
  \label{eq:T_last}
\end{equation}
The two different options account for an even or odd remaining number of \gls{ss} blocks. In the first case, the \gls{ss} blocks are sent in $N_{\rm SS, left}/2$ slots, with total duration $N_{\rm SS, left}/2 T_{slot}$, but the last one is actually received in the 12\textit{th} symbol of the last slot, i.e., 2 symbols before the end of that slot, given the positions of the \gls{ss} blocks in each slot described in~\cite{moto2017ss}. If instead $N_{\rm SS, left}$ is odd, six symbols of slot $\lfloor N_{\rm SS, left} / 2 \rfloor + 1$ are also used.

\begin{figure}[t!]
  \centering
  \setlength\belowcaptionskip{-.5cm}
  \begin{subfigure}[t!]{\columnwidth}
    \flushright
    \setlength\belowcaptionskip{0.1cm}
    \setlength\abovecaptionskip{0cm}
    \setlength\fwidth{0.7\columnwidth}
    \setlength\fheight{0.34\columnwidth}
    \input{figures/nss_eNBAnalogUEAnalog.tex}
    \caption{\gls{gnb} Analog, UE Analog}
    \label{fig:nsseNBAnUeAn}
  \end{subfigure}
  \vspace{0.45cm}

 \begin{subfigure}[t!]{\columnwidth}
 \centering
   \setlength\belowcaptionskip{0.1cm}
   \setlength\abovecaptionskip{0cm}
   \setlength\fwidth{0.7\columnwidth}
   \setlength\fheight{0.36\columnwidth}
   \input{figures/nss_eNBAnalogUEHybrid.tex}
   \caption{\gls{gnb} Analog, UE Hybrid (\gls{dl}-based configuration)}
   \label{fig:nsseNBAnUeHyb}
 \end{subfigure}
   \vspace{0.45cm}

  \begin{subfigure}[t!]{\columnwidth}
  \centering
    \setlength\belowcaptionskip{0.1cm}
    \setlength\abovecaptionskip{0cm}
    \setlength\fwidth{0.7\columnwidth}
    \setlength\fheight{0.34\columnwidth}
    \input{figures/nss_eNBAnalogUEDigital.tex}
    \caption{\gls{gnb} Analog, UE Digital (\gls{dl}-based configuration)}
    \label{fig:nsseNBAnUeDig}
  \end{subfigure}
    \vspace{0.45cm}

  \begin{subfigure}[t!]{\columnwidth}
  \centering
    \setlength\belowcaptionskip{0.1cm}
    \setlength\abovecaptionskip{0cm}
    \setlength\fwidth{0.7\columnwidth}
    \setlength\fheight{0.34\columnwidth}
    \input{figures/nss_eNBDigitalUEAnalog.tex}
    \caption{\gls{gnb} Digital, UE Analog (\gls{ul}-based configuration)}
    \label{fig:nsseNBDigUeAn}
  \end{subfigure}
  \caption{$T_{\rm IA}$ as a function of $N_{\rm SS}$ with $T_{\rm SS} = 20$~ms.}
  \label{fig:nss}
\end{figure}
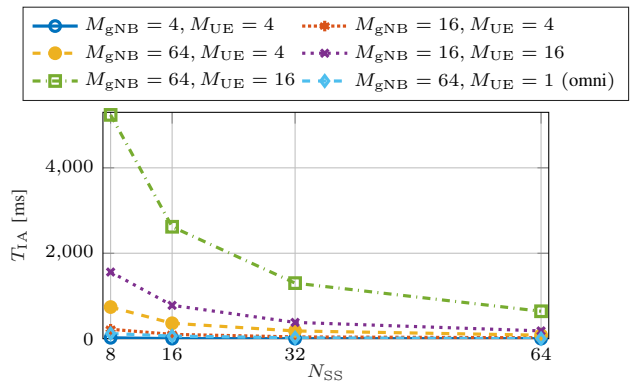
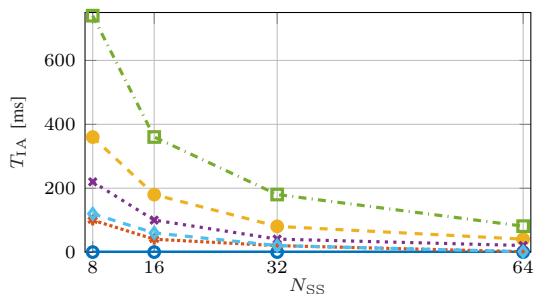
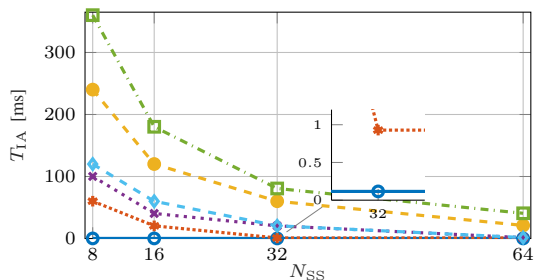
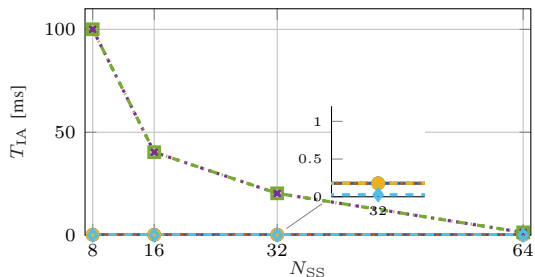

A selection of results is presented in the next paragraphs. 

\textbf{Number of \gls{ss} blocks per burst and beamforming technology -- }In Fig.~\ref{fig:nss} we consider first the impact of the number of \gls{ss} blocks in a burst, with a fixed \gls{ss} burst periodicity $T_{\rm SS} = 20$~ms and  for different beamforming strategies and antenna configurations. 
In particular in Fig.~\ref{fig:nsseNBAnUeAn}, in which both the \gls{ue} and the \gls{gnb} use analog beamforming,  the initial access delay heavily depends on the number of antennas at the transceivers since all the available directions must be scanned one by one. 
It may take from 0.6 s (with $N_{\rm SS} = 64$) to 5.2 s (with $N_{\rm SS} = 8$) to transmit and receive all the possible beams, which makes the scheme infeasible for practical usage.
A reduction in the sweeping time can be achieved either by using an omnidirectional antenna at the \gls{ue} or by decreasing the number of antennas both at the \gls{ue} and at the \gls{gnb}.
In this case, the only configurations that manage to complete a scan in a single \gls{ss} burst are those with 4 antennas at both sides and $N_{\rm SS} \ge 16$, or that with $M_{\rm gNB} = 64$, an omnidirectional \gls{ue} and $N_{\rm SS} = 64$.

Another option is the usage of hybrid or digital beamforming at the \gls{ue} in a downlink-based scheme, or at the \gls{enb} in an uplink-based one. Fig.~\ref{fig:nsseNBAnUeHyb} shows $T_{\rm IA}$ when the \gls{ue} uses hybrid beamforming to receive from half of the available directions at any given time (i.e., $L=2$), while in Fig.~\ref{fig:nsseNBAnUeDig} the \gls{ue} receives from all available directions at any given time. This leads  to an increased number of configurations which are able to complete a sweep in an \gls{ss} block, even with a large number of antennas at the \gls{gnb} and the \gls{ue}. 

Finally, Fig.~\ref{fig:nsseNBDigUeAn} shows the performance of an uplink-based scheme, in which the \glspl{srs} are sent in the same time and frequency resource in which the \gls{ss} blocks would be sent, and the \gls{gnb} uses digital beamforming. It can be seen that there is a gain in performance for most of the configurations, because the \gls{gnb} has to sweep more directions than the \gls{ue} (since it uses narrower beams), thus using digital beamforming at the \gls{gnb}-side makes it possible to reduce $T_{\rm IA}$ even more than when it is used at the \gls{ue}-side.

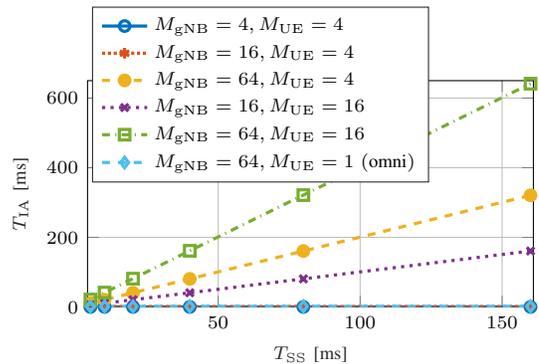
\begin{figure}[t!]
  \centering
      \setlength\belowcaptionskip{0cm}
    \setlength\fwidth{0.7\columnwidth}
    \setlength\fheight{0.34\columnwidth}
    \input{figures/tss_eNBAnalogUEHyb_64.tex}
    \caption{$T_{IA}$ as a function of $T_{SS}$ for the downlink configuration with analog gNB and hybrid UE. $N_{\rm SS} = 64$}
    \label{fig:tss64}
\end{figure}

\begin{figure}[t!]
  \centering
    \setlength\fwidth{0.7\columnwidth}
    \setlength\fheight{0.34\columnwidth}
    \input{figures/nss_eNBAnalogUEAnalog_deltaf.tex}
    \caption{$T_{\rm IA}$ for different antenna configurations and subcarrier spacing $\Delta_f$, with \gls{gnb} Analog, UE Analog.}
    \label{fig:scsTia}
\end{figure}
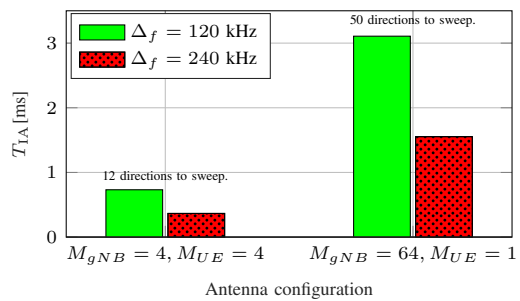

\textbf{\gls{ss} burst periodicity -- } For the setup with hybrid beamforming at the \gls{ue}, that generally requires more than one \gls{ss} burst periodicity, we show in Fig.~\ref{fig:tss64} the dependency of $T_{\rm IA}$ and $T_{\rm SS}$. It can be seen that the highest periodicities are not suited for a mmWave deployment, and that in general it is better to increase the number of \gls{ss} blocks per burst in order to try to complete the sweep in a single burst. 

\begin{figure*}[!b]
  \begin{equation}\label{eq:opt1}
  T_{\rm tr, opt1} = \frac{\sum_{p = 0}^{\left\lfloor\frac{Z_{\rm CSI}}{N_{\rm CSI}}\right\rfloor - 1} \left ( \sum_{i=1}^{N_{\rm CSI}}(pT_{\rm SS} + iT_{\rm CSI})\right) + \sum_{i=1}^{Z_{\rm CSI}\bmod N_{\rm CSI}} \left( \left\lfloor \frac{Z_{\rm CSI}}{N_{\rm CSI}} \right\rfloor T_{\rm SS} + i T_{\rm CSI} \right) }{Z_{\rm CSI}}
  \end{equation}

  \begin{equation}\label{eq:opt2}
  T_{\rm tr, opt2} = \frac{\sum_{p = 0}^{\left\lfloor\frac{Z_{\rm CSI}}{N_{\rm CSI}}\right\rfloor - 1} \left ( \sum_{i=0}^{N_{\rm CSI} - 1}(pT_{\rm SS} + iT_{\rm CSI} + O_{\rm CSI})\right) +
   \sum_{i=0}^{Z_{\rm CSI}\bmod N_{\rm CSI} - 1} \left( \left\lfloor \frac{Z_{\rm CSI}}{N_{\rm CSI}} \right\rfloor T_{\rm SS} + i T_{\rm CSI} + O_{\rm CSI}\right) }{Z_{\rm CSI}}
\end{equation}
\end{figure*}


\textbf{Subcarrier spacing -- }Another parameter that has an impact on $T_{\rm IA}$ is the subcarrier spacing $\Delta_f$. As shown in Fig.~\ref{fig:scsTia}, when the larger spacing is used the \gls{ofdm} symbols have a shorter duration and the transmission of the \gls{ss} blocks in the directions of interest can be completed earlier.

\textbf{Impact of Beam Reporting --} For initial access, in addition to  the time required for  directional sweeping, there is also a delay related to the allocation of the resources in which it is possible to perform initial access, which differs according to the architecture being used.
As introduced in Sec. \ref{sec:meas_frameworks}, 3GPP advocates the implicit reporting of the chosen direction, e.g., the strongest SS block index, through contention-based random access messages, agreeing that  the network should allocate multiple \gls{rach} transmissions and preambles to the UE for conveying the optimal SS block index to the gNB \cite{ericsson2017rach}.
When considering an SA configuration, beam reporting might require an additional sweep at the gNB side while, if an \gls{nsa} architecture is preferred, the beam decision is forwarded through the LTE interface (and requires just a single \gls{rach} opportunity) which makes the beam reporting reactiveness equal to the latency of a legacy LTE connection.
Assuming a 0\% BLER data channel, the uplink latency in legacy LTE, including scheduling
delay, ranges from 10.5 ms to 0.8 ms, according to the latency reduction techniques being implemented \cite{latencyreduction2017}.

\begin{table}[t]
\centering
\footnotesize
\renewcommand{\arraystretch}{1}
\setlength\belowcaptionskip{-0.08cm}
\begin{tabular}{@{}lcccc@{}}
\toprule
& \multicolumn{4}{c}{$T_{\rm BR,SA}$ [ms]}                                       \\ 
                             & \multicolumn{2}{c}{$N_{\rm SS}=8$}      & \multicolumn{2}{c}{$N_{\rm SS}=64$}      \\
$M_{gNB}$                & Analog                 & Digital               &  Analog                 &  Digital                \\ \midrule
4                             & 0.0625               & 0.0625             & 0.0625               & 0.0625              \\
16                            & 0.5               & 0.0625             & 0.5              & 0.0625              \\
64                            & 40.56               & 0.0625             & 1.562               & 0.0625              \\ \midrule
\multicolumn{5}{c}{$T_{\rm BR,\gls{nsa}}\in\{ 10, 4, 0.8\}$ ms, according to \cite{latencyreduction2017}.}  \\ \bottomrule
\end{tabular}
\caption{Reactiveness performance for beam reporting operations considering an SA or an \gls{nsa} architecture. Analog or digital beamforming is implemented at the gNB side, while the UE configures its optimal beamformed direction. $T_{\rm SS}=20$ ms, $\Delta_{f}=120$ KHz.}
\label{tab:react_BR}
\end{table}

In Table \ref{tab:react_BR}, we  analyze the impact of the number of SS blocks (and, consequently, of RACH opportunities) in a burst, with a fixed burst periodicity $T_{\rm SS} = 20$ ms and for a subcarrier spacing of $\Delta_{f}=120$ KHz.
The results are independent of the antenna configuration at the UE side, since the mobile terminal steers its beam through the previously determined optimal direction and does not require a beam sweeping operation to be performed.
It appears clear that the SA scheme presents very good reactiveness for most of the investigated configurations and, most importantly, outperforms the \gls{nsa} solution even when the LTE latency is reduced to $0.8$ ms.
The reason is that, if the network is able to allocate the needed RACH resources within a single SS burst, then it is possible to limit the impact of beam reporting operations on the overall initial access reactiveness, which is instead dominated by the beam sweeping phase.
In particular, when considering small antenna factors and when digital beamforming is employed, beam reporting can be successfully completed through a single RACH allocation, thus guaranteeing very small delays.

\subsection{Reactiveness Results for Beam Tracking}
\label{sec:react_tr}

\textbf{Analysis -- }For tracking, we define the reactiveness as the average time needed to receive the first \gls{csirs} after the end of each \gls{ss} burst. 

We assume that the $N_{\rm user}$ \glspl{ue} are uniformly distributed in the space covered by the $k = N_{\theta,\rm gNB}N_{\phi,\rm gNB}$ beams available at the \gls{gnb}. Moreover, each \gls{ue} has to monitor $N_{\rm CSI, RX}$ directions. Given that a \gls{ue} may or may not be in \gls{los}, it is not obvious that these directions will be associated to the closest beams with respect to the one selected during the initial access. Therefore, we also assume that this scenario is equivalent to a scenario with $n = N_{\rm user}N_{\rm CSI, RX}$ uniformly distributed \glspl{ue}, each of them monitoring a single direction. We will refer to $n$ as the number of measures.

\begin{figure}[t!]
  \centering
    \begin{subfigure}[t]{\columnwidth}
    \centering
    \setlength\fwidth{\columnwidth}
    \setlength\fheight{0.45\columnwidth}
    \input{figures/t_tr_Tss20ms_64antennas.tex}
    \caption{$M_{\rm \gls{gnb}} = 64$, analog beamforming, $T_{\rm SS}= 20$~ms}
    \label{fig:t_tr_64_csiper}
    \end{subfigure}%
    \vspace{.3cm}
    \begin{subfigure}[t]{\columnwidth}
    \centering
    \setlength\fwidth{\columnwidth}
    \setlength\fheight{0.45\columnwidth}
    \input{figures/t_tr_64antennas_changeTss.tex}
    \caption{$M_{\rm \gls{gnb}} = 64$, analog beamforming, $T_{\rm CSI}= 0.625$~ms}
    \label{fig:t_64_antenna}
  \end{subfigure}
  \caption{Performance of tracking using \glspl{csirs} for Option 1 and Option 2, as described in Fig.~\ref{fig:offset_ss_csi}, as a function of different parameters (e.g., $T_{\rm CSI}$, $T_{\rm SS}$), for $\Delta_f=120$~kHz.}
  \label{fig:csi}
\end{figure}
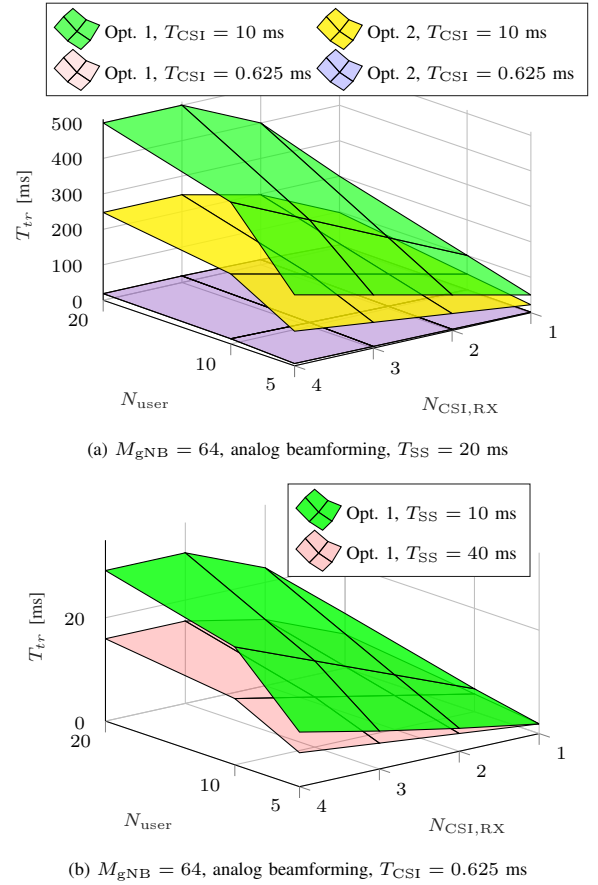

Consequently, on average there are $n / k$ measurements for the area belonging to each beam, if the beams divide the space into equally sized regions. Therefore, if $n \ge k$, a \gls{csirs} is needed in each beam, otherwise it is sufficient to send at least $n$ \glspl{csirs}, and thus the total number of \gls{csirs} that need to be transmitted is on average $Z_{\rm CSI} = \min\{n, k\}$.

Depending on the combination of $T_{\rm SS}$, $T_{\rm CSI} = T_{\rm CSI, slot} T_{\rm slot}$ and $Z_{\rm CSI}$, it may not be possible to allocate all the \gls{csirs} transmissions between two consecutive \gls{ss} bursts. Notice that after the end of an \gls{ss} burst, there are $T_{tot, \rm CSI} = T_{\rm SS} - D_{\max,\rm SS}$~ms available for the \gls{csirs} transmission.
Then, the number $N_{\rm CSI}$ of \gls{csirs} that can be allocated between two \gls{ss} bursts may depend on which of the options shown in Fig.~\ref{fig:offset_ss_csi} is chosen.

\textit{Option 1:} the first \gls{csirs} is transmitted $T_{\rm CSI}$ ms after the transmission of the \gls{ss} burst. In this case, $N_{\rm CSI} = \lfloor T_{tot, \rm CSI} / T_{\rm CSI} \rfloor$, and single periodicity is not enough if $Z_{\rm CSI} > N_{\rm CSI}$. For option 1, the metric $T_{\rm tr, opt1}$ is given by~\eqref{eq:opt1}. 
The last sum accounts for the case $Z_{\rm CSI} < N_{\rm CSI}$ and for the \gls{csirs} in the last \gls{ss} burst periodicity when $Z_{\rm CSI} > N_{\rm CSI}$. The sum over $p$, instead, accounts for $Z_{\rm CSI} \ge N_{\rm CSI}$. 

\textit{Option 2:} thanks to the additional parameter $O_{\rm CSI}$ it is possible to transmit $N_{\rm CSI} = \lceil T_{tot, \rm CSI} / T_{\rm CSI} \rceil$, as shown in Fig.~\ref{fig:csi2}. The offset is computed as
\begin{equation}
  O_{\rm CSI} = \frac{T_{tot, \rm CSI} - (N_{\rm CSI} - 1)T_{\rm CSI}}{2}.
\end{equation}
The metric $T_{\rm tr, opt2}$ is computed as for option 1, but taking into account also $O_{\rm CSI}$, in Eq.~\eqref{eq:opt2}.

Notice that if $Z_{\rm CSI} > N_{\rm CSI}$, a signal in a certain direction could be either received as \gls{ss} block in the next burst, or as \gls{csirs}, depending on how the transmission of \gls{ss} blocks and \glspl{csirs} is scheduled. 

\textbf{Scheduling options, number of users and \gls{csirs} periodicity -- } Fig.~\ref{fig:t_tr_64_csiper} shows the value of $T_{tr}$ for different parameters, such as the different scheduling option 1 or 2, the number of users per \gls{gnb} $N_{\rm user}$ and of directions of interest $N_{\rm CSI, RX}$, for  \gls{ss} burst periodicity $T_{\rm SS} = 20$ ms and 64 antennas at the \gls{gnb}. The fundamental parameter is the periodicity of the \gls{csirs} transmission: only a small \gls{csirs} periodicity makes it possible to sweep all the directions to be covered during a relatively short interval, and to avoid the dependency on $T_{\rm SS}$. Moreover, if the periodicity is small (i.e., $T_{\rm CSI} = 0.625$~ms, or 5 slots with $\Delta_f = 120$ kHz), then there is no difference between the two scheduling options, while this becomes notable for $T_{\rm CSI} = 10$~ms, as expected.

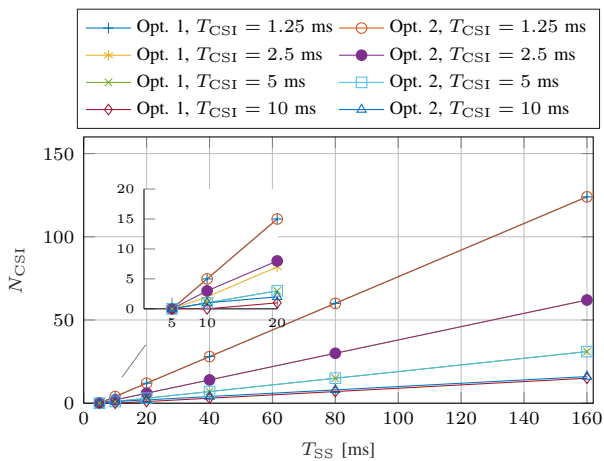
\begin{figure}[t]
      \centering
    \setlength\fwidth{0.8\columnwidth}
    \setlength\fheight{0.4\columnwidth}
    \input{figures/numCsiTss.tex}
    \caption{$N_{\rm CSI}$ as a function of the $T_{\rm SS}$ and $T_{\rm CSI}$ periodicities.}
    \label{fig:num_csi_tss}
\end{figure}

\textbf{\gls{ss} burst periodicity -- }Fig.~\ref{fig:t_64_antenna} compares two different $T_{\rm SS}$ periodicities, i.e., 10 and 40 ms, using the smallest $T_{\rm CSI, slot}$ available (i.e., 5 slots, or $0.625$ ms at $\Delta_f=120$ kHz). 
It can be seen that using a higher $T_{\rm SS}$ would allow a decreased $T_{tr}$, since more \glspl{csirs} can be scheduled between two \gls{ss} bursts and consequently a larger number of directions can be swept. For the sake of completeness, Fig.~\ref{fig:num_csi_tss} shows the number of \glspl{csirs} that can be scheduled in between two \gls{ss} bursts as a function of $T_{\rm SS}$ and of the different scheduling options and periodicities. Since in a mmWave scenario there may be a need to scan a large number of \glspl{csirs},  it is advisable to either use  an adaptive scheme for the scheduling of \glspl{csirs}, which adapts the periodicity according to the number of users in the different directions, or adopt a conservative approach and use a short $T_{\rm CSI}$ interval.

\begin{table*}[t!]
\centering
\small
\begin{tabular}{ccclllclcl}
\toprule
\multicolumn{2}{c|}{Antenna}                                                & \multicolumn{8}{c}{$T_{\rm RLF,SA}$ {[}ms{]}}                                                                                                                                                                                                                                                                                                                                                                              \\
\multirow{2}{*}{$M_{gNB}$} & \multicolumn{1}{c|}{\multirow{2}{*}{$M_{UE}$}} & \multicolumn{4}{c}{\multirow{2}{*}{\begin{tabular}[c]{@{}c@{}}$N_{\rm SS}=8$, $T_{\rm SS}=20$\\ $\gls{gnb}$ ABF, UE ABF\end{tabular}}} & \multicolumn{2}{c}{\multirow{2}{*}{\begin{tabular}[c]{@{}c@{}}$N_{\rm SS}=64$, $T_{\rm SS}=40$\\ $\gls{gnb}$ DBF, UE ABF\end{tabular}}} & \multicolumn{2}{c}{\multirow{2}{*}{\begin{tabular}[c]{@{}c@{}}$N_{\rm SS}=64$, $T_{\rm SS}=80$\\ $\gls{gnb}$ DBF, UE ABF\end{tabular}}} \\
                           & \multicolumn{1}{c|}{}                          & \multicolumn{4}{c}{}                                                                                                                   & \multicolumn{2}{c}{}                                                                                                                    & \multicolumn{2}{c}{}                                                                                                                    \\ \midrule
4                          & \multicolumn{1}{c|}{4}                         & \multicolumn{4}{c}{30.2322}                                                                                                              & \multicolumn{2}{c}{20.3572}                                                                                                                   & \multicolumn{2}{c}{40.3572}                                                                                                                   \\
64                         & \multicolumn{1}{c|}{1}                         & \multicolumn{4}{c}{130.1072}                                                                                                              & \multicolumn{2}{c}{20.0535}                                                                                                                   & \multicolumn{2}{c}{40.0535}                                                                                                                   \\
64                         & \multicolumn{1}{c|}{16}                        & \multicolumn{4}{c}{ 5250}                                                                                                              & \multicolumn{2}{c}{22.6072}                                                                                                                   & \multicolumn{2}{c}{42.6072}                                                                                                                   \\ \midrule
\multicolumn{10}{c}{$T_{\rm RLF,\gls{nsa}}\in\{ 10, 4, 0.8\}$ ms, according to the considerations in \cite{latencyreduction2017}.}                                                                                                                                                                                                                                                                                                                                                                                                                                                    \\ \bottomrule
\end{tabular}
\caption{RLF recovery delay considering the SA or the \gls{nsa} measurement frameworks, for different values of $N_{\rm SS}$, $T_{\rm SS}$ and for different beamforming configurations. $\Delta_f=120$ kHz. ABF stands for Analog Beamforming, and DBF for Digital.}
\label{tab:RLF}
\end{table*}

\textbf{Limits on the \gls{csirs} periodicity -- }Since the \glspl{csirs} that a user receives from multiple base stations should not overlap in time and frequency (otherwise the \gls{rsrp} value would be over-estimated), there is a maximum number of neighboring cells that a \gls{gnb} can support. According to~\cite{qualcomm2017csi}, there are 4 symbols per slot in which a \gls{csirs} can be sent (additional symbols are under discussion), and a \gls{csirs} can last 1, 2 or 4 symbols, each with bandwidth $\rho B$. Assuming a common configuration for the \glspl{gnb} deployed in a certain area,  the total number of orthogonal \gls{csirs} transmission opportunities is
\begin{equation}
  N_{\rm CSI, \perp} = \frac{T_{\rm SS} - D{\max, \rm SS}}{T_{\rm slot}} \frac{4}{N_{\rm symb, CSI}} \left\lfloor\frac{1}{\rho}\right\rfloor,
 \end{equation} 
where the first ratio is the number of slots in the time interval in which \glspl{csirs} can be scheduled, and the second and third express the number of \glspl{csirs} per slot (there are at most 4 \gls{ofdm} symbols per slot for \glspl{csirs}). Then, the maximum number of neighbors that a \gls{gnb} can support is 
\begin{equation}
  N_{\max, \rm neigh} = \left\lfloor \frac{N_{\rm CSI, \perp}}{N_{\rm CSI}} \right\rfloor - 1,
\end{equation}
with $N_{\rm CSI}$ computed as in the previous paragraphs.

\begin{figure}[t]
  \centering
    \setlength\fwidth{0.75\columnwidth}
    \setlength\fheight{0.5\columnwidth}
    \input{figures/numNeigh_tss20.tex}
    \caption{$N_{\max, \rm neigh}$ as a function of $N_{\rm symb, CSI}$ and $\rho$ for different $T_{\rm CSI}$ periodicities, with $T_{\rm SS} = 20$~ms and $\Delta_f = 120$~kHz.}
    \label{fig:max_neigh}
\end{figure}
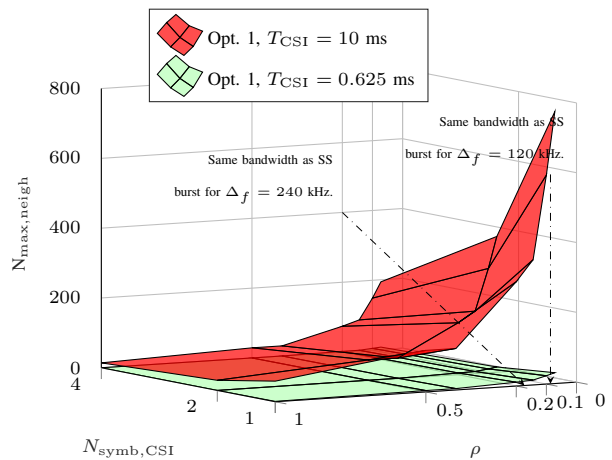

Fig.~\ref{fig:max_neigh} reports the value of $N_{\max, \rm neigh}$ for a different number of \gls{ofdm} symbols for the \glspl{csirs} and bandwidth scaling factor $\rho$, which ranges from 0.1 to 1, and represents also the bandwidth values corresponding to 240 subcarriers with $\Delta_f \in \{120, 240\}$~kHz, i.e., the bandwidth occupied by an \gls{ss} burst. 
Notice that for the frequencies in the mmWave spectrum it is advisable not to use the entire bandwidth for \glspl{csirs}~\cite{samsung2017periodicity}, and  the number of neighbors of a mmWave \gls{gnb} will be limited, given the short propagation distance typical of these frequencies. If $T_{\rm CSI} = 10$~ms, then even when using 4 \gls{ofdm} symbols and the whole bandwidth it is possible to support only 14 neighbors. Instead, when $T_{\rm CSI} = 0.625$~ms it is not feasible to use the whole bandwidth and 4 symbols, but more conservative configurations should be adopted. For example, with $\rho = 0.072$ (i.e., 240 subcarriers with $\Delta_f = 120$~kHz) it is possible to support 15 or 31 neighbors, respectively with 2 or 1 \gls{ofdm} symbols. 

\textbf{Standalone vs non-standalone -- }Notice that when the standalone scheme is used and the \gls{ue} experiences a link failure on all the $N_{\rm CSI, RX}$ directions it is monitoring, then the \gls{ue} has no choice but using the \gls{ss} blocks in the \gls{ss} burst to perform either a link recovery or a new initial access, and meanwhile it is not able to transmit or receive data or control information~\cite{polese2017jsac}. When a non-standalone architecture is used, instead, the \gls{ue} could signal this event to the \gls{ran} on the lower-frequency control link, and the data plane can be switched to the sub-6-GHz \gls{rat}, and faster recovery options could be designed, for example, by instructing the \gls{ue} to monitor additional \glspl{csirs}.

\textbf{Downlink vs uplink and beamforming architecture -- }Finally, we observe that, when a digital architecture is chosen, there exist some specific configurations in which a \gls{ul}-based measurement framework  can ensure more efficient tracking operations than its \gls{dl} counterpart. 
In fact, due to the \gls{gnb}'s less demanding space constraints with respect to a mobile terminal, a larger number of antenna elements can usually be packed at the base station side, resulting in a larger number of directions that can potentially be scanned simultaneously through a digital beamforming scheme.
Moreover,  hybrid or fully digital receivers are more costly in terms
of power consumption, and hence are more likely to be
implemented in a \gls{gnb} rather than in a \gls{ue}.

\textbf{RLF recovery -- } 
Another important factor that affects the reactiveness of beam management schemes is the time it takes to recover from an \gls{rlf}. As assumed by 3GPP \cite{ericsson2017RLF}, \gls{rlf} occurs when the quality of an associated control channel falls below a certain threshold. 
As soon as the failure is detected, mechanisms to recover acceptable communication capabilities (e.g., by determining an alternative suitable direction of transmission or possibly handing over to a stronger and more robust gNB) need to be quickly triggered upon notifying the network. Natural candidates for monitoring the link quality and detect the link failure are the SS blocks in a burst \cite{ericsson2017RLFactions}.
Assume that an object blocks the propagation path of the transceiver at time $T\sim\mathcal{U}[t,t+T_{\rm SS}]$, i.e., on average at time $\bar{T}=T_{\rm SS}/2$ within two consecutive SS bursts.
\begin{itemize}
\item 
When implementing an SA architecture, as soon as an impairment is detected, the UE may no longer be able to communicate with its serving gNB since the optimal directional path  connecting the endpoints is affected by the failure. 
The recovery phase is most likely triggered at the beginning of the subsequent SS burst (i.e., on average after $T_{SS}-\bar{T}=T_{\rm SS}/2$ seconds) and at least after the completion of an IA operation of duration $T_{\rm IA}$ seconds.\footnote{In some circumstances, the UE can autonomously react to an RLF event by selecting an alternative direction of
communication, as a sort of backup solution before the transceiver fully recovers the optimal
beam configuration \cite{giordani2016efficient}. Although having a second available link, when the primary path is obstructed, adds diversity and
robustness to the communication, it may not always guarantee sufficiently good communication performance.}
Table \ref{tab:RLF} reports the \gls{rlf} recovery delay $T_{\rm RLF,SA}$ for some network configurations when an SA architecture is implemented. 
We observe that the latency is quite high for all the investigated settings and is dominated by the IA delay, as illustrated in Fig. \ref{fig:nss}. Moreover, in some circumstances (e.g., $N_{\rm SS}=8$, $T_{\rm SS}=20$ ms, $M_{gNB}=64$, $N_{gNB}=16$ and when analog beamforming is implemented), the RLF recovery delay assumes unacceptably high values.

\item 
Much more responsive RLF recovery operations may be prompted if the failure notification is forwarded through the LTE overlay (i.e., by implementing an \gls{nsa}-based measurement framework), which may also serve the UE's traffic requests until the mmWave directional communication is successfully restored.
If an \gls{nsa}-DL framework is designed, the RLF recovery delay $T_{\rm RLF,\gls{nsa}}$ is equal to the latency of a traditional LTE connection (which depends on the implemented latency reduction technique, as assessed in \cite{latencyreduction2017}). 
Alternatively, the gNB can autonomously declare an RLF event (without the user's notification) and react accordingly by monitoring the \gls{srs} messages.
 Without loss of generality,  assuming that SRSs are uniformly allocated within two  SS bursts with periodicity $T_{\rm SRS}$, an RLF is detected as soon as the gNB is not able to correctly receive $N_{\rm SRS}$ consecutive \glspl{srs} from its reference user.
In this case, the reactiveness of the RLF recovery operation  depends on the periodicity of the sounding signals and is equal to
\begin{equation}
T_{\rm RLF,\gls{nsa}} = \frac{T_{\rm SRS}}{2} + \left( N_{\rm SRS}-1\right)T_{\rm SRS}.
\label{eq:T_RLF}
\end{equation}
Analogously, if an \gls{nsa}-UL framework is designed, the recovery may be immediately triggered by the gNB by switching the traffic to the LTE eNB in $T_{\rm RLF,\gls{nsa}}$ seconds, as given by Eq. \eqref{eq:T_RLF}.
From the results in Table \ref{tab:RLF}, it appears that fast and efficient RLF recovery operations can be guaranteed if an \gls{nsa} solution is preferred over an SA one for all the investigated network configurations.
\end{itemize}

\begin{figure*}[t]
  \centering
  \begin{subfigure}[t]{\columnwidth}
    \setlength\fwidth{0.8\columnwidth}
    \setlength\fheight{0.4\columnwidth}
    \input{figures/omega5.tex}
    \caption{$\Omega_{\rm 5ms}$ as a function of $N_{\rm SS}$, for different subcarrier spacings $\Delta_f$ and repetition strategies.}
    \label{fig:omega5}
  \end{subfigure}\hfill
  \begin{subfigure}[t]{\columnwidth}
    \setlength\fwidth{0.8\columnwidth}
    \setlength\fheight{0.4\columnwidth}
  \input{figures/omegatss.tex}
    \caption{$\Omega_{\rm T_{\rm SS}}$ as a function of $T_{\rm SS}$, for different subcarrier spacings $\Delta_f$ and repetition strategies. $N_{\rm SS}$ is set to the maximum value, i.e., 64.}
    \label{fig:omegatss}
  \end{subfigure}
  \caption{Overhead for  initial access, introduced by the transmission of the \gls{ss} blocks. Notice that the number of repetitions for the different subcarrier spacings $\Delta_f$ is chosen to send as many repetitions of the \gls{ss} blocks as possible.}
  \label{fig:omega_ia}
\end{figure*}
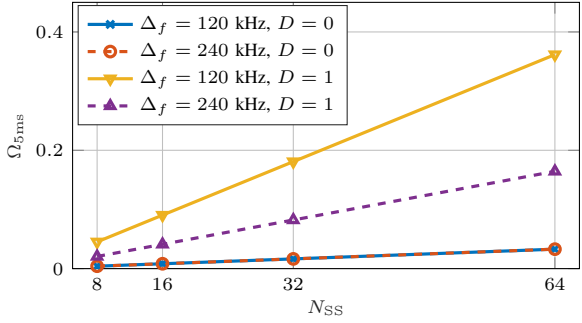
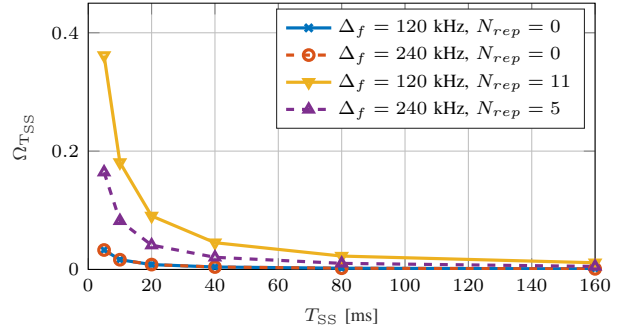

\subsection{Overhead Results}
\label{sec:overhead}
In this section, we characterize the overhead for \gls{ia} and tracking in terms of the ratio between the time and frequency resources that are allocated to \gls{ss} bursts and the maximum duration of the \gls{ss} burst (i.e., 5 ms), or the entire $T_{\rm SS}$ interval. 

\textbf{Analysis -- }The total number of time and frequency resources $R_{\rm SS}$ scheduled for the transmission of $N_{\rm SS}$ \gls{ss} blocks, each spanning 4 \gls{ofdm} symbols and 240 (or multiple of 240) subcarriers, is given by
\begin{equation}
  R_{\rm SS} = N_{\rm SS} \; 4T_{\rm symb} \; 240N_{rep}\Delta_f,
\end{equation}
where $T_{\rm symb}$ is expressed in ms and $\Delta_f$ in kHz.
The overhead for the 5 ms time interval with the \gls{ss} burst transmission and total bandwidth $B$ (in Hz) is then given by
\begin{equation}
  \Omega_{\rm 5ms} = \frac{N_{\rm SS} \; 4T_{\rm symb} \; 240N_{rep}\Delta_f}{5 B},
\end{equation}
and the overhead considering the total burst periodicity $T_{\rm SS}$~is
\begin{equation}\label{eq:ovTss}
  \Omega_{\rm T_{\rm SS}} = \frac{N_{\rm SS} \; 4T_{\rm symb} \; 240N_{rep}\Delta_f}{T_{\rm SS} B}.
\end{equation}

Moreover, additional overhead is introduced by the transmission of \glspl{csirs} after the \gls{ss} burst. The value of the overhead $\Omega_{\rm CSI}$ depends on the number of symbols $N_{\rm symb, CSI}$ and the bandwidth $\rho B$ for each \gls{csirs}, as well as on the number of \glspl{csirs} $N_{\rm CSI}$ computed as in Sec.~\ref{sec:react_tr} for the two \gls{csirs} scheduling options:
\begin{equation}\label{eq:ovCsi}
  \Omega_{\rm CSI} = \frac{N_{\rm CSI} N_{\rm symb, CSI} T_{\rm symb} \rho B}{(T_{\rm SS} - D_{\max, \rm SS})B} = \frac{N_{\rm CSI} N_{\rm symb, CSI} T_{\rm symb} \rho}{(T_{\rm SS} - D_{\max, \rm SS})}.
\end{equation}

Finally, the total overhead $\Omega$ takes into account both the \gls{ss} bursts and the \glspl{csirs} in $T_{\rm SS}$:
\begin{equation}
  \Omega_{tot} = \frac{N_{\rm CSI} N_{\rm symb, CSI} T_{\rm symb} \rho B + R_{\rm SS}}{T_{\rm SS} B}.
\end{equation}

\textbf{Subcarrier spacing and frequency diversity -- } Fig.~\ref{fig:omega_ia} reports the overhead related  to the maximum duration of the \gls{ss} burst (i.e., 5 ms) for different subcarrier spacings and repetition strategies. 
It can be seen that if no repetitions are used (i.e., $D=0$) then the overheads for the configurations with $\Delta_f = 120$~kHz and $\Delta_f = 240$~kHz are equivalent.
In fact, when configuring large subcarrier spacings (i.e., $\Delta_f = 240$~kHz), the \gls{ofdm} symbols used for the \gls{ss} blocks have  half the duration, but they occupy twice the bandwidth of the systems with narrower subcarrier spacings  (i.e., $\Delta_f = 120$~kHz), given that the same number of subcarriers are used. Instead, when a repetition strategy is used (i.e., $D=1$), the overhead is different. As mentioned in Sec.~\ref{sec:3gpp_params}, we consider 5 repetitions for $\Delta_f = 240$ kHz and 11 for $\Delta_f = 120$ kHz. Therefore, the actual amount of bandwidth that is used is comparable, but since the \gls{ofdm} symbols with $\Delta_f= 120$~kHz last twice as long as those with the larger subcarrier spacing,  the overhead in terms of resources used for the \gls{ss} burst is higher with $\Delta_f= 120$~kHz. 

\textbf{\gls{ss} burst periodicity -- }Fig.~\ref{fig:omegatss} shows the dependency of the overhead for  initial access on $T_{\rm SS}$, which follows an inverse proportionality law. In particular, for very small $T_{\rm SS}$ (i.e., 5 ms) the impact of the \gls{ss} bursts with repetitions in frequency is massive, with up to 43\% of the resources allocated to the \gls{ss} blocks. For $T_{\rm SS} = 20$~ms or higher, instead, the overhead is always below 10\%.

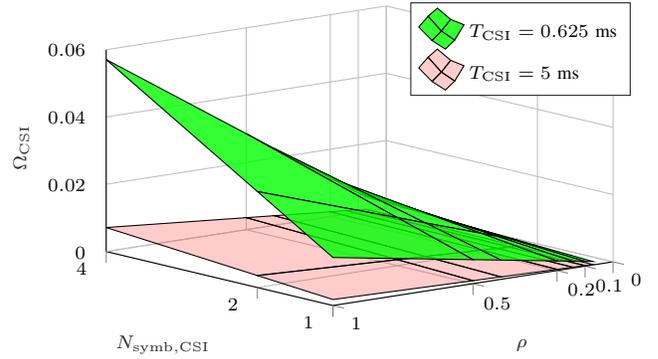
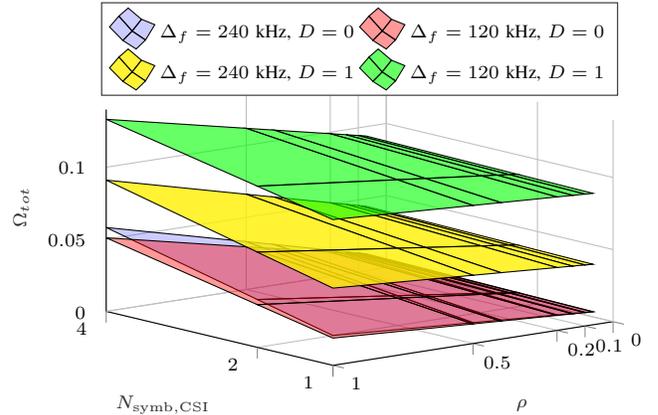
\begin{figure}[t]
  \centering
  \begin{subfigure}[t]{\columnwidth}
    \setlength\fwidth{0.8\columnwidth}
    \setlength\fheight{0.45\columnwidth}
    \input{figures/omegacsi_tss20ms.tex}
    \caption{Overhead $\Omega_{\rm CSI}$ as a function of $N_{\rm symb, CSI}$ and $\rho$, for different $T_{\rm CSI}$ periodicities, with $T_{\rm SS} = 20$~ms.}
    \label{fig:omegacsi20ms}
  \end{subfigure}%
  \vspace{.5cm}
  \begin{subfigure}[t]{\columnwidth}
    \setlength\fwidth{0.8\columnwidth}
    \setlength\fheight{0.45\columnwidth}
    \input{figures/omegatot_tss20ms.tex}
    \caption{Overhead $\Omega_{tot}$ as a function of $N_{\rm symb, CSI}$ and $\rho$, for different subcarrier spacings $\Delta_f$ and repetition strategies. $N_{\rm SS}$ is set to the maximum value, i.e., 64, and $T_{\rm CSI, slot} = 5$ slot.}
    \label{fig:omegatot20ms}
  \end{subfigure}
  \caption{Overhead for the \gls{csirs} transmission and total overhead, with $T_{\rm SS} = 20$~ms. Notice that the number of repetitions for the different subcarrier spacings $\Delta_f$ is chosen to send as many repetitions of the \gls{ss} blocks as possible.}
  \label{fig:omega_tot}
\end{figure}

\textbf{\gls{csirs} periodicity -- }The overhead due to the transmission of \glspl{csirs} is shown in Fig.~\ref{fig:omegacsi20ms} for different $T_{\rm CSI}$ periodicities and time and frequency resource allocation to the \glspl{csirs}. It is always below 0.008 with $T_{\rm CSI} = 5$~ms, and below 0.06 for $T_{\rm CSI} = 0.625$~ms. However, for practical values of the configuration of the \glspl{csirs}, in which the bandwidth for the reference signal is smaller than half of the entire bandwidth, then also for $T_{\rm CSI} = 0.625$~ms the overhead reaches very small values, i.e., below 0.028.

\textbf{Impact of \gls{ia} and tracking -- }The trend of $\Omega_{tot}$ is shown in Fig.~\ref{fig:omegatot20ms}, where it can be immediately seen that the largest impact is given by the term $R_{\rm SS}$ at the numerator and not by the \gls{csirs}-related overhead. The parameters on the $x$ and $y$ axes have indeed a limited effect on the gradient of the surfaces, which are almost horizontal. The main difference is introduced by the different subcarrier spacings and repetition strategies. 
Notice that, contrary to what is shown in Fig.~\ref{fig:omega5}, there is a difference between the two different subcarrier spacings for the total overhead $\Omega_{tot}$ and for the \gls{csirs}-related overhead $\Omega_{\rm CSI}$, because we consider a different $T_{symb}$ in Eq.~\eqref{eq:ovCsi}, but the same $\rho$ factor, thus a different number of subcarriers for the different values of $\Delta_f$. 

\textbf{Impact of beam reporting --}
 For the SA case, as reported in Table \ref{tab:ov_BR}, the completion of the beam reporting procedure for initial access may require an additional overhead, due to the need for the system to allocate possibly multiple RACH resources\footnote{According to the 3GPP agreements \cite{ls2017prach}, a bandwidth of $10$ MHz (for $\Delta_{f,\rm RACH}=60$ kHz) or a bandwidth of $20$ MHz (for $\Delta_{f,\rm RACH}=120$ kHz) is reserved for the RACH resources.} for the reporting operations.
 Conversely, for the \gls{nsa} case, the beam decision is forwarded through the LTE overlay and requires
a single RACH opportunity, with a total overhead of $0.0894\cdot10^{-3}$. 
Nevertheless, from Table \ref{tab:ov_BR}, we
notice that the SA additional reporting overhead  is quite limited due to the relatively small number of directions that need to be investigated at this stage, especially when designing digital beamforming solutions.

\begin{table}[t!]
\centering
\footnotesize
  \renewcommand{\arraystretch}{1.1}
\begin{tabular}{lcccc}
\toprule
& \multicolumn{4}{c}{$\Omega_{\rm BR,SA}$ $\cdot10^{-3}$}                                       \\ 
                             & \multicolumn{2}{c}{$ \Delta_{f,\rm RACH}=60$ kHz}      & \multicolumn{2}{c}{$  \Delta_{f,\rm RACH}=120$ kHz}      \\
$M_{gNB}$                             & Analog                  & Digital                & Analog                  & Digital                 \\ \midrule
\multicolumn{1}{c}{4}                             & 0.0894               & 0.0894              &  0.0894               & 0.0894           \\
\multicolumn{1}{c}{16}                            &  0.7149              & 0.0894              & 0.7149              & 0.0894            \\
\multicolumn{1}{c}{64}                            & 2.2341               & 0.0894              & 2.2341               & 0.0894           \\ \bottomrule
\end{tabular}
\caption{Overhead for beam reporting operations considering an SA architecture. Analog or digital beamforming is implemented at the gNB side, for different antenna array structures.}
\label{tab:ov_BR}
\end{table}

\section{Final Considerations}
\label{sec:considerations}
In the following paragraphs we will provide some insights on the trade-offs related to the different parameters we investigated and the three metrics considered, and some suggestions and guidelines to optimally design a measurement framework for \gls{nr} at mmWave frequencies.

\paragraph{\textbf{Subcarrier spacing} $\Delta_f$}
When using a smaller subcarrier spacing (i.e., $\Delta_f = 120$~kHz) it is possible to achieve a higher accuracy (i.e., smaller misdetection probability), either because the impact of the noise is less relevant, when frequency diversity is not used, or because it is possible to allocate a larger number of repetitions, when frequency diversity is used. This last option comes however at the price of an increase in the overhead in the order of 2 times, while the accuracy gain for the configuration with $\lambda = 30$~$\text{gNB/km}^\text{2}$ and the $4\times 4$ antenna arrays is in the order of 23\%, according to Fig. \ref{fig:D_4x4}. A smaller subcarrier spacing has also a negative effect on the reactiveness, as shown in Fig.~\ref{fig:scsTia}, since the \gls{ofdm} symbols last longer and the \gls{ss} blocks sweep takes more time. 

\paragraph{\textbf{Frequency diversity}}
The repetition in frequency of multiple \gls{ss} signals for the same \gls{ofdm} symbol results in an increased accuracy (e.g., up to 45\%, when $\lambda = 60$~$\text{gNB/km}^\text{2}$ and considering the $4\times 4$ array configuration). The overhead is, however, from 5 to 11 times higher in our setup (according to the $\Delta_f$ used), thus there is a trade-off between the amount of resources to allocate to the users that are already connected (which is higher if frequency diversity is not used) and the opportunity to discover new users (which increases with frequency diversity for the \gls{ss} blocks).
However, notice that the accuracy gain reduces when increasing the array dimension (e.g., when $\lambda = 60$~$\text{gNB/km}^\text{2}$ and considering the $64\times 4$ array configuration, a gain of just 15\% is achieved, as seen from Fig. \ref{fig:D_4x4}). 
In those circumstances, it may not be desirable to adopt a frequency diversity scheme which would inevitably increase the overhead while only providing marginal accuracy gain.

\paragraph{\textbf{Number of \gls{ss} blocks in a burst} $N_{\rm SS}$}
This parameter has a fundamental impact on the reactiveness, since a higher number of \gls{ss} blocks per burst increases the probability of completing the sweep in a single burst and thus prevents $T_{\rm IA}$ from being dependent on  $T_{\rm SS}$.
The number of \gls{ss} blocks per burst, however, increases also the overhead linearly. $N_{\rm SS}$ has a strict relationship with the number of directions to be swept, i.e., with both the beamforming architecture and the number of antennas: if, for example, hybrid or digital beamforming is used at the receiver, a larger number of antennas (i.e., narrower beams) can be supported even with a smaller $N_{\rm SS}$, as shown in Fig.~\ref{fig:nss}

\paragraph{\textbf{\gls{ss} burst periodicity }$T_{\rm SS}$}
The periodicity of a burst has an impact on the reactiveness for  initial access, since a smaller $T_{\rm SS}$ enables a larger number of opportunities in which a \gls{ue} can receive synchronization signals. However, if the beam sweeping procedure is completed in a single burst, $T_{\rm SS}$ does not impact $T_{\rm IA}$ as previously defined. 
The overhead is inversely proportional to $T_{\rm SS}$, which has a major impact also on the reactiveness related to the tracking and the transmission of \glspl{csirs}, as shown in Fig.~\ref{fig:csi}. Overall, if the sweep can be completed in a single burst, a higher $T_{\rm SS}$ would decrease the overhead and increase the reactiveness for the \glspl{csirs}.

\paragraph{\textbf{\gls{csirs} periodicity} $T_{\rm CSI}$}
A short $T_{\rm CSI}$ allows an improved reactiveness for the beam tracking of connected users. In particular, when the number of users per \gls{gnb} is high then a short \glspl{csirs} periodicity enables a much shorter $T_{tr}$. On the other hand, the overhead related to the \glspl{csirs} is small if compared with that of the \gls{ss} bursts. 

\paragraph{\textbf{Number of \glspl{csirs} to be monitored at the \gls{ue} side} $N_{\rm CSI, RX}$}
The impact of this parameter on the reactiveness is related to both the number of users per \gls{gnb} and the total number of directions to be swept with the reference signals. If there is a limited number of directions and a large number of users, uniformly distributed in the available directions, then the monitoring of additional \glspl{csirs} does not impact $T_{tr}$ or the overhead at the network side. The \gls{ue} may, however, be impacted by the energy consumption related to the monitoring of too many directions, i.e., by a needlessly high $N_{\rm CSI, RX}$.

\paragraph{\textbf{\gls{gnb} density} $\lambda_b$ }
As the network density increases, the accuracy and the average received power increase, and this allows a larger number of users to be served by a mmWave network. Besides the cost in terms of equipment and energy, a higher density has also a negative effect on the interference~\cite{rebato2016understanding}. Moreover, the number of neighbors of each single \gls{gnb} increases, and this limits the available configurations for the \glspl{csirs}.

\paragraph{\textbf{Beamforming architecture} $K_{\rm BF}$} 
A digital beamforming architecture at the receiver side would improve the reactiveness of the measurement scheme and decrease the overhead, without penalizing the accuracy. The same improvement in terms of reactiveness and accuracy can be achieved with an omnidirectional receiver, but the accuracy would decrease with a loss of around 30\% (when $\lambda = 30$~$\text{gNB/km}^\text{2}$) with respect to the $M_{\gls{gnb}} = 64$ configuration, as displayed in Fig. \ref{fig:P_err_N}. The complexity of the transceiver implementation and the energy consumption~\cite{abbas2017adc} are, however, two important parameters that must be taken into account. 
A hybrid configuration could represent a trade-off between an improved reactiveness and a simpler and less consuming transceiver design. Finally, notice that a digital architecture allows a higher gain with respect to the reactiveness if used at the \gls{gnb} in an uplink-based framework, since the directions to be swept at the \gls{gnb} are usually more than at the \gls{ue}.

\begin{figure}[t]
  \centering
    \input{figures/kiviat_diagram.tex}
    \caption{Comparison of three different configurations for accuracy, reactiveness and overhead. The common parameters are $\Delta_f=120$~kHz, $N_{\rm user}=10$, $N_{\rm symb, CSI} = 2$, $\rho = 0.072$, $\lambda_b=30$~$\text{gNB/km}^\text{2}$, $N_{\rm CSI, RX} = 3$, $T_{\rm SS} = 20$~ms, $T_{\rm CSI}=0.625$~ms.}
    \label{fig:kiviat}
\end{figure}
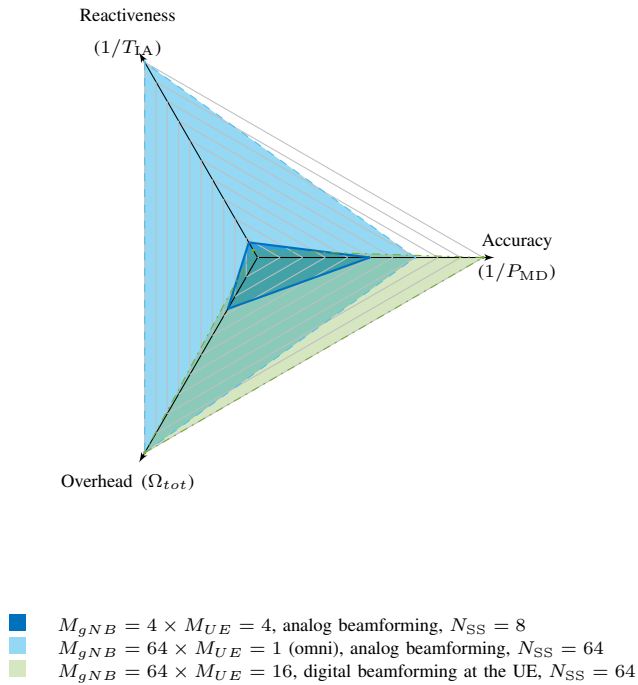

\paragraph{\textbf{Antenna Arrays} $M_{\gls{gnb}}, M_{\gls{ue}}$}
The antenna array is one of the parameters that has the largest impact on the accuracy. A larger number of antennas enable narrower beams and  higher accuracy, since the received power at the \gls{ue} (in downlink) or at the \gls{gnb} (in uplink) increases. 
The width of the beam has, however, an inverse relationship with the number of directions to scan, thus configurations that provide a higher accuracy perform worse in terms of reactiveness and overhead. Notice that the choice of the antenna array and of the beam design is strictly tied to the beamforming architecture (if digital or hybrid beamforming is used then narrower beams can be supported without penalizing reactiveness and overhead) and the configuration of the \gls{ss} bursts (a large number of directions to be swept with a limited number of \gls{ss} blocks per bursts has a negative impact on the reactiveness). 
In Fig.~\ref{fig:kiviat} a direct comparison among three different schemes is shown. 
It can be seen that the $M_{\gls{gnb}} = 4 \times M_{\gls{ue}} = 4$ configuration presents a smaller overhead and an improved reactiveness with respect to the $M_{\gls{gnb}} = 64 \times M_{\gls{ue}} = 16$ configuration. Moreover, both configurations with 64 antennas at the \gls{gnb} have the same overhead, but there is a trade-off between the reactiveness (the configuration with the omnidirectional \gls{ue} has the best reactiveness) and accuracy (using 16 antennas at the \gls{ue} provides the best accuracy, at the cost of a higher energy consumption because of digital beamforming).

\paragraph{\textbf{Measurement Framework}}
As far as initial access is concerned, the implementation of a standalone scheme generally guarantees more reactive access capabilities.
The reason is that faster beam reporting operations are ensured if multiple \gls{ss} blocks and \gls{rach} opportunities can be allocated within a single SS burst. 
On the other hand, a non-standalone framework may be preferable to:
(i) reduce the impact of the overhead in the beam reporting phase; 
(ii) in connected mode, implement efficient and reactive recovery operations as soon as a radio link failure event is detected;
(iii) guarantee a more robust control signaling exchange (e.g., when forwarding the beam reporting messages). 
Moreover, a non-standalone architecture is also better than an SA one when it is not possible to allocate in the same \gls{ss} burst the \gls{ss} blocks for the first sweep and the subsequent \gls{rach} opportunities, because for example there are too many directions to monitor at the~\gls{gnb}. 
Finally, \gls{nsa} enables a centralized beam decision: unlike in traditional attachment policies based on pathloss measurements, by leveraging on the presence of an eNB operating at sub-6 GHz frequencies, an \gls{nsa}-based beam association can be  performed by taking into account the instantaneous load conditions of the surrounding cells, thereby promoting fairness in the whole cellular network \cite{giordani2016efficient}.

Overall, it is possible to identify some guidelines for the configuration of the measurement framework and the deployment of a \gls{nr} network at mmWave frequencies. First, a setup of  $N_{\rm SS}$, the \gls{rach} resources, the beamforming and the antenna array architectures that allows the completion of the beam sweeping and reporting procedures in a single burst is preferable, so that it is possible to increase $T_{\rm SS}$ (e.g., to 20 or 40 ms), and consequently allocate a larger number of \glspl{csirs} for the connected users (to guarantee more reactive tracking operations) and reduce the overhead of the \gls{ss} blocks. 

Second, the adoption of a frequency diversity scheme for the \gls{ss} blocks depends on the load of the \glspl{gnb}: if many users are connected to a certain \gls{gnb}, this could disable the frequency diversity to both reduce the overhead and avoid discovering new users. Third, with low network density, larger antenna arrays make it possible to detect farther users, and provide a wider coverage but, as $\lambda_b$ increases, it is possible to use a configuration with wide beams for \gls{ss} bursts (so that it is more likely to complete a sweep in a single burst) and narrow ones for  \gls{csirs}, to refine the pointing directions for the data transmission and achieve higher gains.

Finally, when considering stable and dense scenarios which are marginally affected by the variability of the mmWave channel, a standalone architecture is preferable for the design of fast initial access procedures, since it enables rapid beam reporting operations. 
Conversely, an \gls{nsa} configuration should be employed by users in connected mode to guarantee higher resilience and an improved reactiveness in case a radio link failure occurs.
 A downlink configuration is in line with the 3GPP design for \gls{nr} and reduces the energy consumption at the \gls{ue} side (since it has just to receive the synchronization or reference signals), but  is less reactive because the \glspl{gnb} have a larger number of directions to sweep with downlink \gls{ss} blocks or \glspl{csirs}. 

\section{Conclusions}
\label{sec:conclusions}

In this paper, we have presented  a tutorial on beam management frameworks for mmWave communications in \gls{3gpp} \gls{nr}. The harsh propagation at mmWave frequencies requires the implementation  of directional transmissions supported by beamforming techniques to increase the link budget. 
Therefore, control procedures such as initial access must be updated to account for the lack of an omnidirectional broadcast channel, and the optimal beam pair with which a base station and a \gls{ue} communicate should be tracked when needed. 
Consequently, the design and configuration of efficient \gls{ia} and tracking procedures is of extreme importance in cellular networks operating at mmWaves. 

After a brief overview of the literature on beam management at mmWave frequencies, we described the frame structure and reference signals in 3GPP \gls{nr}, focusing on the settings for communication at frequencies above 6 GHz. 
Then, we described several beam management procedures according to different network architectures (standalone and non-standalone) and signal transmission directions (downlink or uplink). 
We also evaluated the impact of several parameters
(specified by 3GPP for \gls{nr}) on their performance.
We showed that there exist trade-offs among
better detection accuracy, improved reactiveness and reduced
overhead. 
Finally, we provide insights and guidelines for
determining the optimal initial access and tracking strategies in different
network deployments, according to the need of the network
operator and the specific environment in which the nodes
are deployed.

\renewcommand{\arraystretch}{1}
\footnotesize
\setlength{\glsdescwidth}{0.75\columnwidth}
\printglossary[style=index]

\bibliographystyle{IEEEtran}
\bibliography{bibl.bib}

\end{document}

%% file: acronymsSorted.tex
\newacronym{3gpp}{3GPP}{3rd Generation Partnership Project}
\newacronym{adc}{ADC}{Analog to Digital Converter}
\newacronym{5g}{5G}{5th generation}
\newacronym{aimd}{AIMD}{Additive Increase Multiplicative Decrease}
\newacronym{am}{AM}{Acknowledged Mode}
\newacronym{amc}{AMC}{Adaptive Modulation and Coding}
\newacronym{aqm}{AQM}{Active Queue Management}
\newacronym{awgn}{AGWN}{Additive White Gaussian Noise}
\newacronym{balia}{BALIA}{Balanced Link Adaptation}
\newacronym{bdp}{BDP}{Bandwidth-Delay Product}
\newacronym{bf}{BF}{Beamforming}
\newacronym{cc}{CC}{Congestion Control}
\newacronym{cdf}{CDF}{Cumulative Distribution Function}
\newacronym{cn}{CN}{Core Network}
\newacronym{cqi}{CQI}{Channel Quality Information}
\newacronym{cp}{CP}{Control Plane}
\newacronym{csirs}{CSI-RS}{Channel State Information - Reference Signal}
\newacronym{dc}{DC}{Dual Connectivity}
\newacronym{dce}{DCE}{Direct Code Execution}
\newacronym{dci}{DCI}{Downlink Control Information}
\newacronym{dl}{DL}{Downlink}
\newacronym{dmr}{DMR}{Deadline Miss Ratio}
\newacronym{dmrs}{DMRS}{DeModulation Reference Signal}
\newacronym{e2e}{E2E}{End-to-End}
\newacronym{ecn}{ECN}{Explicit Congestion Notification}
\newacronym{edf}{EDF}{Earliest Deadline First}
\newacronym{enb}{eNB}{evolved Node Base}
\newacronym{epc}{EPC}{Evolved Packet Core}
\newacronym{es}{ES}{Edge Server}
\newacronym{fdma}{FDMA}{Frequency Division Multiple Access}
\newacronym{fdd}{FDD}{Frequency Division Duplexing}
\newacronym[firstplural=Radio Access Technologies (RATs)]{rat}{RAT}{Radio Access Technology}
\newacronym{fs}{FS}{Fast Switching}
\newacronym{ftp}{FTP}{File Transfer Protocol}
\newacronym{gnb}{gNB}{Next Generation Node Base}
\newacronym{harq}{HARQ}{Hybrid Automatic Repeat reQuest}
\newacronym{hetnet}{HetNet}{Heterogeneous Network}
\newacronym{hh}{HH}{Hard Handover}
\newacronym{hol}{HOL}{Head-of-Line}
\newacronym{ia}{IA}{Initial Access}
\newacronym{imt}{IMT}{International Mobile Telecommunication}
\newacronym{iot}{IoT}{Internet of Things}
\newacronym{los}{LOS}{Line of Sight}
\newacronym{lte}{LTE}{Long Term Evolution}
\newacronym{m2m}{M2M}{Machine to Machine}
\newacronym{mac}{MAC}{Medium Access Control}
\newacronym{mc}{MC}{Multi-Connectivity}
\newacronym{mcs}{MCS}{Modulation and Coding Scheme}
\newacronym{mec}{MEC}{Mobile Edge Cloud}
\newacronym{mi}{MI}{Mutual Information}
\newacronym{mimo}{MIMO}{Multiple Input, Multiple Output}
\newacronym{mmwave}{mmWave}{millimeter wave}
\newacronym{mptcp}{MPTCP}{Multipath TCP}
\newacronym{mr}{MR}{Maximum Rate}
\newacronym{mss}{MSS}{Maximum Segment Size}
\newacronym{mtd}{MTD}{Machine-Type Device}
\newacronym{mtu}{MTU}{Maximum Transmission Unit}
\newacronym{nfv}{NFV}{Network Function Virtualization}
\newacronym{nlos}{NLOS}{Non Line of Sight}
\newacronym{nr}{NR}{NR}
\newacronym{ofdm}{OFDM}{Orthogonal Frequency Division Multiplexing}
\newacronym{pdcch}{PDCCH}{Physical Downlonk Control Channel}
\newacronym{pdcp}{PDCP}{Packet Data Convergence Protocol}
\newacronym{pdsch}{PDSCH}{Physical Downlink Shared Channel}
\newacronym{pdu}{PDU}{Packet Data Unit}
\newacronym{pf}{PF}{Proportional Fair}
\newacronym{pgw}{PGW}{Packet Gateway}
\newacronym{phy}{PHY}{Physical}
\newacronym{pbch}{PBCH}{Physical Broadcast Channel}
\newacronym[plural=\gls{mme}s,firstplural=Mobility Management Entities (MMEs)]{mme}{MME}{Mobility Management Entity}
\newacronym{prb}{PRB}{Physical Resource Block}
\newacronym{pss}{PSS}{Primary Synchronization Signal}
\newacronym{pucch}{PUCCH}{Physical Uplink Control Channel}
\newacronym{pusch}{PUSCH}{Physical Uplink Shared Channel}
\newacronym{rach}{RACH}{Random Access Channel}
\newacronym{ran}{RAN}{Radio Access Network}
\newacronym{red}{RED}{Random Early Detection}
\newacronym{rf}{RF}{Radio Frequency}
\newacronym{rlc}{RLC}{Radio Link Control}
\newacronym{rlf}{RLF}{Radio Link Failure}
\newacronym{rrc}{RRC}{Radio Resource Control}
\newacronym{rrm}{RRM}{Radio Resource Management}
\newacronym{rr}{RR}{Round Robin}
\newacronym{rs}{RS}{Remote Server}
\newacronym{rsrp}{RSRP}{Reference Signal Received Power}
\newacronym{rss}{RSS}{Received Signal Strength}
\newacronym{rtt}{RTT}{Round Trip Time}
\newacronym{rw}{RW}{Receive Window}
\newacronym{rx}{RX}{Receiver}
\newacronym{sa}{SA}{standalone}
\newacronym{sack}{SACK}{Selective Acknowledgment}
\newacronym{sap}{SAP}{Service Access Point}
\newacronym{sch}{SCH}{Secondary Cell Handover}
\newacronym{scoot}{SCOOT}{Split Cycle Offset Optimization Technique}
\newacronym{sdma}{SDMA}{Spatial Division Multiple Access}
\newacronym{sinr}{SINR}{Signal to Interference plus Noise Ratio}
\newacronym{sm}{SM}{Saturation Mode}
\newacronym{snr}{SNR}{Signal to Noise Ratio}
\newacronym{son}{SON}{Self-Organizing Network}
\newacronym{ss}{SS}{Synchronization Signal}
\newacronym{srs}{SRS}{Sounding Reference Signal}
\newacronym{sss}{SSS}{Secondary Synchronization Signal}
\newacronym{tb}{TB}{Transport Block}
\newacronym{tcp}{TCP}{Transmission Control Protocol}
\newacronym{tdd}{TDD}{Time Division Duplexing}
\newacronym{tdma}{TDMA}{Time Division Multiple Access}
\newacronym{tfl}{TfL}{Transport for London}
\newacronym{tm}{TM}{Transparent Mode}
\newacronym{trp}{TRP}{Transmitter Receiver Pair}
\newacronym{tti}{TTI}{Transmission Time Interval}
\newacronym{ttt}{TTT}{Time-to-Trigger}
\newacronym{tx}{TX}{Transmitter}
\newacronym{ue}{UE}{User Equipment}
\newacronym{ul}{UL}{Uplink}
\newacronym{uml}{UML}{Unified Modeling Language}
\newacronym{um}{UM}{Unacknowledged Mode}
\newacronym{utc}{UTC}{Urban Traffic Control}
\newacronym{vm}{VM}{Virtual Machine}
\newacronym{rsrq}{RSRQ}{Reference Signal Received Quality}
\newacronym{rssi}{RSSI}{Received Signal Strength Indicator}
\newacronym{crs}{CRS}{Cell Reference Signal}
\newacronym{nsa}{NSA}{Non-Standalone}
\newacronym{mrdc}{MR-DC}{Multi \gls{rat} \gls{dc}}
\newacronym{endc}{EN-DC}{E-UTRAN-\gls{nr} \gls{dc}}
\newacronym{5gc}{5GC}{5G Core}

%% file: myTikz.tex
\tikzstyle{startstop} = [rectangle, rounded corners, minimum width=2cm, minimum height=0.5cm,text centered, draw=black]
\tikzstyle{io} = [trapezium, trapezium left angle=70, trapezium right angle=110, minimum width=3cm, minimum height=1cm, text centered, draw=black]
\tikzstyle{process} = [rectangle, minimum width=2cm, minimum height=0.5cm, text centered, draw=black, alignb=center]
\tikzstyle{decision} = [ellipse, minimum width=2cm, minimum height=1cm, text centered, draw=black]
\tikzstyle{arrow} = [thick,<->,>=stealth]
\tikzstyle{line} = [thick,>=stealth]
\tikzstyle{darrow} = [thick,<->,>=stealth,dashed]
\tikzstyle{sarrow} = [thick,->,>=stealth]
\tikzstyle{larrow} = [line width=0.1mm,dashdotted,->,>=stealth]

\makeatletter
\def\grd@save@target#1{%
  \def\grd@target{#1}}
\def\grd@save@start#1{%
  \def\grd@start{#1}}
\tikzset{
  grid with coordinates/.style={
    to path={%
      \pgfextra{%
        \edef\grd@@target{(\tikztotarget)}%
        \tikz@scan@one@point\grd@save@target\grd@@target\relax
        \edef\grd@@start{(\tikztostart)}%
        \tikz@scan@one@point\grd@save@start\grd@@start\relax
        \draw[minor help lines] (\tikztostart) grid (\tikztotarget);
        \draw[major help lines] (\tikztostart) grid (\tikztotarget);
        \grd@start
        \pgfmathsetmacro{\grd@xa}{\the\pgf@x/1cm}
        \pgfmathsetmacro{\grd@ya}{\the\pgf@y/1cm}
        \grd@target
        \pgfmathsetmacro{\grd@xb}{\the\pgf@x/1cm}
        \pgfmathsetmacro{\grd@yb}{\the\pgf@y/1cm}
        \pgfmathsetmacro{\grd@xc}{\grd@xa + \pgfkeysvalueof{/tikz/grid with coordinates/major step x}}
        \pgfmathsetmacro{\grd@yc}{\grd@ya + \pgfkeysvalueof{/tikz/grid with coordinates/major step y}}
        \foreach \x in {\grd@xa,\grd@xc,...,\grd@xb}
        \node[anchor=north] at (\x,\grd@ya) {\pgfmathprintnumber{\x}};
        \foreach \y in {\grd@ya,\grd@yc,...,\grd@yb}
        \node[anchor=east] at (\grd@xa,\y) {\pgfmathprintnumber{\y}};
      }
    }
  },
  minor help lines/.style={
    help lines,
    gray,
    line cap =round,
    xstep=\pgfkeysvalueof{/tikz/grid with coordinates/minor step x},
    ystep=\pgfkeysvalueof{/tikz/grid with coordinates/minor step y}
  },
  major help lines/.style={
    help lines,
    line cap =round,
    line width=\pgfkeysvalueof{/tikz/grid with coordinates/major line width},
    xstep=\pgfkeysvalueof{/tikz/grid with coordinates/major step x},
    ystep=\pgfkeysvalueof{/tikz/grid with coordinates/major step y}
  },
  grid with coordinates/.cd,
  minor step x/.initial=.5,
  minor step y/.initial=.2,
  major step x/.initial=1,
  major step y/.initial=1,
  major line width/.initial=1pt,
}
\makeatother

%% file: figures/beamwidth.tex
%
%
\definecolor{mycolor1}{rgb}{0.00000,0.44700,0.74100}%
\definecolor{mycolor2}{rgb}{0.85000,0.32500,0.09800}%
\definecolor{mycolor3}{rgb}{0.92900,0.69400,0.12500}%
\definecolor{mycolor4}{rgb}{0.49400,0.18400,0.55600}%
\definecolor{mycolor5}{rgb}{0.46600,0.67400,0.18800}%
\definecolor{mycolor6}{rgb}{0.30100,0.74500,0.93300}

\pgfplotsset{
tick label style={font=\scriptsize},
label style={font=\scriptsize},
legend  style={font=\scriptsize}
}

\begin{tikzpicture}

\begin{axis}[%
width=0.956\fwidth,
height=\fheight,
at={(0\fwidth,0\fheight)},
scale only axis,
xmin=-60,
xmax=60,
ymin=-40,
ymax=30,
ylabel={ Directivity (dBi)},
axis background/.style={fill=white},
title style={font=\bfseries},
xlabel={Azimuth Angle (degrees)},
xmajorgrids,
ymajorgrids,
ylabel shift=-6pt,
mark repeat={10},
legend columns=3,
label style={font=\scriptsize},
legend style={font=\scriptsize, at={(0.5, 0)},anchor=south,legend cell align=left, align=left, draw=white!15!black}
]
\addplot [color=mycolor1,mark=o, mark size=2.0pt, mark options={solid, mycolor1}]
  table[row sep=crcr]{%
-180	7.08272879180339\\
-179	7.07946449899839\\
-178	7.06967069033215\\
-177	7.0533445705891\\
-176	7.03048146618374\\
-175	7.00107480274778\\
-174	6.96511607360566\\
-173	6.92259479898996\\
-172	6.87349847580367\\
-171	6.81781251768984\\
-170	6.7555201851219\\
-169	6.68660250517553\\
-168	6.61103818058933\\
-167	6.52880348766519\\
-166	6.43987216249474\\
-165	6.34421527493561\\
-164	6.24180108968471\\
-163	6.1325949137206\\
-162	6.01655892929961\\
-161	5.89365201159649\\
-160	5.76382952997761\\
-159	5.62704313177875\\
-158	5.48324050733532\\
-157	5.33236513486986\\
-156	5.17435600368719\\
-155	5.00914731395096\\
-154	4.83666815112122\\
-153	4.65684213291235\\
-152	4.46958702638375\\
-151	4.27481433249773\\
-150	4.07242883516358\\
-149	3.86232811143236\\
-148	3.64440199909961\\
-147	3.41853201751553\\
-146	3.18459073687569\\
-145	2.9424410906642\\
-144	2.6919356252356\\
-143	2.43291567972636\\
-142	2.1652104885809\\
-141	1.8886361979192\\
-140	1.60299478575695\\
-139	1.30807287467098\\
-138	1.00364042385337\\
-137	0.689449285567491\\
-136	0.365231608761121\\
-135	0.0306980699315318\\
-134	-0.314464091798413\\
-133	-0.67059326214769\\
-132	-1.0380558928569\\
-131	-1.41724923301479\\
-130	-1.8086043932712\\
-129	-2.21258979326285\\
-128	-2.62971505173886\\
-127	-3.06053538999288\\
-126	-3.505656632767\\
-125	-3.96574090740626\\
-124	-4.44151316251708\\
-123	-4.93376865274553\\
-122	-5.44338156790576\\
-121	-5.97131502433751\\
-120	-6.51863268642808\\
-119	-7.08651234987911\\
-118	-7.67626189982133\\
-117	-8.28933816214368\\
-116	-8.92736930347507\\
-115	-9.59218161537353\\
-114	-10.2858317572761\\
-113	-11.0106458532177\\
-112	-11.7692672718667\\
-111	-12.5647155158795\\
-110	-13.4004594761466\\
-109	-14.2805094769718\\
-108	-15.2095342157703\\
-107	-16.1930111468473\\
-106	-17.2374224933888\\
-105	-18.3505145870155\\
-104	-19.541646801027\\
-103	-20.8222700033243\\
-102	-22.2065968897742\\
-101	-23.7125646679278\\
-100	-25.3632578535633\\
-99	-27.1890832469269\\
-98	-29.2312313168101\\
-97	-31.5474610602099\\
-96	-34.2223743998037\\
-95	-37.3871430124251\\
-94	-41.2615283937757\\
-93	-46.2575198474673\\
-92	-53.3000623429118\\
-91	-65.3405994500866\\
-90	-317.177654087045\\
-89	-65.3405994500866\\
-88	-53.3000623429118\\
-87	-46.2575198474673\\
-86	-41.2615283937757\\
-85	-37.3871430124251\\
-84	-34.2223743998037\\
-83	-31.5474610602099\\
-82	-29.2312313168101\\
-81	-27.1890832469269\\
-80	-25.3632578535633\\
-79	-23.7125646679278\\
-78	-22.2065968897742\\
-77	-20.8222700033243\\
-76	-19.541646801027\\
-75	-18.3505145870155\\
-74	-17.2374224933888\\
-73	-16.1930111468473\\
-72	-15.2095342157703\\
-71	-14.2805094769718\\
-70	-13.4004594761466\\
-69	-12.5647155158795\\
-68	-11.7692672718667\\
-67	-11.0106458532177\\
-66	-10.2858317572761\\
-65	-9.59218161537353\\
-64	-8.92736930347507\\
-63	-8.28933816214368\\
-62	-7.67626189982133\\
-61	-7.08651234987911\\
-60	-6.51863268642808\\
-59	-5.97131502433751\\
-58	-5.44338156790576\\
-57	-4.93376865274553\\
-56	-4.44151316251708\\
-55	-3.96574090740626\\
-54	-3.505656632767\\
-53	-3.06053538999288\\
-52	-2.62971505173886\\
-51	-2.21258979326285\\
-50	-1.8086043932712\\
-49	-1.41724923301479\\
-48	-1.0380558928569\\
-47	-0.67059326214769\\
-46	-0.314464091798413\\
-45	0.0306980699315318\\
-44	0.365231608761121\\
-43	0.689449285567491\\
-42	1.00364042385337\\
-41	1.30807287467098\\
-40	1.60299478575695\\
-39	1.8886361979192\\
-38	2.1652104885809\\
-37	2.43291567972636\\
-36	2.6919356252356\\
-35	2.9424410906642\\
-34	3.18459073687569\\
-33	3.41853201751553\\
-32	3.64440199909961\\
-31	3.86232811143236\\
-30	4.07242883516358\\
-29	4.27481433249773\\
-28	4.46958702638375\\
-27	4.65684213291235\\
-26	4.83666815112122\\
-25	5.00914731395096\\
-24	5.17435600368719\\
-23	5.33236513486986\\
-22	5.48324050733532\\
-21	5.62704313177875\\
-20	5.76382952997761\\
-19	5.89365201159649\\
-18	6.01655892929961\\
-17	6.1325949137206\\
-16	6.24180108968471\\
-15	6.34421527493561\\
-14	6.43987216249474\\
-13	6.52880348766519\\
-12	6.61103818058933\\
-11	6.68660250517553\\
-10	6.7555201851219\\
-9	6.81781251768984\\
-8	6.87349847580367\\
-7	6.92259479898996\\
-6	6.96511607360566\\
-5	7.00107480274778\\
-4	7.03048146618374\\
-3	7.0533445705891\\
-2	7.06967069033215\\
-1	7.07946449899839\\
0	7.08272879180339\\
1	7.07946449899839\\
2	7.06967069033215\\
3	7.0533445705891\\
4	7.03048146618374\\
5	7.00107480274778\\
6	6.96511607360566\\
7	6.92259479898996\\
8	6.87349847580367\\
9	6.81781251768984\\
10	6.7555201851219\\
11	6.68660250517553\\
12	6.61103818058933\\
13	6.52880348766519\\
14	6.43987216249474\\
15	6.34421527493561\\
16	6.24180108968471\\
17	6.1325949137206\\
18	6.01655892929961\\
19	5.89365201159649\\
20	5.76382952997761\\
21	5.62704313177875\\
22	5.48324050733532\\
23	5.33236513486986\\
24	5.17435600368719\\
25	5.00914731395096\\
26	4.83666815112122\\
27	4.65684213291235\\
28	4.46958702638375\\
29	4.27481433249773\\
30	4.07242883516358\\
31	3.86232811143236\\
32	3.64440199909961\\
33	3.41853201751553\\
34	3.18459073687569\\
35	2.9424410906642\\
36	2.6919356252356\\
37	2.43291567972636\\
38	2.1652104885809\\
39	1.8886361979192\\
40	1.60299478575695\\
41	1.30807287467098\\
42	1.00364042385337\\
43	0.689449285567491\\
44	0.365231608761121\\
45	0.0306980699315318\\
46	-0.314464091798413\\
47	-0.67059326214769\\
48	-1.0380558928569\\
49	-1.41724923301479\\
50	-1.8086043932712\\
51	-2.21258979326285\\
52	-2.62971505173886\\
53	-3.06053538999288\\
54	-3.505656632767\\
55	-3.96574090740626\\
56	-4.44151316251708\\
57	-4.93376865274553\\
58	-5.44338156790576\\
59	-5.97131502433751\\
60	-6.51863268642808\\
61	-7.08651234987911\\
62	-7.67626189982133\\
63	-8.28933816214368\\
64	-8.92736930347507\\
65	-9.59218161537353\\
66	-10.2858317572761\\
67	-11.0106458532177\\
68	-11.7692672718667\\
69	-12.5647155158795\\
70	-13.4004594761466\\
71	-14.2805094769718\\
72	-15.2095342157703\\
73	-16.1930111468473\\
74	-17.2374224933888\\
75	-18.3505145870155\\
76	-19.541646801027\\
77	-20.8222700033243\\
78	-22.2065968897742\\
79	-23.7125646679278\\
80	-25.3632578535633\\
81	-27.1890832469269\\
82	-29.2312313168101\\
83	-31.5474610602099\\
84	-34.2223743998037\\
85	-37.3871430124251\\
86	-41.2615283937757\\
87	-46.2575198474673\\
88	-53.3000623429118\\
89	-65.3405994500866\\
90	-317.177654087045\\
91	-65.3405994500866\\
92	-53.3000623429118\\
93	-46.2575198474673\\
94	-41.2615283937757\\
95	-37.3871430124251\\
96	-34.2223743998037\\
97	-31.5474610602099\\
98	-29.2312313168101\\
99	-27.1890832469269\\
100	-25.3632578535633\\
101	-23.7125646679278\\
102	-22.2065968897742\\
103	-20.8222700033243\\
104	-19.541646801027\\
105	-18.3505145870155\\
106	-17.2374224933888\\
107	-16.1930111468473\\
108	-15.2095342157703\\
109	-14.2805094769718\\
110	-13.4004594761466\\
111	-12.5647155158795\\
112	-11.7692672718667\\
113	-11.0106458532177\\
114	-10.2858317572761\\
115	-9.59218161537353\\
116	-8.92736930347507\\
117	-8.28933816214368\\
118	-7.67626189982133\\
119	-7.08651234987911\\
120	-6.51863268642808\\
121	-5.97131502433751\\
122	-5.44338156790576\\
123	-4.93376865274553\\
124	-4.44151316251708\\
125	-3.96574090740626\\
126	-3.505656632767\\
127	-3.06053538999288\\
128	-2.62971505173886\\
129	-2.21258979326285\\
130	-1.8086043932712\\
131	-1.41724923301479\\
132	-1.0380558928569\\
133	-0.67059326214769\\
134	-0.314464091798413\\
135	0.0306980699315318\\
136	0.365231608761121\\
137	0.689449285567491\\
138	1.00364042385337\\
139	1.30807287467098\\
140	1.60299478575695\\
141	1.8886361979192\\
142	2.1652104885809\\
143	2.43291567972636\\
144	2.6919356252356\\
145	2.9424410906642\\
146	3.18459073687569\\
147	3.41853201751553\\
148	3.64440199909961\\
149	3.86232811143236\\
150	4.07242883516358\\
151	4.27481433249773\\
152	4.46958702638375\\
153	4.65684213291235\\
154	4.83666815112122\\
155	5.00914731395096\\
156	5.17435600368719\\
157	5.33236513486986\\
158	5.48324050733532\\
159	5.62704313177875\\
160	5.76382952997761\\
161	5.89365201159649\\
162	6.01655892929961\\
163	6.1325949137206\\
164	6.24180108968471\\
165	6.34421527493561\\
166	6.43987216249474\\
167	6.52880348766519\\
168	6.61103818058933\\
169	6.68660250517553\\
170	6.7555201851219\\
171	6.81781251768984\\
172	6.87349847580367\\
173	6.92259479898996\\
174	6.96511607360566\\
175	7.00107480274778\\
176	7.03048146618374\\
177	7.0533445705891\\
178	7.06967069033215\\
179	7.07946449899839\\
180	7.08272879180339\\
};
\addlegendentry{$M = 4$}

\addplot [color=mycolor2,mark=square*, mark size=2.0pt, mark options={solid, mycolor2}]
  table[row sep=crcr]{%
-180	13.5049084275736\\
-179	13.4885820527687\\
-178	13.4395391587373\\
-177	13.357586985967\\
-176	13.2423993645069\\
-175	13.0935091122501\\
-174	12.9102968804449\\
-173	12.6919758827411\\
-172	12.4375717017322\\
-171	12.1458960488538\\
-170	11.8155129228579\\
-169	11.4446950141075\\
-168	11.0313673528758\\
-167	10.573033968475\\
-166	10.0666815019112\\
-165	9.50865095152307\\
-164	8.89446444301632\\
-163	8.21858708149966\\
-162	7.47409272249814\\
-161	6.65218344160979\\
-160	5.74147883388099\\
-159	4.72692911978987\\
-158	3.58808495239697\\
-157	2.29620539893642\\
-156	0.809121071583263\\
-155	-0.938629034498661\\
-154	-3.05680739478095\\
-153	-5.75129684198856\\
-152	-9.48446445613109\\
-151	-15.7319330609349\\
-150	-293.528755569493\\
-149	-16.2318812387636\\
-148	-10.4841963201348\\
-147	-7.25048394990716\\
-146	-5.05495844089785\\
-145	-3.43509149308341\\
-144	-2.18484126893736\\
-143	-1.19428909617631\\
-142	-0.39782120077939\\
-141	0.246880493860544\\
-140	0.768700836172229\\
-139	1.18822757552846\\
-138	1.52064258492163\\
-137	1.77745119328807\\
-136	1.96756814392677\\
-135	2.09802727284443\\
-134	2.1744609169528\\
-133	2.20143290598776\\
-132	2.1826753411986\\
-131	2.12126030388964\\
-130	2.01972641308498\\
-129	1.88017331396281\\
-128	1.70433288605483\\
-127	1.49362319139323\\
-126	1.24918935186901\\
-125	0.971934305407558\\
-124	0.662541531288582\\
-123	0.321491224321846\\
-122	-0.0509290487931935\\
-121	-0.45461851543206\\
-120	-0.889661757093734\\
-119	-1.35633145905535\\
-118	-1.85509534603\\
-117	-2.38662749535615\\
-116	-2.95182442462243\\
-115	-3.55182658310281\\
-114	-4.18804615581513\\
-113	-4.8622024409238\\
-112	-5.57636652036967\\
-111	-6.33301755943398\\
-110	-7.13511391583739\\
-109	-7.98618342161132\\
-108	-8.89043888822357\\
-107	-9.85292733917339\\
-106	-10.8797251150734\\
-105	-11.9781965164775\\
-104	-13.1573422205608\\
-103	-14.428277370939\\
-102	-15.8049016774306\\
-101	-17.3048619740067\\
-100	-18.950974977679\\
-99	-20.7734022971915\\
-98	-22.813111903164\\
-97	-25.1276631123833\\
-96	-27.8014811347417\\
-95	-30.96558406364\\
-94	-34.8396031040225\\
-93	-39.8354207164366\\
-92	-46.8778986133134\\
-91	-58.9184208086069\\
-90	-310.755474451275\\
-89	-58.9184208086069\\
-88	-46.8778986133134\\
-87	-39.8354207164366\\
-86	-34.8396031040225\\
-85	-30.96558406364\\
-84	-27.8014811347417\\
-83	-25.1276631123833\\
-82	-22.813111903164\\
-81	-20.7734022971915\\
-80	-18.950974977679\\
-79	-17.3048619740067\\
-78	-15.8049016774306\\
-77	-14.428277370939\\
-76	-13.1573422205608\\
-75	-11.9781965164775\\
-74	-10.8797251150734\\
-73	-9.85292733917339\\
-72	-8.89043888822357\\
-71	-7.98618342161132\\
-70	-7.13511391583739\\
-69	-6.33301755943398\\
-68	-5.57636652036967\\
-67	-4.8622024409238\\
-66	-4.18804615581513\\
-65	-3.55182658310281\\
-64	-2.95182442462243\\
-63	-2.38662749535615\\
-62	-1.85509534603\\
-61	-1.35633145905535\\
-60	-0.889661757093734\\
-59	-0.45461851543206\\
-58	-0.0509290487931935\\
-57	0.321491224321846\\
-56	0.662541531288582\\
-55	0.971934305407558\\
-54	1.24918935186901\\
-53	1.49362319139323\\
-52	1.70433288605483\\
-51	1.88017331396281\\
-50	2.01972641308498\\
-49	2.12126030388964\\
-48	2.1826753411986\\
-47	2.20143290598776\\
-46	2.1744609169528\\
-45	2.09802727284443\\
-44	1.96756814392677\\
-43	1.77745119328807\\
-42	1.52064258492163\\
-41	1.18822757552846\\
-40	0.768700836172229\\
-39	0.246880493860544\\
-38	-0.39782120077939\\
-37	-1.19428909617631\\
-36	-2.18484126893736\\
-35	-3.43509149308341\\
-34	-5.05495844089785\\
-33	-7.25048394990716\\
-32	-10.4841963201348\\
-31	-16.2318812387636\\
-30	-293.528755569493\\
-29	-15.7319330609349\\
-28	-9.48446445613109\\
-27	-5.75129684198856\\
-26	-3.05680739478095\\
-25	-0.938629034498661\\
-24	0.809121071583263\\
-23	2.29620539893642\\
-22	3.58808495239697\\
-21	4.72692911978987\\
-20	5.74147883388099\\
-19	6.65218344160979\\
-18	7.47409272249814\\
-17	8.21858708149966\\
-16	8.89446444301632\\
-15	9.50865095152307\\
-14	10.0666815019112\\
-13	10.573033968475\\
-12	11.0313673528758\\
-11	11.4446950141075\\
-10	11.8155129228579\\
-9	12.1458960488538\\
-8	12.4375717017322\\
-7	12.6919758827411\\
-6	12.9102968804449\\
-5	13.0935091122501\\
-4	13.2423993645069\\
-3	13.357586985967\\
-2	13.4395391587373\\
-1	13.4885820527687\\
0	13.5049084275736\\
1	13.4885820527687\\
2	13.4395391587373\\
3	13.357586985967\\
4	13.2423993645069\\
5	13.0935091122501\\
6	12.9102968804449\\
7	12.6919758827411\\
8	12.4375717017322\\
9	12.1458960488538\\
10	11.8155129228579\\
11	11.4446950141075\\
12	11.0313673528758\\
13	10.573033968475\\
14	10.0666815019112\\
15	9.50865095152307\\
16	8.89446444301632\\
17	8.21858708149966\\
18	7.47409272249814\\
19	6.65218344160979\\
20	5.74147883388099\\
21	4.72692911978987\\
22	3.58808495239697\\
23	2.29620539893642\\
24	0.809121071583263\\
25	-0.938629034498661\\
26	-3.05680739478095\\
27	-5.75129684198856\\
28	-9.48446445613109\\
29	-15.7319330609349\\
30	-293.528755569493\\
31	-16.2318812387636\\
32	-10.4841963201348\\
33	-7.25048394990716\\
34	-5.05495844089785\\
35	-3.43509149308341\\
36	-2.18484126893736\\
37	-1.19428909617631\\
38	-0.39782120077939\\
39	0.246880493860544\\
40	0.768700836172229\\
41	1.18822757552846\\
42	1.52064258492163\\
43	1.77745119328807\\
44	1.96756814392677\\
45	2.09802727284443\\
46	2.1744609169528\\
47	2.20143290598776\\
48	2.1826753411986\\
49	2.12126030388964\\
50	2.01972641308498\\
51	1.88017331396281\\
52	1.70433288605483\\
53	1.49362319139323\\
54	1.24918935186901\\
55	0.971934305407558\\
56	0.662541531288582\\
57	0.321491224321846\\
58	-0.0509290487931935\\
59	-0.45461851543206\\
60	-0.889661757093734\\
61	-1.35633145905535\\
62	-1.85509534603\\
63	-2.38662749535615\\
64	-2.95182442462243\\
65	-3.55182658310281\\
66	-4.18804615581513\\
67	-4.8622024409238\\
68	-5.57636652036967\\
69	-6.33301755943398\\
70	-7.13511391583739\\
71	-7.98618342161132\\
72	-8.89043888822357\\
73	-9.85292733917339\\
74	-10.8797251150734\\
75	-11.9781965164775\\
76	-13.1573422205608\\
77	-14.428277370939\\
78	-15.8049016774306\\
79	-17.3048619740067\\
80	-18.950974977679\\
81	-20.7734022971915\\
82	-22.813111903164\\
83	-25.1276631123833\\
84	-27.8014811347417\\
85	-30.96558406364\\
86	-34.8396031040225\\
87	-39.8354207164366\\
88	-46.8778986133134\\
89	-58.9184208086069\\
90	-310.755474451275\\
91	-58.9184208086069\\
92	-46.8778986133134\\
93	-39.8354207164366\\
94	-34.8396031040225\\
95	-30.96558406364\\
96	-27.8014811347417\\
97	-25.1276631123833\\
98	-22.813111903164\\
99	-20.7734022971915\\
100	-18.950974977679\\
101	-17.3048619740067\\
102	-15.8049016774306\\
103	-14.428277370939\\
104	-13.1573422205608\\
105	-11.9781965164775\\
106	-10.8797251150734\\
107	-9.85292733917339\\
108	-8.89043888822357\\
109	-7.98618342161132\\
110	-7.13511391583739\\
111	-6.33301755943398\\
112	-5.57636652036967\\
113	-4.8622024409238\\
114	-4.18804615581513\\
115	-3.55182658310281\\
116	-2.95182442462243\\
117	-2.38662749535615\\
118	-1.85509534603\\
119	-1.35633145905535\\
120	-0.889661757093734\\
121	-0.45461851543206\\
122	-0.0509290487931935\\
123	0.321491224321846\\
124	0.662541531288582\\
125	0.971934305407558\\
126	1.24918935186901\\
127	1.49362319139323\\
128	1.70433288605483\\
129	1.88017331396281\\
130	2.01972641308498\\
131	2.12126030388964\\
132	2.1826753411986\\
133	2.20143290598776\\
134	2.1744609169528\\
135	2.09802727284443\\
136	1.96756814392677\\
137	1.77745119328807\\
138	1.52064258492163\\
139	1.18822757552846\\
140	0.768700836172229\\
141	0.246880493860544\\
142	-0.39782120077939\\
143	-1.19428909617631\\
144	-2.18484126893736\\
145	-3.43509149308341\\
146	-5.05495844089785\\
147	-7.25048394990716\\
148	-10.4841963201348\\
149	-16.2318812387636\\
150	-293.528755569493\\
151	-15.7319330609349\\
152	-9.48446445613109\\
153	-5.75129684198856\\
154	-3.05680739478095\\
155	-0.938629034498661\\
156	0.809121071583263\\
157	2.29620539893642\\
158	3.58808495239697\\
159	4.72692911978987\\
160	5.74147883388099\\
161	6.65218344160979\\
162	7.47409272249814\\
163	8.21858708149966\\
164	8.89446444301632\\
165	9.50865095152307\\
166	10.0666815019112\\
167	10.573033968475\\
168	11.0313673528758\\
169	11.4446950141075\\
170	11.8155129228579\\
171	12.1458960488538\\
172	12.4375717017322\\
173	12.6919758827411\\
174	12.9102968804449\\
175	13.0935091122501\\
176	13.2423993645069\\
177	13.357586985967\\
178	13.4395391587373\\
179	13.4885820527687\\
180	13.5049084275736\\
};
\addlegendentry{$M=16$}

\addplot [color=mycolor3, dashed]
  table[row sep=crcr]{%
-180	19.7368004166453\\
-179	19.6681469042629\\
-178	19.4609109128497\\
-177	19.1111454467818\\
-176	18.6118444875687\\
-175	17.9522021227026\\
-174	17.1163353391544\\
-173	16.0811060734463\\
-172	14.8122949010681\\
-171	13.2574818719154\\
-170	11.3316298302124\\
-169	8.88413488870634\\
-168	5.60861072959517\\
-167	0.70842338968157\\
-166	-9.59536965310866\\
-165	-9.39187381373574\\
-164	-0.77017790312118\\
-163	2.85274521432626\\
-162	4.88914664413221\\
-161	6.08559823291279\\
-160	6.7251796075721\\
-159	6.93858961233184\\
-158	6.78815364052309\\
-157	6.29655484057412\\
-156	5.45562791330781\\
-155	4.22341933117013\\
-154	2.50693221000029\\
-153	0.112480752780755\\
-152	-3.41335338859477\\
-151	-9.53966356377742\\
-150	-306.535433840008\\
-149	-10.0388200025844\\
-148	-4.40669482343156\\
-147	-1.36481142399765\\
-146	0.561815274933053\\
-145	1.83356741150692\\
-144	2.65280737457408\\
-143	3.12392034369935\\
-142	3.30462214756318\\
-141	3.22616729341109\\
-140	2.90193627918114\\
-139	2.33047670598722\\
-138	1.49478672198848\\
-137	0.357422844013968\\
-136	-1.15165028016463\\
-135	-3.16411413716031\\
-134	-5.95234131587927\\
-133	-10.2111047114391\\
-132	-18.904150867825\\
-131	-22.2248740103855\\
-130	-11.6930082719974\\
-129	-7.29238833774463\\
-128	-4.58477828353398\\
-127	-2.70742665473244\\
-126	-1.33575672649692\\
-125	-0.311512321746079\\
-124	0.454703127218977\\
-123	1.01848043510154\\
-122	1.41680490443781\\
-121	1.67553309026692\\
-120	1.8134221370085\\
-119	1.84445639886843\\
-118	1.77926441192523\\
-117	1.62601805093095\\
-116	1.39102069100022\\
-115	1.07909946733747\\
-114	0.693868392510922\\
-113	0.237902216341695\\
-112	-0.287154816730492\\
-111	-0.880530893417586\\
-110	-1.54231128819777\\
-109	-2.27346254540504\\
-108	-3.07589698473237\\
-107	-3.95258232001657\\
-106	-4.90770586348344\\
-105	-5.94690902519511\\
-104	-7.07761688978798\\
-103	-8.30950167099473\\
-102	-9.65514153648229\\
-101	-11.1309745912347\\
-100	-12.7587152369273\\
-99	-14.5675245303866\\
-98	-16.5974684015033\\
-97	-18.9053004887745\\
-96	-21.5747353907698\\
-95	-24.7361749999506\\
-94	-28.608728528816\\
-93	-33.6038507493081\\
-92	-40.6460702490309\\
-91	-52.686532796651\\
-90	-304.523582462203\\
-89	-52.686532796651\\
-88	-40.6460702490309\\
-87	-33.6038507493081\\
-86	-28.608728528816\\
-85	-24.7361749999506\\
-84	-21.5747353907698\\
-83	-18.9053004887745\\
-82	-16.5974684015033\\
-81	-14.5675245303866\\
-80	-12.7587152369273\\
-79	-11.1309745912347\\
-78	-9.65514153648229\\
-77	-8.30950167099473\\
-76	-7.07761688978798\\
-75	-5.94690902519511\\
-74	-4.90770586348344\\
-73	-3.95258232001657\\
-72	-3.07589698473237\\
-71	-2.27346254540504\\
-70	-1.54231128819777\\
-69	-0.880530893417586\\
-68	-0.287154816730492\\
-67	0.237902216341695\\
-66	0.693868392510922\\
-65	1.07909946733747\\
-64	1.39102069100022\\
-63	1.62601805093095\\
-62	1.77926441192523\\
-61	1.84445639886843\\
-60	1.8134221370085\\
-59	1.67553309026692\\
-58	1.41680490443781\\
-57	1.01848043510154\\
-56	0.454703127218977\\
-55	-0.311512321746079\\
-54	-1.33575672649692\\
-53	-2.70742665473244\\
-52	-4.58477828353398\\
-51	-7.29238833774463\\
-50	-11.6930082719974\\
-49	-22.2248740103855\\
-48	-18.904150867825\\
-47	-10.2111047114391\\
-46	-5.95234131587927\\
-45	-3.16411413716031\\
-44	-1.15165028016463\\
-43	0.357422844013968\\
-42	1.49478672198848\\
-41	2.33047670598722\\
-40	2.90193627918114\\
-39	3.22616729341109\\
-38	3.30462214756318\\
-37	3.12392034369935\\
-36	2.65280737457408\\
-35	1.83356741150692\\
-34	0.561815274933053\\
-33	-1.36481142399765\\
-32	-4.40669482343156\\
-31	-10.0388200025844\\
-30	-306.535433840008\\
-29	-9.53966356377742\\
-28	-3.41335338859477\\
-27	0.112480752780755\\
-26	2.50693221000029\\
-25	4.22341933117013\\
-24	5.45562791330781\\
-23	6.29655484057412\\
-22	6.78815364052309\\
-21	6.93858961233184\\
-20	6.7251796075721\\
-19	6.08559823291279\\
-18	4.88914664413221\\
-17	2.85274521432626\\
-16	-0.77017790312118\\
-15	-9.39187381373574\\
-14	-9.59536965310866\\
-13	0.70842338968157\\
-12	5.60861072959517\\
-11	8.88413488870634\\
-10	11.3316298302124\\
-9	13.2574818719154\\
-8	14.8122949010681\\
-7	16.0811060734463\\
-6	17.1163353391544\\
-5	17.9522021227026\\
-4	18.6118444875687\\
-3	19.1111454467818\\
-2	19.4609109128497\\
-1	19.6681469042629\\
0	19.7368004166453\\
1	19.6681469042629\\
2	19.4609109128497\\
3	19.1111454467818\\
4	18.6118444875687\\
5	17.9522021227026\\
6	17.1163353391544\\
7	16.0811060734463\\
8	14.8122949010681\\
9	13.2574818719154\\
10	11.3316298302124\\
11	8.88413488870634\\
12	5.60861072959517\\
13	0.70842338968157\\
14	-9.59536965310866\\
15	-9.39187381373574\\
16	-0.77017790312118\\
17	2.85274521432626\\
18	4.88914664413221\\
19	6.08559823291279\\
20	6.7251796075721\\
21	6.93858961233184\\
22	6.78815364052309\\
23	6.29655484057412\\
24	5.45562791330781\\
25	4.22341933117013\\
26	2.50693221000029\\
27	0.112480752780755\\
28	-3.41335338859477\\
29	-9.53966356377742\\
30	-306.535433840008\\
31	-10.0388200025844\\
32	-4.40669482343156\\
33	-1.36481142399765\\
34	0.561815274933053\\
35	1.83356741150692\\
36	2.65280737457408\\
37	3.12392034369935\\
38	3.30462214756318\\
39	3.22616729341109\\
40	2.90193627918114\\
41	2.33047670598722\\
42	1.49478672198848\\
43	0.357422844013968\\
44	-1.15165028016463\\
45	-3.16411413716031\\
46	-5.95234131587927\\
47	-10.2111047114391\\
48	-18.904150867825\\
49	-22.2248740103855\\
50	-11.6930082719974\\
51	-7.29238833774463\\
52	-4.58477828353398\\
53	-2.70742665473244\\
54	-1.33575672649692\\
55	-0.311512321746079\\
56	0.454703127218977\\
57	1.01848043510154\\
58	1.41680490443781\\
59	1.67553309026692\\
60	1.8134221370085\\
61	1.84445639886843\\
62	1.77926441192523\\
63	1.62601805093095\\
64	1.39102069100022\\
65	1.07909946733747\\
66	0.693868392510922\\
67	0.237902216341695\\
68	-0.287154816730492\\
69	-0.880530893417586\\
70	-1.54231128819777\\
71	-2.27346254540504\\
72	-3.07589698473237\\
73	-3.95258232001657\\
74	-4.90770586348344\\
75	-5.94690902519511\\
76	-7.07761688978798\\
77	-8.30950167099473\\
78	-9.65514153648229\\
79	-11.1309745912347\\
80	-12.7587152369273\\
81	-14.5675245303866\\
82	-16.5974684015033\\
83	-18.9053004887745\\
84	-21.5747353907698\\
85	-24.7361749999506\\
86	-28.608728528816\\
87	-33.6038507493081\\
88	-40.6460702490309\\
89	-52.686532796651\\
90	-304.523582462203\\
91	-52.686532796651\\
92	-40.6460702490309\\
93	-33.6038507493081\\
94	-28.608728528816\\
95	-24.7361749999506\\
96	-21.5747353907698\\
97	-18.9053004887745\\
98	-16.5974684015033\\
99	-14.5675245303866\\
100	-12.7587152369273\\
101	-11.1309745912347\\
102	-9.65514153648229\\
103	-8.30950167099473\\
104	-7.07761688978798\\
105	-5.94690902519511\\
106	-4.90770586348344\\
107	-3.95258232001657\\
108	-3.07589698473237\\
109	-2.27346254540504\\
110	-1.54231128819777\\
111	-0.880530893417586\\
112	-0.287154816730492\\
113	0.237902216341695\\
114	0.693868392510922\\
115	1.07909946733747\\
116	1.39102069100022\\
117	1.62601805093095\\
118	1.77926441192523\\
119	1.84445639886843\\
120	1.8134221370085\\
121	1.67553309026692\\
122	1.41680490443781\\
123	1.01848043510154\\
124	0.454703127218977\\
125	-0.311512321746079\\
126	-1.33575672649692\\
127	-2.70742665473244\\
128	-4.58477828353398\\
129	-7.29238833774463\\
130	-11.6930082719974\\
131	-22.2248740103855\\
132	-18.904150867825\\
133	-10.2111047114391\\
134	-5.95234131587927\\
135	-3.16411413716031\\
136	-1.15165028016463\\
137	0.357422844013968\\
138	1.49478672198848\\
139	2.33047670598722\\
140	2.90193627918114\\
141	3.22616729341109\\
142	3.30462214756318\\
143	3.12392034369935\\
144	2.65280737457408\\
145	1.83356741150692\\
146	0.561815274933053\\
147	-1.36481142399765\\
148	-4.40669482343156\\
149	-10.0388200025844\\
150	-306.535433840008\\
151	-9.53966356377742\\
152	-3.41335338859477\\
153	0.112480752780755\\
154	2.50693221000029\\
155	4.22341933117013\\
156	5.45562791330781\\
157	6.29655484057412\\
158	6.78815364052309\\
159	6.93858961233184\\
160	6.7251796075721\\
161	6.08559823291279\\
162	4.88914664413221\\
163	2.85274521432626\\
164	-0.77017790312118\\
165	-9.39187381373574\\
166	-9.59536965310866\\
167	0.70842338968157\\
168	5.60861072959517\\
169	8.88413488870634\\
170	11.3316298302124\\
171	13.2574818719154\\
172	14.8122949010681\\
173	16.0811060734463\\
174	17.1163353391544\\
175	17.9522021227026\\
176	18.6118444875687\\
177	19.1111454467818\\
178	19.4609109128497\\
179	19.6681469042629\\
180	19.7368004166453\\
};
\addlegendentry{$M = 64$}

\end{axis}
\end{tikzpicture}%

%% file: figures/params_table.tex
\begin{table}
  \centering 
  \begin{tabularx}{\columnwidth}{@{}lX@{}}
  \toprule
  Symbol & Meaning \\ \midrule
  $\Delta_f$ & Subcarrier spacing \\
  $T_{\rm slot}$ & Duration of a slot \\
  $T_{\rm symb}$ & Duration of a symbol\\
  $B$ & Bandwidth \\
  $D$ & Usage of frequency diversity \\
  $N_{rep}$ & Number of repetitions in frequency of an \gls{ss} block \\

  \midrule

  $P_{\rm MD}$ & Probability of misdetection \\
  $\Gamma$ & SNR threshold for the misdetection \\
  $\lambda_b$ & \gls{gnb} density \\

  \midrule

  $N_{\rm SS}$ & Number of \gls{ss} blocks per burst \\
  $L$ & Maximum number of \gls{ss} blocks per burst \\
  $D_{\max, \rm SS}$ & Maximum duration of an \gls{ss} burst \\
  $T_{\rm SS}$ & \gls{ss} burst periodicity \\
  $S_D$ & Number of \gls{ss} blocks for a complete sweep \\
  $T_{\rm IA}$ & Time required to perform \gls{ia} \\
  $T_{last}$ & Time to transmit the \gls{ss} blocks in the last (or only) burst \\
  $T_{\rm BR}$ & Time to perform beam reporting during \gls{ia} \\

  \midrule

  $N_{\rm CSI}$ & Number of \glspl{csirs} per \gls{ss} burst periodicity \\
  $T_{\rm CSI}$ & \gls{csirs} periodicity \\
  $T_{\rm CSI,slot}$ & \gls{csirs} periodicity in slot \\
  $O_{\rm CSI}$ & Time offset between the end of the \gls{ss} burst and the first \gls{csirs} \\
  $N_{\rm symb, CSI}$ & Number of \gls{ofdm} symbols for a \gls{csirs} \\
  $\rho$ & Portion of bandwidth $B$ for \glspl{csirs} \\
  $N_{\rm CSI, RX}$ & Number of directions that a \gls{ue} monitors \\
  $Z_{\rm CSI}$ & Number of \glspl{csirs} to be transmitted \\
  $T_{tot, \rm CSI}$ & Time available for the \gls{csirs} transmission between two \gls{ss} bursts\\
  $N_{\rm CSI}$ & Number of \gls{csirs} that can be transmitted between two bursts \\ 
  $T_{\rm tr}$ & Average time needed to receive the first \gls{csirs} \\
  $N_{\rm CSI, \perp}$ & Number of orthogonal \glspl{csirs} between two \gls{ss} bursts \\
  $N_{\max, \rm neigh}$ & Number of neighbors that can be supported with orthogonal \glspl{csirs} \\
  $T_{\rm RLF}$ & \gls{rlf} recovery delay \\

  \midrule

  $M$ & Number of antenna elements at the transceiver \\
  $\theta$ & Azimuth angle \\
  $\phi$ & Elevation angle \\
  $\Delta_{\theta}$ & Angular range for the azimuth \\
  $\Delta_{\phi}$ & Angular range for the elevation \\
  $N_{\theta}$ & Number of directions to cover in azimuth \\
  $N_{\phi}$ & Number of directions to cover in elevation \\
  $\Delta_{\rm beam}$ & Beamwidth at 3 dB \\
  $K_{\rm BF}$ & Number of beams that the transceiver can handle simultaneously \\
  $N_{\rm user}$ & Number of users \\

  \midrule

  $R_{\rm SS}$ & Time and frequency resources occupied by \gls{ss} blocks \\
  $\Omega_{\rm 5ms}$ & SS blocks overhead in 5 ms \\
  $\Omega_{T_{\rm SS}}$ & SS blocks overhead in $T_{\rm SS}$ \\
  $\Omega_{\rm CSI}$ & \gls{csirs} overhead in $T_{\rm SS}$ \\
  $\Omega_{tot}$ & Total overhead in $T_{\rm SS}$ \\

  \midrule
  $\mathcal{U}[a,b]$ & Uniform random variable in the interval $[a,b]$ \\
  \bottomrule
  \end{tabularx}
  \caption{Notation.}
  \label{table:symbols}
\end{table}

%% file: figures/Perr_N.tex
%
%
\usetikzlibrary{spy}
\definecolor{mycolor1}{rgb}{0.00000,0.44700,0.74100}%
\definecolor{mycolor2}{rgb}{0.85000,0.32500,0.09800}%
\definecolor{mycolor3}{rgb}{0.92900,0.69400,0.12500}%
\definecolor{mycolor4}{rgb}{0.49400,0.18400,0.55600}%
\definecolor{mycolor5}{rgb}{0.46600,0.67400,0.18800}%
\definecolor{mycolor6}{rgb}{0.30100,0.74500,0.93300}

\pgfplotsset{
tick label style={font=\scriptsize},
label style={font=\scriptsize},
legend  style={font=\scriptsize}
}

\begin{tikzpicture}[spy using outlines=
	{circle, magnification=2, connect spies}]
	
\begin{axis}[%
width=0.956\fwidth,
height=\fheight,
at={(0\fwidth,0\fheight)},
scale only axis,
xmin=10,
xmin=10,
xmax=60,
xlabel style={font=\color{white!15!black}},
xlabel={$\lambda_b$ $[\text{gNB/km}^\text{2}]$},
ylabel={Misdetection probability},
ymode=log,
ymin=0.001,
ymax=1,
xmajorgrids,
ymajorgrids,
xlabel shift=-5pt,
label style={font=\scriptsize},
legend columns=2,
legend style={font=\scriptsize, at={(0.5,1.05)},anchor=south,legend cell align=left, align=left, draw=white!15!black}
]
\addplot  [ color=mycolor1, mark=o, mark size=2.0pt, mark options={solid, mycolor1}]
  table[row sep=crcr]{%
10	0.41647\\
20	0.18026\\
30	0.07491\\
40	0.03094\\
50	0.01316\\
60	0.00515\\
};
\addlegendentry{$ M_{\rm gNB} = 4, M_{\rm UE} = 4$}

\addplot  [ color=mycolor2, densely dotted, mark=asterisk, mark size=2.0pt, mark options={solid, mycolor2}]
  table[row sep=crcr]{%
10	0.36243\\
20	0.13468\\
30	0.04889\\
40	0.01719\\
50	0.0064\\
60	0.00243\\
};
\addlegendentry{$ M_{\rm gNB} = 16, M_{\rm UE} = 4$}

\addplot [ color=mycolor3,mark=*, dashed, mark size=2.0pt, mark options={solid, mycolor3}]
  table[row sep=crcr]{%
10	0.33909\\
20	0.11887\\
30	0.04093\\
40	0.01379\\
50	0.00465\\
60	0.00145\\
};
\addlegendentry{$ M_{\rm gNB} = 64, M_{\rm UE} = 4$}

\addplot [color=mycolor5,dashdotted, mark=square, mark size=2.0pt, mark options={solid, mycolor5}]
  table[row sep=crcr]{%
10	0.32903\\
20	0.11071\\
30	0.0373\\
40	0.01281\\
50	0.00419\\
60	0.00132\\
};
\addlegendentry{$ M_{\rm gNB} = 64, M_{\rm UE} = 16$}

\addplot [color=mycolor6, mark=diamond, dashed, mark size=2.0pt, mark options={solid, mycolor6}]
  table[row sep=crcr]{%
10	0.36753\\
20	0.14157\\
30	0.05337\\
40	0.01975\\
50	0.00757\\
60	0.00265\\
};
\addlegendentry{$ M_{\rm gNB} = 64, M_{\rm UE} = 1 \text{ (omni)}$}

 \coordinate (spypoint) at (axis cs:40,0.02);
  \coordinate (magnifyglass) at (axis cs:22,0.008);
\end{axis}

\spy [blue, size=2cm] on (spypoint)
   in node[fill=white] at (magnifyglass);
\end{tikzpicture}%

%% file: figures/D_4x4.tex
%
%
\usetikzlibrary{spy}
\definecolor{mycolor1}{rgb}{0.00000,0.44700,0.74100}%
\definecolor{mycolor2}{rgb}{0.85000,0.32500,0.09800}%
\definecolor{mycolor3}{rgb}{0.92900,0.69400,0.12500}%
\definecolor{mycolor4}{rgb}{0.49400,0.18400,0.55600}%
\definecolor{mycolor5}{rgb}{0.46600,0.67400,0.18800}%
\definecolor{mycolor6}{rgb}{0.30100,0.74500,0.93300}

\pgfplotsset{
tick label style={font=\scriptsize},
label style={font=\scriptsize},
legend  style={font=\scriptsize}
}

\begin{tikzpicture}[spy using outlines=
	{circle, magnification=5, connect spies}]

\begin{axis}[%
width=0.956\fwidth,
height=\fheight,
at={(0\fwidth,0\fheight)},
scale only axis,
xmin=10,
xmax=60,
xlabel style={font=\color{white!15!black}},
xlabel={$\lambda_b$ $[\text{BS/km}^\text{2}]$},
ymode=log,
ymin=0.001,
ymax=1,
yminorticks=true,
xmajorgrids,
ymajorgrids,
ylabel style={font=\color{white!15!black}},
ylabel={Misdetection probability},
axis background/.style={fill=white},
legend columns=2,
xlabel shift=-5pt,
label style={font=\scriptsize},
legend style={font=\scriptsize, at={(0.5,1.05)},anchor=south,legend cell align=left, align=left, draw=white!15!black}
]

\addplot [color=black, mark=x,mark size=4pt, mark options={solid, fill=red, red}]
  table[row sep=crcr]{%
10	0.41647\\\\
};
\addlegendentry{ $\Delta_f=120$ kHz, $D=0$}

\addplot [color=black, mark=square*,mark size=2pt, mark options={solid, fill=black, black}]
  table[row sep=crcr]{%
10	0.42265\\
};
\addlegendentry{ $\Delta_f=120$ kHz, $D=1$}

\addplot [color=black, mark=o,mark size=2pt, mark options={solid, fill=black, black}]
  table[row sep=crcr]{%
10	0.44092\\
};
\addlegendentry{ $\Delta_f=240$ kHz, $D=1$}

\addplot [color=black, mark=triangle*,mark size=3pt, mark options={solid, fill=black, black}]
  table[row sep=crcr]{%
10	0.46164\\
};
\addlegendentry{ $\Delta_f=240$ kHz, $D=0$}

\addplot  [ forget plot, color=mycolor1, mark=square*, mark size=2.0pt, mark options={solid, mycolor1}]
  table[row sep=crcr]{%
10	0.3772\\
20	0.14623\\
30	0.05532\\
40	0.02132\\
50	0.00812\\
60	0.00319\\
};
\addlegendentry{4x4, Spacing = 120 kHz, freq. diversity}

\addplot[forget plot,dashed,color=mycolor1, mark=x, mark size=2.0pt, mark options={solid, red}]
  table[row sep=crcr]{%
10	0.41709\\
20	0.17705\\
30	0.07494\\
40	0.03177\\
50	0.01358\\
60	0.00569\\
};
\addlegendentry{4x4, Spacing = 120 kHz, NO freq. diversity}

\addplot [forget plot,color=mycolor1, mark=o, mark size=4.0pt, mark options={solid}]
  table[row sep=crcr]{%
10	0.41954\\
20	0.17947\\
30	0.07494\\
40	0.03228\\
50	0.01345\\
60	0.0056\\
};
\addlegendentry{4x4, Spacing = 240 kHz, freq. diversity}

\addplot [forget plot,color=mycolor1, mark=triangle*, mark size=2.0pt, mark options={solid, mycolor1}]
  table[row sep=crcr]{%
10	0.45907\\
20	0.21563\\
30	0.09824\\
40	0.04558\\
50	0.02133\\
60	0.00967\\
};
\addlegendentry{4x4, Spacing = 240 kHz, NO freq. diversity}

\addplot[ forget plot, color=mycolor3, mark=square*, mark size=2.0pt, mark options={solid, mycolor3}]
  table[row sep=crcr]{%
10	0.33103\\
20	0.11234\\
30	0.03773\\
40	0.01269\\
50	0.00406\\
60	0.00159\\
};
\addlegendentry{64x4, Spacing = 120 kHz, freq. diversity}

\addplot [forget plot,dashed,color=mycolor3, mark=x, mark size=2.0pt, mark options={solid, red}]
  table[row sep=crcr]{%
10	0.33887\\
20	0.11752\\
30	0.04032\\
40	0.01402\\
50	0.0046\\
60	0.00183\\
};
\addlegendentry{64x4, Spacing = 120 kHz, NO freq. diversity}

\addplot [forget plot,color=mycolor3, mark=o, mark size=4.0pt, mark options={solid}]
  table[row sep=crcr]{%
10	0.33715\\
20	0.11816\\
30	0.03913\\
40	0.01359\\
50	0.0047\\
60	0.00153\\
};
\addlegendentry{64x4, Spacing = 240 kHz, freq. diversity}

\addplot [forget plot,color=mycolor3, mark=triangle*, mark size=2.0pt, mark options={solid, mycolor3}]
  table[row sep=crcr]{%
10	0.34845\\
20	0.12596\\
30	0.04325\\
40	0.01556\\
50	0.00561\\
60	0.00189\\
};
\addlegendentry{64x4, Spacing = 240 kHz, NO freq. diversity}

\node[coordinate] (A) at (axis cs:35,0.049) {};                       
\node[coordinate,pin=above:{\footnotesize $M_{\rm gNB} = 4,  M_{\rm UE} = 4$}] at (axis cs:35,0.085){};  

\node[coordinate] (B) at (axis cs:45,0.008) {};                       
\node[coordinate,pin=left:{\footnotesize $M_{\rm gNB} = 64,  M_{\rm UE} = 4$}] at (axis cs:42,0.008){};  

\end{axis}
\draw[mycolor1] (A) ellipse (0.3 and 0.45);                                
\draw[mycolor3] (B) ellipse (0.3 and 0.2);                                
\end{tikzpicture}%

%% file: figures/nss_eNBAnalogUEAnalog.tex
%
%
\definecolor{mycolor1}{rgb}{0.00000,0.44700,0.74100}%
\definecolor{mycolor2}{rgb}{0.85000,0.32500,0.09800}%
\definecolor{mycolor3}{rgb}{0.92900,0.69400,0.12500}%
\definecolor{mycolor4}{rgb}{0.49400,0.18400,0.55600}%
\definecolor{mycolor5}{rgb}{0.46600,0.67400,0.18800}%
\definecolor{mycolor6}{rgb}{0.30100,0.74500,0.93300}%
\pgfplotsset{scaled y ticks=false}
\begin{tikzpicture}
\pgfplotsset{every tick label/.append style={font=\scriptsize}}

\begin{axis}[%
width=0.956\fwidth,
height=\fheight,
at={(0\fwidth,0\fheight)},
scale only axis,
xmin=7,
xmax=65,
xtick=data,
xlabel style={font=\scriptsize\color{white!15!black}},
xlabel={$N_{\rm SS}$},
ymin=0,
ymax=5300,
ylabel style={font=\scriptsize\color{white!15!black}},
ylabel={$T_{\rm IA}$ [ms]},
axis background/.style={fill=white},
xmajorgrids,
ymajorgrids,
xlabel shift=-5pt,
legend columns=2,
legend style={at={(0.5, 1.05)},anchor=south,font=\scriptsize,legend cell align=left, align=left, draw=white!15!black}
]
\addplot [color=mycolor1, line width=1.2pt, mark=o, mark options={solid, mycolor1}]
  table[row sep=crcr]{%
8	20.11608125\\
16	0.36608125\\
32	0.36608125\\
64	0.36608125\\
};
\addlegendentry{$ M_{\rm gNB} = 4, M_{\rm UE} = 4$}

\addplot [color=mycolor2,  densely dotted, line width=1.2pt, mark=asterisk, mark options={solid, mycolor2}]
  table[row sep=crcr]{%
8	220.05358125\\
16	100.30358125\\
32	40.80358125\\
64	20.80358125\\
};
\addlegendentry{$ M_{\rm gNB} = 16, M_{\rm UE} = 4$}

\addplot [color=mycolor3, dashed, line width=1.2pt, mark=*, mark options={solid, mycolor3}]
  table[row sep=crcr]{%
8	740.11608125\\
16	360.36608125\\
32	180.36608125\\
64	81.36608125\\
};
\addlegendentry{$ M_{\rm gNB} = 64, M_{\rm UE} = 4$}

\addplot [color=mycolor4, dotted, line width=1.2pt, mark=x, mark options={solid, mycolor4}]
  table[row sep=crcr]{%
8	1560.17858125\\
16	780.17858125\\
32	380.67858125\\
64	181.67858125\\
};
\addlegendentry{$ M_{\rm gNB} = 16, M_{\rm UE} = 16$}

\addplot [color=mycolor5, dashdotted, line width=1.2pt, mark=square, mark options={solid, mycolor5}]
  table[row sep=crcr]{%
8	5240.11608125\\
16	2620.11608125\\
32	1300.61608125\\
64	641.61608125\\
};
\addlegendentry{$ M_{\rm gNB} = 64, M_{\rm UE} = 16$}

\addplot [color=mycolor6, dashed, line width=1.2pt, mark=diamond, mark options={solid, mycolor6}]
  table[row sep=crcr]{%
8	120.05358125\\
16	60.05358125\\
32	20.55358125\\
64	1.55358125\\
};
\addlegendentry{$ M_{\rm gNB} = 64, M_{\rm UE} = 1$ $(\mbox{omni})$}

\end{axis}
\end{tikzpicture}%

%% file: figures/nss_eNBAnalogUEHybrid.tex
%
%
\definecolor{mycolor1}{rgb}{0.00000,0.44700,0.74100}%
\definecolor{mycolor2}{rgb}{0.85000,0.32500,0.09800}%
\definecolor{mycolor3}{rgb}{0.92900,0.69400,0.12500}%
\definecolor{mycolor4}{rgb}{0.49400,0.18400,0.55600}%
\definecolor{mycolor5}{rgb}{0.46600,0.67400,0.18800}%
\definecolor{mycolor6}{rgb}{0.30100,0.74500,0.93300}%
\pgfplotsset{scaled y ticks=false}
\begin{tikzpicture}
\pgfplotsset{every tick label/.append style={font=\scriptsize}}
\begin{axis}[%
width=0.956\fwidth,
height=\fheight,
at={(0\fwidth,0\fheight)},
scale only axis,
xmin=7,
xmax=65,
xtick=data,
xlabel style={font=\scriptsize\color{white!15!black}},
xlabel={$N_{\rm SS}$},
ymin=0,
ymax=750,
ylabel style={font=\scriptsize\color{white!15!black}},
ylabel={$T_{\rm IA}$ [ms]},
axis background/.style={fill=white},
xlabel shift=-5pt,
xmajorgrids,
ymajorgrids,
legend style={font=\scriptsize,legend cell align=left, align=left, draw=white!15!black}
]
\addplot [color=mycolor1, line width=1.2pt, mark=o, mark options={solid, mycolor1}]
  table[row sep=crcr]{%
8	0.17858125\\
16	0.17858125\\
32	0.17858125\\
64	0.17858125\\
};
\addlegendentry{$ M_{\rm gNB} = 4, M_{\rm UE} = 4$}

\addplot [color=mycolor2, densely dotted, line width=1.2pt, mark=asterisk, mark options={solid, mycolor2}]
  table[row sep=crcr]{%
8	100.15175625\\
16	40.40175625\\
32	20.40175625\\
64	1.40175625\\
};
\addlegendentry{$ M_{\rm gNB} = 16, M_{\rm UE} = 4$}

\addplot [color=mycolor3, dashed, line width=1.2pt, mark=*, mark options={solid, mycolor3}]
  table[row sep=crcr]{%
8	360.17858125\\
16	180.17858125\\
32	80.67858125\\
64	40.67858125\\
};
\addlegendentry{$ M_{\rm gNB} = 64, M_{\rm UE} = 4$}

\addplot [color=mycolor4, dotted, line width=1.2pt, mark=x, mark options={solid, mycolor4}]
  table[row sep=crcr]{%
8	220.05358125\\
16	100.30358125\\
32	40.80358125\\
64	20.80358125\\
};
\addlegendentry{$ M_{\rm gNB} = 16, M_{\rm UE} = 16$}

\addplot [color=mycolor5, dashdotted, line width=1.2pt, mark=square, mark options={solid, mycolor5}]
  table[row sep=crcr]{%
8	740.11608125\\
16	360.36608125\\
32	180.36608125\\
64	81.36608125\\
};
\addlegendentry{$ M_{\rm gNB} = 64, M_{\rm UE} = 16$}

\addplot [color=mycolor6, dashed, line width=1.2pt, mark=diamond, mark options={solid, mycolor6}]
  table[row sep=crcr]{%
8	120.05358125\\
16	60.05358125\\
32	20.55358125\\
64	1.55358125\\
};
\addlegendentry{$ M_{\rm gNB} = 64, M_{\rm UE} = 1$ $(\mbox{omni})$}

\legend{};

\end{axis}
\end{tikzpicture}%

%% file: figures/nss_eNBAnalogUEDigital.tex
%
%
\definecolor{mycolor1}{rgb}{0.00000,0.44700,0.74100}%
\definecolor{mycolor2}{rgb}{0.85000,0.32500,0.09800}%
\definecolor{mycolor3}{rgb}{0.92900,0.69400,0.12500}%
\definecolor{mycolor4}{rgb}{0.49400,0.18400,0.55600}%
\definecolor{mycolor5}{rgb}{0.46600,0.67400,0.18800}%
\definecolor{mycolor6}{rgb}{0.30100,0.74500,0.93300}%
\pgfplotsset{scaled y ticks=false}
\begin{tikzpicture}
\pgfplotsset{every tick label/.append style={font=\scriptsize}}

\begin{axis}[%
width=0.956\fwidth,
height=\fheight,
at={(0\fwidth,0\fheight)},
scale only axis,
xmin=7,
xmax=65,
xtick=data,
xlabel style={font=\scriptsize\color{white!15!black}},
xlabel={$N_{\rm SS}$},
ymin=0,
ymax=365,
ylabel style={font=\scriptsize\color{white!15!black}},
ylabel={$T_{\rm IA}$ [ms]},
axis background/.style={fill=white},
xmajorgrids,
ymajorgrids,
xlabel shift=-5pt,
legend style={font=\scriptsize,legend cell align=left, align=left, draw=white!15!black,at={(0.99,0.6)},anchor=south east}
]
\addplot [color=mycolor1, line width=1.2pt, mark=o, mark options={solid, mycolor1}]
  table[row sep=crcr]{%
8 0.11608125\\
16  0.11608125\\
32  0.11608125\\
64  0.11608125\\
};
\addlegendentry{$ M_{\rm gNB} = 4, M_{\rm UE} = 4$}

\addplot [color=mycolor2, densely dotted, line width=1.2pt, mark=asterisk, mark options={solid, mycolor2}]
  table[row sep=crcr]{%
8 60.17858125\\
16  20.42858125\\
32  0.92858125\\
64  0.92858125\\
};
\addlegendentry{$ M_{\rm gNB} = 16, M_{\rm UE} = 4$}

\addplot [color=mycolor3, dashed, line width=1.2pt, mark=*, mark options={solid, mycolor3}]
  table[row sep=crcr]{%
8 240.11608125\\
16  120.11608125\\
32  60.11608125\\
64  21.11608125\\
};
\addlegendentry{$ M_{\rm gNB} = 64, M_{\rm UE} = 4$}

\addplot [color=mycolor4, dotted, line width=1.2pt, mark=x, mark options={solid, mycolor4}]
  table[row sep=crcr]{%
8 100.15175625\\
16  40.40175625\\
32  20.40175625\\
64  1.40175625\\
};
\addlegendentry{$ M_{\rm gNB} = 16, M_{\rm UE} = 16$}

\addplot [color=mycolor5, dashdotted, line width=1.2pt, mark=square, mark options={solid, mycolor5}]
  table[row sep=crcr]{%
8 360.17858125\\
16  180.17858125\\
32  80.67858125\\
64  40.67858125\\
};
\addlegendentry{$ M_{\rm gNB} = 64, M_{\rm UE} = 16$}

\addplot [color=mycolor6, dashed, line width=1.2pt, mark=diamond, mark options={solid, mycolor6}]
  table[row sep=crcr]{%
8 120.05358125\\
16  60.05358125\\
32  20.55358125\\
64  1.55358125\\
};
\addlegendentry{$ M_{\rm gNB} = 64, M_{\rm UE} = 1$ $(\mbox{omni})$}

\coordinate (pt) at (axis cs:32,0);

\legend{};

\end{axis}

\node[pin=3:{%
    \begin{tikzpicture}[trim axis left,trim axis right]
    \pgfplotsset{every tick label/.append style={font=\tiny}}
    \begin{axis}[
    scale only axis,
bar shift auto,
      xmin=30,xmax=34,
      ymin=0,ymax=1.2,
width=0.2\fwidth,
height=0.4\fheight,
axis x line*=bottom,
axis y line*=left, 
xtick=data,
axis background/.style={fill=white},
    ]
\addplot [color=mycolor1, line width=1.2pt, mark=o, mark options={solid, mycolor1}]
  table[row sep=crcr]{%
8 0.11608125\\
16  0.11608125\\
32  0.11608125\\
64  0.11608125\\
};

\addplot [color=mycolor2, densely dotted, line width=1.2pt, mark=asterisk, mark options={solid, mycolor2}]
  table[row sep=crcr]{%
8 60.17858125\\
16  20.42858125\\
32  0.92858125\\
64  0.92858125\\
};

\addplot [color=mycolor3, dashed, line width=1.2pt, mark=*, mark options={solid, mycolor3}]
  table[row sep=crcr]{%
8 240.11608125\\
16  120.11608125\\
32  60.11608125\\
64  21.11608125\\
};

\addplot [color=mycolor4, dotted, line width=1.2pt, mark=x, mark options={solid, mycolor4}]
  table[row sep=crcr]{%
8 100.15175625\\
16  40.40175625\\
32  20.40175625\\
64  1.40175625\\
};

\addplot [color=mycolor5, dashdotted, line width=1.2pt, mark=square, mark options={solid, mycolor5}]
  table[row sep=crcr]{%
8 360.17858125\\
16  180.17858125\\
32  80.67858125\\
64  40.67858125\\
};

\addplot [color=mycolor6, dashed, line width=1.2pt, mark=diamond, mark options={solid, mycolor6}]
  table[row sep=crcr]{%
8 120.05358125\\
16  60.05358125\\
32  20.55358125\\
64  1.55358125\\
};

    \end{axis}
    \end{tikzpicture}%
}] at (pt) {};

\end{tikzpicture}%

%% file: figures/nss_eNBDigitalUEAnalog.tex
%
%
\definecolor{mycolor1}{rgb}{0.00000,0.44700,0.74100}%
\definecolor{mycolor2}{rgb}{0.85000,0.32500,0.09800}%
\definecolor{mycolor3}{rgb}{0.92900,0.69400,0.12500}%
\definecolor{mycolor4}{rgb}{0.49400,0.18400,0.55600}%
\definecolor{mycolor5}{rgb}{0.46600,0.67400,0.18800}%
\definecolor{mycolor6}{rgb}{0.30100,0.74500,0.93300}%
\pgfplotsset{scaled y ticks=false}
\begin{tikzpicture}
\pgfplotsset{every tick label/.append style={font=\scriptsize}}

\begin{axis}[%
width=0.956\fwidth,
height=\fheight,
at={(0\fwidth,0\fheight)},
scale only axis,
xmin=7,
xmax=65,
xtick=data,
xlabel style={font=\scriptsize\color{white!15!black}},
xlabel={$N_{\rm SS}$},
ymin=0,
ymax=110,
ylabel style={font=\scriptsize\color{white!15!black}},
ylabel={$T_{\rm IA}$ [ms]},
axis background/.style={fill=white},
xmajorgrids,
ymajorgrids,
xlabel shift=-5pt,
legend style={font=\scriptsize,legend cell align=left, align=left, draw=white!15!black, at={(0.99,0.6)},anchor=south east}
]
\addplot [color=mycolor1, line width=1.2pt, mark=o, mark options={solid, mycolor1}]
  table[row sep=crcr]{%
8 0.17858125\\
16  0.17858125\\
32  0.17858125\\
64  0.17858125\\
};
\addlegendentry{$ M_{\rm gNB} = 4, M_{\rm UE} = 4$}

\addplot [color=mycolor2, densely dotted, line width=1.2pt, mark=asterisk, mark options={solid, mycolor2}]
  table[row sep=crcr]{%
8 0.17858125\\
16  0.17858125\\
32  0.17858125\\
64  0.17858125\\
};
\addlegendentry{$ M_{\rm gNB} = 16, M_{\rm UE} = 4$}

\addplot [color=mycolor3, dashed, line width=1.2pt, mark=*, mark options={solid, mycolor3}]
  table[row sep=crcr]{%
8 0.17858125\\
16  0.17858125\\
32  0.17858125\\
64  0.17858125\\
};
\addlegendentry{$ M_{\rm gNB} = 64, M_{\rm UE} = 4$}

\addplot [color=mycolor4, dotted, line width=1.2pt, mark=x, mark options={solid, mycolor4}]
  table[row sep=crcr]{%
8 100.05358125\\
16  40.30358125\\
32  20.30358125\\
64  1.30358125\\
};
\addlegendentry{$ M_{\rm gNB} = 16, M_{\rm UE} = 16$}

\addplot [color=mycolor5, dashdotted, line width=1.2pt, mark=square, mark options={solid, mycolor5}]
  table[row sep=crcr]{%
8 100.05358125\\
16  40.30358125\\
32  20.30358125\\
64  1.30358125\\
};
\addlegendentry{$ M_{\rm gNB} = 64, M_{\rm UE} = 16$}

\addplot [color=mycolor6, dashed, line width=1.2pt, mark=diamond, mark options={solid, mycolor6}]
  table[row sep=crcr]{%
8 0.02675625\\
16  0.02675625\\
32  0.02675625\\
64  0.02675625\\
};
\addlegendentry{$ M_{\rm gNB} = 64, M_{\rm UE} = 1$ $(\mbox{omni})$}

\coordinate (pt) at (axis cs:32,0);

\legend{};

\end{axis}

\node[pin=3:{%
    \begin{tikzpicture}[trim axis left,trim axis right]
    \pgfplotsset{every tick label/.append style={font=\tiny}}
    \begin{axis}[
    scale only axis,
bar shift auto,
      xmin=30,xmax=34,
      ymin=0,ymax=1.2,
width=0.2\fwidth,
height=0.4\fheight,
axis x line*=bottom,
axis y line*=left, 
xtick=data,
axis background/.style={fill=white},
    ]
\addplot [color=mycolor1, line width=1.2pt, mark=o, mark options={solid, mycolor1}, forget plot]
  table[row sep=crcr]{%
8	0.17858125\\
16	0.17858125\\
32	0.17858125\\
64	0.17858125\\
};

\addplot [color=mycolor2, densely dotted, line width=1.2pt, mark=asterisk, mark options={solid, mycolor2}, forget plot]
  table[row sep=crcr]{%
8	0.17858125\\
16	0.17858125\\
32	0.17858125\\
64	0.17858125\\
};

\addplot [color=mycolor3, dashed, line width=1.2pt, mark=*, mark options={solid, mycolor3}, forget plot]
  table[row sep=crcr]{%
8	0.17858125\\
16	0.17858125\\
32	0.17858125\\
64	0.17858125\\
};

\addplot [color=mycolor4, dotted,  line width=1.2pt, mark=x, mark options={solid, mycolor4}, forget plot]
 table[row sep=crcr]{%
8 100.05358125\\
16  40.30358125\\
32  20.30358125\\
64  1.30358125\\
};

\addplot [color=mycolor5, dashdotted, line width=1.2pt, mark=square, mark options={solid, mycolor5}, forget plot]
  table[row sep=crcr]{%
8 100.05358125\\
16  40.30358125\\
32  20.30358125\\
64  1.30358125\\
};

\addplot [color=mycolor6, dashed, line width=1.2pt, mark=diamond, mark options={solid, mycolor6}, forget plot]
  table[row sep=crcr]{%
8	0.02675625\\
16	0.02675625\\
32	0.02675625\\
64	0.02675625\\
};

    \end{axis}
    \end{tikzpicture}%
}] at (pt) {};

\end{tikzpicture}%

%% file: figures/tss_eNBAnalogUEHyb_64.tex
%
%
\definecolor{mycolor1}{rgb}{0.00000,0.44700,0.74100}%
\definecolor{mycolor2}{rgb}{0.85000,0.32500,0.09800}%
\definecolor{mycolor3}{rgb}{0.92900,0.69400,0.12500}%
\definecolor{mycolor4}{rgb}{0.49400,0.18400,0.55600}%
\definecolor{mycolor5}{rgb}{0.46600,0.67400,0.18800}%
\definecolor{mycolor6}{rgb}{0.30100,0.74500,0.93300}%
\pgfplotsset{scaled y ticks=false}
\begin{tikzpicture}
\pgfplotsset{every tick label/.append style={font=\scriptsize}}
\begin{axis}[%
width=0.956\fwidth,
height=\fheight,
at={(0\fwidth,0\fheight)},
scale only axis,
xmin=4,
xmax=161,
xlabel style={font=\scriptsize\color{white!15!black}},
xlabel={$T_{\rm SS}$ [ms]},
ymin=0,
ymax=650,
ylabel style={font=\scriptsize\color{white!15!black}},
ylabel={$T_{\rm IA}$ [ms]},
axis background/.style={fill=white},
xmajorgrids,
ymajorgrids,
legend style={font=\scriptsize,at={(0.01, 0.55)},anchor=south west,legend cell align=left, align=left, draw=white!15!black}
]
\addplot [color=mycolor1, line width=1.2pt, mark=o, mark options={solid, mycolor1}]
  table[row sep=crcr]{%
5	0.17858125\\
10	0.17858125\\
20	0.17858125\\
40	0.17858125\\
80	0.17858125\\
160	0.17858125\\
};
\addlegendentry{$ M_{\rm gNB} = 4, M_{\rm UE} = 4$}

\addplot [color=mycolor2, densely dotted, line width=1.2pt, mark=asterisk, mark options={solid, mycolor2}]
  table[row sep=crcr]{%
5	1.40175625\\
10	1.40175625\\
20	1.40175625\\
40	1.40175625\\
80	1.40175625\\
160	1.40175625\\
};
\addlegendentry{$ M_{\rm gNB} = 16, M_{\rm UE} = 4$}

\addplot [color=mycolor3, dashed, line width=1.2pt, mark=*, mark options={solid, mycolor3}]
  table[row sep=crcr]{%
5	10.67858125\\
10	20.67858125\\
20	40.67858125\\
40	80.67858125\\
80	160.67858125\\
160	320.67858125\\
};
\addlegendentry{$ M_{\rm gNB} = 64, M_{\rm UE} = 4$}

\addplot [color=mycolor4, dotted, line width=1.2pt, mark=x, mark options={solid, mycolor4}]
  table[row sep=crcr]{%
5	5.80358125\\
10	10.80358125\\
20	20.80358125\\
40	40.80358125\\
80	80.80358125\\
160	160.80358125\\
};
\addlegendentry{$ M_{\rm gNB} = 16, M_{\rm UE} = 16$}

\addplot [color=mycolor5, dashdotted, line width=1.2pt, mark=square, mark options={solid, mycolor5}]
  table[row sep=crcr]{%
5	21.36608125\\
10	41.36608125\\
20	81.36608125\\
40	161.36608125\\
80	321.36608125\\
160	641.36608125\\
};
\addlegendentry{$ M_{\rm gNB} = 64, M_{\rm UE} = 16$}

\addplot [color=mycolor6, dashed, line width=1.2pt, mark=diamond, mark options={solid, mycolor6}]
  table[row sep=crcr]{%
5	1.55358125\\
10	1.55358125\\
20	1.55358125\\
40	1.55358125\\
80	1.55358125\\
160	1.55358125\\
};
\addlegendentry{$ M_{\rm gNB} = 64, M_{\rm UE} = 1$  $(\mbox{omni})$}

\end{axis}
\end{tikzpicture}%

%% file: figures/nss_eNBAnalogUEAnalog_deltaf.tex
%
%
\definecolor{mycolor1}{rgb}{0.24220,0.15040,0.66030}%
\definecolor{mycolor2}{rgb}{0.97690,0.98390,0.08050}%
\begin{tikzpicture}
\pgfplotsset{every tick label/.append style={font=\scriptsize}}

\begin{axis}[%
width=0.956\fwidth,
height=\fheight,
at={(0\fwidth,0\fheight)},
scale only axis,
bar shift auto,
xmin=0.6,
xmax=2.4,
xtick={1, 2},
xticklabels={{$M_{\gls{gnb}} = 4, M_{\gls{ue}} = 4$}, {$M_{\gls{gnb}} = 64, M_{\gls{ue}} = 1$}},
xlabel style={font=\scriptsize\color{white!15!black}},
xlabel={Antenna configuration},
xmajorgrids,
ymajorgrids,
yminorgrids,
ymin=0,
ymax=3.5,
ylabel style={font=\scriptsize\color{white!15!black}},
ylabel={$T_{\rm IA}$[ms]},
axis background/.style={fill=white},
legend style={font=\scriptsize,at={(0.01, 0.99)},anchor=north west,legend cell align=left, align=left, draw=white!15!black}
]
\addplot[ybar, bar width=0.229, fill=green, draw=black, area legend] table[row sep=crcr] {%
1	0.7321625\\
2	3.1071625\\
};
\addlegendentry{$\Delta_f = 120$~kHz}

\node[anchor=south, align=center] at (2, 3.1) (50dir) {\tiny 50 directions to sweep.};
\node[anchor=south, align=center] at (1, 0.69) (12dir) {\tiny 12 directions to sweep.};

\addplot[forget plot, color=white!15!black] table[row sep=crcr] {%
0.6	0\\
2.4	0\\
};
\addplot[ybar, bar width=0.229, postaction={pattern=crosshatch dots}, fill=red, draw=black, area legend] table[row sep=crcr] {%
1	0.36608125\\
2	1.55358125\\
};
\addlegendentry{$\Delta_f = 240$~kHz}

\addplot[forget plot, color=white!15!black] table[row sep=crcr] {%
0.6	0\\
2.4	0\\
};
\end{axis}
\end{tikzpicture}%

%% file: figures/t_tr_Tss20ms_64antennas.tex
%
%
\begin{tikzpicture}
\pgfplotsset{every tick label/.append style={font=\scriptsize}}

\begin{axis}[%
width=0.642\fwidth,
height=\fheight,
at={(0\fwidth,0\fheight)},
scale only axis,
xmin=5,
xmax=20,
xtick=data,
tick align=outside,
xlabel style={font=\scriptsize\color{white!15!black}},
xlabel={$N_{\rm user}$},
ymin=1,
ymax=4,
ytick=data,
ylabel style={font=\scriptsize\color{white!15!black}},
ylabel={$N_{\rm CSI, RX}$},
zmin=0,
zmax=510,
ztick = {0, 100, 200, 300, 400, 500},
zlabel style={font=\scriptsize\color{white!15!black}},
zlabel={$T_{tr}$ [ms]},
view={-129}{25},
axis background/.style={fill=white},
axis x line*=bottom,
axis y line*=left,
axis z line*=left,
xmajorgrids,
ymajorgrids,
zmajorgrids,
reverse legend,
legend columns=2,
legend style={font=\scriptsize, at={(0.5,0.9)}, anchor=south, legend cell align=left, align=left, draw=white!15!black}
]

\addplot3[%
surf,
fill opacity=0.2, shader=flat corner, color=blue, draw=black, z buffer=sort, mesh/rows=3]
table[row sep=crcr, point meta=\thisrow{c}] {%
x	y	z	c\\
5	1	1.5625	3\\
5	2	3.125	3\\
5	3	4.6875	3\\
5	4	6.25	3\\
10	1	3.125	3\\
10	2	6.25	3\\
10	3	10.375	3\\
10	4	14.5	3\\
20	1	6.25	3\\
20	2	14.5	3\\
20	3	18.425	3\\
20	4	18.425	3\\
};
\addlegendentry{Opt. 2, $T_{\rm CSI} = 0.625$~ms}

\addplot3[%
surf,
fill opacity=0.1, shader=flat corner, color=red, draw=black, z buffer=sort, colormap/redyellow, mesh/rows=3]
table[row sep=crcr, point meta=\thisrow{c}] {%
x	y	z	c\\
5	1	1.875	1\\
5	2	3.4375	1\\
5	3	5	1\\
5	4	6.5625	1\\
10	1	3.4375	1\\
10	2	6.5625	1\\
10	3	10.6875	1\\
10	4	14.8125	1\\
20	1	6.5625	1\\
20	2	14.8125	1\\
20	3	18.7375	1\\
20	4	18.7375	1\\
};
\addlegendentry{Opt. 1, $T_{\rm CSI} = 0.625$~ms}

\addplot3[%
surf,
fill opacity=0.8, shader=flat corner, color=yellow, draw=black, z buffer=sort, mesh/rows=3]
table[row sep=crcr, point meta=\thisrow{c}] {%
x	y	z	c\\
5	1	22.5	7\\
5	2	47.5	7\\
5	3	72.5	7\\
5	4	97.5	7\\
10	1	47.5	7\\
10	2	97.5	7\\
10	3	147.5	7\\
10	4	197.5	7\\
20	1	97.5	7\\
20	2	197.5	7\\
20	3	247.5	7\\
20	4	247.5	7\\
};
\addlegendentry{Opt. 2, $T_{\rm CSI} = 10$~ms}

\addplot3[%
surf,
fill opacity=0.6, shader=flat corner, color=green, draw=black, z buffer=sort, mesh/rows=3]
table[row sep=crcr, point meta=\thisrow{c}] {%
x	y	z	c\\
5	1	50	5\\
5	2	100	5\\
5	3	150	5\\
5	4	200	5\\
10	1	100	5\\
10	2	200	5\\
10	3	300	5\\
10	4	400	5\\
20	1	200	5\\
20	2	400	5\\
20	3	500	5\\
20	4	500	5\\
};
\addlegendentry{Opt. 1, $T_{\rm CSI} = 10$~ms}

\end{axis}
\end{tikzpicture}%

%% file: figures/t_tr_64antennas_changeTss.tex
%
%
\begin{tikzpicture}
\pgfplotsset{every tick label/.append style={font=\scriptsize}}

\begin{axis}[%
width=0.651\fwidth,
height=\fheight,
at={(0\fwidth,0\fheight)},
scale only axis,
xmin=5,
xmax=20,
xtick=data,
tick align=outside,
xlabel style={font=\scriptsize\color{white!15!black}},
xlabel={$N_{\rm user}$},
ymin=1,
ymax=4,
ytick=data,
ylabel style={font=\scriptsize\color{white!15!black}},
ylabel={$N_{\rm CSI, RX}$},
zmin=0,
zmax=35,
ztick = {0, 20, 40, 60},
zlabel style={font=\scriptsize\color{white!15!black}},
zlabel={$T_{tr}$ [ms]},
view={-129}{25},
axis background/.style={fill=white},
axis x line*=bottom,
axis y line*=left,
axis z line*=left,
xmajorgrids,
ymajorgrids,
zmajorgrids,
reverse legend,
legend style={font=\scriptsize, at={(0.7,0.7)}, anchor=south, legend cell align=left, align=left, draw=white!15!black}
]


\addplot3[%
surf,
fill opacity=0.2, shader=flat corner, color=red, draw=black, z buffer=sort, mesh/rows=3]
table[row sep=crcr, point meta=\thisrow{c}] {%
x	y	z	c\\
5	1	1.875	5\\
5	2	3.4375	5\\
5	3	5	5\\
5	4	6.5625	5\\
10	1	3.4375	5\\
10	2	6.5625	5\\
10	3	9.6875	5\\
10	4	12.8125	5\\
20	1	6.5625	5\\
20	2	12.8125	5\\
20	3	15.9375	5\\
20	4	15.9375	5\\
};
\addlegendentry{Opt. 1, $T_{\rm SS} = 40$~ms}


\addplot3[%
surf,
fill opacity=0.8, shader=flat corner, color=green, draw=black, z buffer=sort, mesh/rows=3]
table[row sep=crcr, point meta=\thisrow{c}] {%
x	y	z	c\\
5	1	1.875	1\\
5	2	4.4375	1\\
5	3	7.33333333333333	1\\
5	4	10.5625	1\\
10	1	4.4375	1\\
10	2	10.5625	1\\
10	3	16.6875	1\\
10	4	22.8125	1\\
20	1	10.5625	1\\
20	2	22.8125	1\\
20	3	29.1375	1\\
20	4	29.1375	1\\
};
\addlegendentry{Opt. 1, $T_{\rm SS} = 10$~ms}

\end{axis}
\end{tikzpicture}%

%% file: figures/numCsiTss.tex
%
%
\definecolor{mycolor1}{rgb}{0.00000,0.44700,0.74100}%
\definecolor{mycolor2}{rgb}{0.85000,0.32500,0.09800}%
\definecolor{mycolor3}{rgb}{0.92900,0.69400,0.12500}%
\definecolor{mycolor4}{rgb}{0.49400,0.18400,0.55600}%
\definecolor{mycolor5}{rgb}{0.46600,0.67400,0.18800}%
\definecolor{mycolor6}{rgb}{0.30100,0.74500,0.93300}%
\definecolor{mycolor7}{rgb}{0.63500,0.07800,0.18400}%
\begin{tikzpicture}
\pgfplotsset{every tick label/.append style={font=\scriptsize}}

\begin{axis}[%
width=0.956\fwidth,
height=\fheight,
at={(0\fwidth,0\fheight)},
scale only axis,
xmin=0,
xmax=162,
xlabel style={font=\scriptsize\color{white!15!black}},
xlabel={$T_{\rm SS}$~[ms]},
ymin=0,
ymax=160,
ylabel style={font=\scriptsize\color{white!15!black}},
ylabel={$N_{\rm CSI}$},
axis background/.style={fill=white},
xmajorgrids,
ymajorgrids,
legend columns=2,
legend style={font=\scriptsize, at={(0.5, 1.03)}, anchor=south, legend cell align=left, align=left, draw=white!15!black}
]
\addplot [color=mycolor1, mark=+, mark options={solid, mycolor1}]
  table[row sep=crcr]{%
5 0\\
10  4\\
20  12\\
40  28\\
80  60\\
160 124\\
};
\addlegendentry{Opt. 1, $T_{\rm CSI}=1.25$~ms}

\addplot [color=mycolor2, mark=o, mark options={solid, mycolor2}]
  table[row sep=crcr]{%
5 0\\
10  4\\
20  12\\
40  28\\
80  60\\
160 124\\
};
\addlegendentry{Opt. 2, $T_{\rm CSI}=1.25$~ms}

\addplot [color=mycolor3, mark=asterisk, mark options={solid, mycolor3}]
  table[row sep=crcr]{%
5 0\\
10  2\\
20  6\\
40  14\\
80  30\\
160 62\\
};
\addlegendentry{Opt. 1, $T_{\rm CSI}=2.5$~ms}

\addplot [color=mycolor4, mark=*, mark options={solid, mycolor4}]
  table[row sep=crcr]{%
5 0\\
10  2\\
20  6\\
40  14\\
80  30\\
160 62\\
};
\addlegendentry{Opt. 2, $T_{\rm CSI}=2.5$~ms}

\addplot [color=mycolor5, mark=x, mark options={solid, mycolor5}]
  table[row sep=crcr]{%
5	0\\
10	1\\
20	3\\
40	7\\
80	15\\
160	31\\
};
\addlegendentry{Opt. 1, $T_{\rm CSI}=5$~ms}

\addplot [color=mycolor6, mark=square, mark options={solid, mycolor6}]
  table[row sep=crcr]{%
5	0\\
10	1\\
20	3\\
40	7\\
80	15\\
160	31\\
};
\addlegendentry{Opt. 2, $T_{\rm CSI}=5$~ms}

\addplot [color=mycolor7, mark=diamond, mark options={solid, mycolor7}]
  table[row sep=crcr]{%
5	0\\
10	0\\
20	1\\
40	3\\
80	7\\
160	15\\
};
\addlegendentry{Opt. 1, $T_{\rm CSI}=10$~ms}

\addplot [color=mycolor1, mark=triangle, mark options={solid, mycolor1}]
  table[row sep=crcr]{%
5	0\\
10	1\\
20	2\\
40	4\\
80	8\\
160	16\\
};
\addlegendentry{Opt. 2, $T_{\rm CSI}=10$~ms}

\coordinate (pt) at (axis cs:10,10);

\end{axis}

\node[pin=65:{%
    \begin{tikzpicture}[trim axis left,trim axis right]
    \pgfplotsset{every tick label/.append style={font=\tiny}}
    \begin{axis}[
    scale only axis,
bar shift auto,
      xmin=1,xmax=20,
      ymin=0,ymax=20,
width=0.25\fwidth,
height=0.45\fheight,
axis x line*=bottom,
axis y line*=left, 
xtick=data,
axis background/.style={fill=white},
    ]
\addplot [color=mycolor1, mark=+, mark options={solid, mycolor1}, forget plot]
  table[row sep=crcr]{%
5	0\\
10	5\\
20	15\\
40	35\\
80	75\\
160	155\\
};

\addplot [color=mycolor2, mark=o, mark options={solid, mycolor2}, forget plot]
  table[row sep=crcr]{%
5	0\\
10	5\\
20	15\\
40	35\\
80	75\\
160	155\\
};

\addplot [color=mycolor3, mark=asterisk, mark options={solid, mycolor3}, forget plot]
  table[row sep=crcr]{%
5	0\\
10	2\\
20	7\\
40	17\\
80	37\\
160	77\\
};

\addplot [color=mycolor4, mark=*, mark options={solid, mycolor4}, forget plot]
  table[row sep=crcr]{%
5	0\\
10	3\\
20	8\\
40	18\\
80	38\\
160	78\\
};

\addplot [color=mycolor5, mark=x, mark options={solid, mycolor5}, forget plot]
  table[row sep=crcr]{%
5	0\\
10	1\\
20	3\\
40	7\\
80	15\\
160	31\\
};

\addplot [color=mycolor6, mark=square, mark options={solid, mycolor6}, forget plot]
  table[row sep=crcr]{%
5	0\\
10	1\\
20	3\\
40	7\\
80	15\\
160	31\\
};

\addplot [color=mycolor7, mark=diamond, mark options={solid, mycolor7}, forget plot]
  table[row sep=crcr]{%
5	0\\
10	0\\
20	1\\
40	3\\
80	7\\
160	15\\
};

\addplot [color=mycolor1, mark=triangle, mark options={solid, mycolor1}, forget plot]
  table[row sep=crcr]{%
5	0\\
10	1\\
20	2\\
40	4\\
80	8\\
160	16\\
};

    \end{axis}
    \end{tikzpicture}%
}] at (pt) {};

\end{tikzpicture}%

%% file: figures/numNeigh_tss20.tex
%
%
\begin{tikzpicture}
\pgfplotsset{every tick label/.append style={font=\scriptsize}}

\begin{axis}[%
width=0.951\fwidth,
height=\fheight,
at={(0\fwidth,0\fheight)},
scale only axis,
xmin=1,
xmax=4,
xtick=data,
tick align=outside,
xlabel style={font=\scriptsize\color{white!15!black}},
xlabel={$N_{\rm symb, CSI}$},
ymin=0,
ymax=1,
ylabel style={font=\scriptsize\color{white!15!black}},
ylabel={$\rho$},
ytick={0, 0.1, 0.2, 0.5, 1},
zmin=0,
zmax=800,
zlabel style={font=\scriptsize\color{white!15!black}},
zlabel={N$_{\max, \rm neigh}$},
view={-120}{8},
axis background/.style={fill=white},
axis x line*=bottom,
axis y line*=left,
axis z line*=left,
xmajorgrids,
ymajorgrids,
zmajorgrids,
reverse legend,
legend style={font=\scriptsize, at={(0.1, 0.9, 0.9)}, anchor=south west, legend cell align=left, align=left, draw=white!15!black}
]

\addplot3[%
surf,
fill opacity=0.2, shader=flat corner, color=green, draw=black, z buffer=sort, mesh/rows=3]
table[row sep=crcr, point meta=\thisrow{c}] {%
x	y	z	c\\
1	0.072	31	1\\
1	0.1	24	1\\
1	0.144	14	1\\
1	0.2	11	1\\
1	0.4	4	1\\
1	0.5	4	1\\
1	1	1	1\\
2	0.072	15	1\\
2	0.1	11	1\\
2	0.144	6	1\\
2	0.2	5	1\\
2	0.4	1	1\\
2	0.5	1	1\\
2	1	0	1\\
4	0.072	7	1\\
4	0.1	5	1\\
4	0.144	2	1\\
4	0.2	2	1\\
4	0.4	0	1\\
4	0.5	0	1\\
4	1	0	1\\
};
\addlegendentry{Opt. 1, $T_{\rm CSI} = 0.625$~ms}

\addplot3[%
surf,
fill opacity=0.7, shader=flat corner, color=red, draw=black, z buffer=sort, mesh/rows=3]
table[row sep=crcr, point meta=\thisrow{c}] {%
x	y	z	c\\
1	0.072	779	1\\
1	0.1	599	1\\
1	0.144	359	1\\
1	0.2	299	1\\
1	0.4	119	1\\
1	0.5	119	1\\
1	1	59	1\\
2	0.072	389	1\\
2	0.1	299	1\\
2	0.144	179	1\\
2	0.2	149	1\\
2	0.4	59	1\\
2	0.5	59	1\\
2	1	29	1\\
4	0.072	194	1\\
4	0.1	149	1\\
4	0.144	89	1\\
4	0.2	74	1\\
4	0.4	29	1\\
4	0.5	29	1\\
4	1	14	1\\
};
\addlegendentry{Opt. 1, $T_{\rm CSI} = 10$~ms}

\draw[larrow] (1, 0.0864, 600) -- (1, 0.0864, 0);
\node[anchor=south east, align=right] at (1, 0.01, 595) (ltelabel) {\tiny Same bandwidth as \gls{ss}\\\tiny burst for $\Delta_f = 120$~kHz.};

\draw[larrow] (4, 0.2, 400) -- (1, 0.1728, 0);
\node[anchor=south east, align=right] at (4, 0.2, 400) (ltelabel) {\tiny Same bandwidth as \gls{ss}\\\tiny burst for $\Delta_f = 240$~kHz.};

\end{axis}
\end{tikzpicture}%

%% file: figures/omega5.tex
%
%
\definecolor{mycolor1}{rgb}{0.00000,0.44700,0.74100}%
\definecolor{mycolor2}{rgb}{0.85000,0.32500,0.09800}%
\definecolor{mycolor3}{rgb}{0.92900,0.69400,0.12500}%
\definecolor{mycolor4}{rgb}{0.49400,0.18400,0.55600}%
\begin{tikzpicture}
\pgfplotsset{every tick label/.append style={font=\scriptsize}}

\begin{axis}[%
width=0.951\fwidth,
height=\fheight,
at={(0\fwidth,0\fheight)},
scale only axis,
xtick=data,
xmin=5,
xmax=67,
xlabel style={font=\scriptsize\color{white!15!black}},
xlabel={$N_{\rm SS}$},
ymin=0,
ymax=0.45,
ylabel style={font=\scriptsize\color{white!15!black}},
ylabel={$\Omega_{\rm 5ms}$},
ylabel shift=-2pt,
xlabel shift=-3pt,
axis background/.style={fill=white},
xmajorgrids,
ymajorgrids,
legend style={font=\scriptsize, at={(0.01, 0.99)}, anchor=north west, legend cell align=left, align=left, draw=white!15!black}
]
\addplot [color=mycolor1, line width=1.2, mark=x, mark options={solid, mycolor1}]
  table[row sep=crcr]{%
8	0.00410976\\
16	0.00821952\\
32	0.01643904\\
64	0.03287808\\
};
\addlegendentry{$\Delta_f = 120$~kHz, $D = 0$}

\addplot [color=mycolor2, line width=1.2, dashed, mark=o, mark options={solid, mycolor2}]
  table[row sep=crcr]{%
8	0.00410976\\
16	0.00821952\\
32	0.01643904\\
64	0.03287808\\
};
\addlegendentry{$\Delta_f = 240$~kHz, $D = 0$}

\addplot [color=mycolor3, line width=1.2, mark=triangle, mark options={solid, rotate=180, mycolor3}]
  table[row sep=crcr]{%
8	0.04520736\\
16	0.09041472\\
32	0.18082944\\
64	0.36165888\\
};
\addlegendentry{$\Delta_f = 120$~kHz, $D = 1$}

\addplot [color=mycolor4, line width=1.2, dashed, mark=triangle, mark options={solid, mycolor4}]
  table[row sep=crcr]{%
8	0.0205488\\
16	0.0410976\\
32	0.0821952\\
64	0.1643904\\
};
\addlegendentry{$\Delta_f = 240$~kHz, $D = 1$}

\end{axis}
\end{tikzpicture}%

%% file: figures/omegatss.tex
%
%
\definecolor{mycolor1}{rgb}{0.00000,0.44700,0.74100}%
\definecolor{mycolor2}{rgb}{0.85000,0.32500,0.09800}%
\definecolor{mycolor3}{rgb}{0.92900,0.69400,0.12500}%
\definecolor{mycolor4}{rgb}{0.49400,0.18400,0.55600}%
\begin{tikzpicture}
\pgfplotsset{every tick label/.append style={font=\scriptsize}}

\begin{axis}[%
width=0.951\fwidth,
height=\fheight,
at={(0\fwidth,0\fheight)},
scale only axis,
xmin=0,
xmax=160,
xlabel style={font=\scriptsize\color{white!15!black}},
xlabel={$T_{\rm SS}$ [ms]},
ymin=0,
ymax=0.45,
ylabel style={font=\scriptsize\color{white!15!black}},
ylabel={$\Omega_{\rm T_{\rm SS}}$},
axis background/.style={fill=white},
xmajorgrids,
ymajorgrids,
legend style={font=\scriptsize, legend cell align=left, align=left, draw=white!15!black}
]
\addplot [color=mycolor1, line width=1.2, mark=x, mark options={solid, mycolor1}]
  table[row sep=crcr]{%
5	0.03287808\\
10	0.01643904\\
20	0.00821952\\
40	0.00410976\\
80	0.00205488\\
160	0.00102744\\
};
\addlegendentry{$\Delta_f = 120$~kHz, $N_{rep} = 0$}

\addplot [color=mycolor2, line width=1.2, dashed, mark=o, mark options={solid, mycolor2}]
  table[row sep=crcr]{%
5	0.03287808\\
10	0.01643904\\
20	0.00821952\\
40	0.00410976\\
80	0.00205488\\
160	0.00102744\\
};
\addlegendentry{$\Delta_f = 240$~kHz, $N_{rep} = 0$}

\addplot [color=mycolor3, line width=1.2, mark=triangle, mark options={solid, rotate=180, mycolor3}]
  table[row sep=crcr]{%
5	0.36165888\\
10	0.18082944\\
20	0.09041472\\
40	0.04520736\\
80	0.02260368\\
160	0.01130184\\
};
\addlegendentry{$\Delta_f = 120$~kHz, $N_{rep} =11$}

\addplot [color=mycolor4, line width=1.2, dashed, mark=triangle, mark options={solid, mycolor4}]
  table[row sep=crcr]{%
5	0.1643904\\
10	0.0821952\\
20	0.0410976\\
40	0.0205488\\
80	0.0102744\\
160	0.0051372\\
};
\addlegendentry{$\Delta_f = 240$~kHz, $N_{rep} = 5$}

\end{axis}
\end{tikzpicture}%

%% file: figures/omegacsi_tss20ms.tex
%
%
\begin{tikzpicture}
\pgfplotsset{every tick label/.append style={font=\scriptsize}}
\pgfplotsset{scaled z ticks=false}

\begin{axis}[%
width=0.951\fwidth,
height=\fheight,
at={(0\fwidth,0\fheight)},
scale only axis,
xmin=1,
xmax=4,
xtick=data,
tick align=outside,
xlabel style={font=\scriptsize\color{white!15!black}},
xlabel={$N_{\rm symb, CSI}$},
ymin=0,
ymax=1,
ylabel style={font=\scriptsize\color{white!15!black}},
ylabel={$\rho$},
ytick={0, 0.1, 0.2, 0.5, 1},
zmin=0,
zmax=0.06,
zlabel style={font=\scriptsize\color{white!15!black}},
zlabel={$\Omega_{\rm CSI}$},
view={-129}{19},
axis background/.style={fill=white},
axis x line*=bottom,
axis y line*=left,
axis z line*=left,
xmajorgrids,
ymajorgrids,
zmajorgrids,
zticklabel style={
        /pgf/number format/fixed,
        /pgf/number format/precision=2
},
scaled z ticks=false,
reverse legend,
legend style={font=\scriptsize, at={(0.6,0.7)}, anchor=south west, legend cell align=left, align=left, draw=white!15!black}
]

\addplot3[%
surf,
fill opacity=0.2, shader=flat corner, color = red, draw=black, z buffer=sort, mesh/rows=3]
table[row sep=crcr, point meta=\thisrow{c}] {%
x	y	z	c\\
1	0.072	0.00012843	1\\
1	0.1	0.000178375	1\\
1	0.144	0.00025686	1\\
1	0.2	0.00035675	1\\
1	0.4	0.0007135	1\\
1	0.5	0.000891875	1\\
1	1	0.00178375	1\\
2	0.072	0.00025686	1\\
2	0.1	0.00035675	1\\
2	0.144	0.00051372	1\\
2	0.2	0.0007135	1\\
2	0.4	0.001427	1\\
2	0.5	0.00178375	1\\
2	1	0.0035675	1\\
4	0.072	0.00051372	1\\
4	0.1	0.0007135	1\\
4	0.144	0.00102744	1\\
4	0.2	0.001427	1\\
4	0.4	0.002854	1\\
4	0.5	0.0035675	1\\
4	1	0.007135	1\\
};
\addlegendentry{$T_{\rm CSI} = 5$~ms}

\addplot3[%
surf,
fill opacity=0.8, shader=flat corner, color=green, draw=black, z buffer=sort, mesh/rows=3]
table[row sep=crcr, point meta=\thisrow{c}] {%
x	y	z	c\\
1	0.072	0.00102744	1\\
1	0.1	0.001427	1\\
1	0.144	0.00205488	1\\
1	0.2	0.002854	1\\
1	0.4	0.005708	1\\
1	0.5	0.007135	1\\
1	1	0.01427	1\\
2	0.072	0.00205488	1\\
2	0.1	0.002854	1\\
2	0.144	0.00410976	1\\
2	0.2	0.005708	1\\
2	0.4	0.011416	1\\
2	0.5	0.01427	1\\
2	1	0.02854	1\\
4	0.072	0.00410976	1\\
4	0.1	0.005708	1\\
4	0.144	0.00821952	1\\
4	0.2	0.011416	1\\
4	0.4	0.022832	1\\
4	0.5	0.02854	1\\
4	1	0.05708	1\\
};
\addlegendentry{$T_{\rm CSI} = 0.625$~ms}

\end{axis}
\end{tikzpicture}%

%% file: figures/omegatot_tss20ms.tex
%
%
\begin{tikzpicture}
\pgfplotsset{every tick label/.append style={font=\scriptsize}}

\begin{axis}[%
width=0.951\fwidth,
height=\fheight,
at={(0\fwidth,0\fheight)},
scale only axis,
xmin=1,
xmax=4,
xtick=data,
tick align=outside,
xlabel style={font=\scriptsize\color{white!15!black}},
xlabel={$N_{\rm symb, CSI}$},
ymin=0,
ymax=1,
ylabel style={font=\scriptsize\color{white!15!black}},
ylabel={$\rho$},
ytick={0, 0.1, 0.2, 0.5, 1},
zmin=0,
zmax=0.14,
zlabel style={font=\scriptsize\color{white!15!black}},
zlabel={$\Omega_{tot}$},
view={-129}{19},
axis background/.style={fill=white},
axis x line*=bottom,
axis y line*=left,
axis z line*=left,
xmajorgrids,
ymajorgrids,
zmajorgrids,
legend columns=2,
zticklabel style={
        /pgf/number format/fixed,
        /pgf/number format/precision=2
},
scaled z ticks=false,
legend style={font=\scriptsize, at={(0.5, 0.9)}, anchor=south, legend cell align=left, align=left, draw=white!15!black}
]

\addplot3[%
surf,
fill opacity=0.2, shader=flat corner, color=blue, draw=black, z buffer=sort, mesh/rows=3]
table[row sep=crcr, point meta=\thisrow{c}] {%
x	y	z	c\\
1	0.072	0.00911853	5\\
1	0.1	0.009468145	5\\
1	0.144	0.01001754	5\\
1	0.2	0.01071677	5\\
1	0.4	0.01321402	5\\
1	0.5	0.014462645	5\\
1	1	0.02070577	5\\
2	0.072	0.01001754	5\\
2	0.1	0.01071677	5\\
2	0.144	0.01181556	5\\
2	0.2	0.01321402	5\\
2	0.4	0.01820852	5\\
2	0.5	0.02070577	5\\
2	1	0.03319202	5\\
4	0.072	0.01181556	5\\
4	0.1	0.01321402	5\\
4	0.144	0.0154116	5\\
4	0.2	0.01820852	5\\
4	0.4	0.02819752	5\\
4	0.5	0.03319202	5\\
4	1	0.05816452	5\\
};
\addlegendentry{$\Delta_f = 240$~kHz, $D = 0$}

\addplot3[%
surf,
fill opacity=0.4, shader=flat corner, color=red, draw=black, z buffer=sort, mesh/rows=3]
table[row sep=crcr, point meta=\thisrow{c}] {%
x	y	z	c\\
1	0.072	0.0089901	3\\
1	0.1	0.00928977	3\\
1	0.144	0.00976068	3\\
1	0.2	0.01036002	3\\
1	0.4	0.01250052	3\\
1	0.5	0.01357077	3\\
1	1	0.01892202	3\\
2	0.072	0.00976068	3\\
2	0.1	0.01036002	3\\
2	0.144	0.01130184	3\\
2	0.2	0.01250052	3\\
2	0.4	0.01678152	3\\
2	0.5	0.01892202	3\\
2	1	0.02962452	3\\
4	0.072	0.01130184	3\\
4	0.1	0.01250052	3\\
4	0.144	0.01438416	3\\
4	0.2	0.01678152	3\\
4	0.4	0.02534352	3\\
4	0.5	0.02962452	3\\
4	1	0.05102952	3\\
};
\addlegendentry{$\Delta_f = 120$~kHz, $D = 0$}

\addplot3[%
surf,
fill opacity=0.8, shader=flat corner, color=yellow, draw=black, z buffer=sort, mesh/rows=3]
table[row sep=crcr, point meta=\thisrow{c}] {%
x	y	z	c\\
1	0.072	0.04199661	9\\
1	0.1	0.042346225	9\\
1	0.144	0.04289562	9\\
1	0.2	0.04359485	9\\
1	0.4	0.0460921	9\\
1	0.5	0.047340725	9\\
1	1	0.05358385	9\\
2	0.072	0.04289562	9\\
2	0.1	0.04359485	9\\
2	0.144	0.04469364	9\\
2	0.2	0.0460921	9\\
2	0.4	0.0510866	9\\
2	0.5	0.05358385	9\\
2	1	0.0660701	9\\
4	0.072	0.04469364	9\\
4	0.1	0.0460921	9\\
4	0.144	0.04828968	9\\
4	0.2	0.0510866	9\\
4	0.4	0.0610756	9\\
4	0.5	0.0660701	9\\
4	1	0.0910426	9\\
};
\addlegendentry{$\Delta_f = 240$~kHz, $D = 1$}

\addplot3[%
surf,
fill opacity=0.6, shader=flat corner, color=green, draw=black, z buffer=sort, mesh/rows=3]
table[row sep=crcr, point meta=\thisrow{c}] {%
x	y	z	c\\
1	0.072	0.0911853	5\\
1	0.1	0.09148497	5\\
1	0.144	0.09195588	5\\
1	0.2	0.09255522	5\\
1	0.4	0.09469572	5\\
1	0.5	0.09576597	5\\
1	1	0.10111722	5\\
2	0.072	0.09195588	5\\
2	0.1	0.09255522	5\\
2	0.144	0.09349704	5\\
2	0.2	0.09469572	5\\
2	0.4	0.09897672	5\\
2	0.5	0.10111722	5\\
2	1	0.11181972	5\\
4	0.072	0.09349704	5\\
4	0.1	0.09469572	5\\
4	0.144	0.09657936	5\\
4	0.2	0.09897672	5\\
4	0.4	0.10753872	5\\
4	0.5	0.11181972	5\\
4	1	0.13322472	5\\
};
\addlegendentry{$\Delta_f = 120$~kHz, $D = 1$}

\end{axis}
\end{tikzpicture}%

%% file: figures/kiviat_diagram.tex
\definecolor{mycolor1}{rgb}{0.00000,0.44700,0.74100}%
\definecolor{mycolor2}{rgb}{0.85000,0.32500,0.09800}%
\definecolor{mycolor3}{rgb}{0.92900,0.69400,0.12500}%
\definecolor{mycolor4}{rgb}{0.49400,0.18400,0.55600}%
\definecolor{mycolor5}{rgb}{0.46600,0.67400,0.18800}%
\definecolor{mycolor6}{rgb}{0.30100,0.74500,0.93300}

\pgfplotsset{
tick label style={font=\scriptsize},
label style={font=\scriptsize},
legend  style={font=\scriptsize}
}
\begin{tikzpicture}
\tkzKiviatDiagram[scale=0.3,label distance=.5cm,
        radial  = 3,
        gap     = 1,  
        lattice = 10]{\scriptsize Accuracy ($1/P_{\rm MD}$), \scriptsize Reactiveness ($1/T_{\rm IA}$), \scriptsize Overhead ($\Omega_{tot}$)}
\tkzKiviatLine[thick,color=mycolor1,
               fill=mycolor1,opacity=1](4.97930850353758, 0.772308100515353, 2.631578947368421);\label{44ana8}
\tkzKiviatLine[dashed,mark=x,
                 mark size=2pt,color=mycolor6,fill=mycolor6,opacity=0.6](6.98894510024358,	10,	10);\label{641ana64} 
\tkzKiviatLine[dashdotted,color=mycolor5,mark=asterisk,
               fill=mycolor5,opacity=.3](10, 0.38191628180247, 10);\label{6416dig64}


\node[anchor=north,xshift=0pt,yshift=-40pt] at (current bounding box.south) 
{
\scriptsize
\begin{tabular}{@{}ll@{}} 
\tikz\node[color=mycolor1, inner sep=0pt, opacity=1]{\rule{2ex}{2ex}}; & $M_{\gls{gnb}} = 4 \times M_{\gls{ue}} = 4$, analog beamforming, $N_{\rm SS}=8$ \\
\tikz\node[color=mycolor6, inner sep=0pt, opacity=0.6]{\rule{2ex}{2ex}}; & $M_{\gls{gnb}} = 64 \times M_{\gls{ue}} = 1$ (omni), analog beamforming, $N_{\rm SS}=64$ \\
\tikz\node[color=mycolor5, inner sep=0pt, opacity=0.3]{\rule{2ex}{2ex}}; & $M_{\gls{gnb}} = 64 \times M_{\gls{ue}} = 16$, digital beamforming at the \gls{ue}, $N_{\rm SS}=64$ \\
\end{tabular}
};
\end{tikzpicture}